\documentclass[onecolumn,authoryear]{els-mrw} 
\usepackage{amsmath,amssymb,amsfonts,amsthm,makeidx,graphicx}
\usepackage{appendix}
\usepackage{txfonts}
\usepackage{helvet}
\usepackage{braket}
\usepackage{comment}
\usepackage{subcaption}
\usepackage{multirow}
\usepackage{tensor}
\usepackage{hyperref}
\hypersetup{
    unicode=false,          
    colorlinks=true,       
    linkcolor=magenta,          
    citecolor=blue,        
    filecolor=magenta,      
    urlcolor=blue           
}
\begin{document}

\chapter{Modified theories of gravity at different curvature scales}\label{MGravity:SS}

\author[1]{Susobhan Mandal}%
\author[2]{S. Shankaranarayanan}%


\address[1]{\orgname{Indian Institute of Technology Bombay}, \orgdiv{Department of Physics}, \orgaddress{Mumbai}}
\address[2]{\orgname{Indian Institute of Technology Bombay}, \orgdiv{Department of Physics}, \orgaddress{Mumbai}}

\articletag{Chapter Article tagline: update of previous edition,, reprint..}

\maketitle

\begin{glossary}[Glossary]

\term{Einstein's equivalence principle (EEP)} states that locally, the effects of gravity 
are indistinguishable from acceleration, meaning free-falling observers cannot 
distinguish between uniform gravitational fields and acceleration in the absence 
of gravity.

\term{Black holes} regions of spacetime where the curvature becomes infinitely steep, 
leading to a singularity where gravitational forces are so intense that not even light 
can escape.

\term{Late time accelerating expansion of Universe} driven by dark energy, a hypothetical 
form of energy that causes the rate of expansion to increase over time.

\term{Violations of EEP} are motivated by attempts to unify General Relativity with quantum mechanics or explain phenomena like dark energy and dark matter. Testing for EEP violations could reveal new interactions, particles, or fundamental forces beyond the standard framework of physics, providing insights into the nature of gravity at extreme scales.

\term{Modified theories of gravity} extend or alter Einstein's General Relativity to address
phenomena such as dark energy, dark matter, or inconsistencies at extreme scales like those of
black holes or the early universe. These theories, including \(f(R)\) gravity, quadratic gravity, scalar-tensor theories, and dynamical Chern-Simons gravity, aim to provide insights into unresolved questions about the universe's structure and evolution.
\end{glossary}

\begin{glossary}[Nomenclature]
\begin{tabular}{@{}lp{34pc}@{}}
BH & Black Hole\\
CMB & Cosmic Microwave Background\\
CPT & Charge, Parity, and Time reversal\\
DE & Dark Energy\\
DM & Dark Matter\\ 
DoF & Degrees of Freedom\\
EEP & Einstein's Equivalence Principle\\
EP & Equivalence Principle\\
EFT & Effective Field Theory\\
EHT & Event Horizon Telescope\\
EM & Electromagnetic\\
FRB & Fast Radio Bursts\\
FLRW & Friedmann-Lema\^itre-Robertson-Walker\\
GRB & Gravitational Radio Bursts\\
GR & General Relativity\\
GUP & Generalized Uncertainty Principle\\
GW & Gravitation Wave\\
HL  & Horava-Lifshitz  \\
$\Lambda$CDM & Lambda Cold Dark Matter\\
LLI & Local Lorentz Invariance \\
LPI & Local Position Invariance\\
NEC & Null Energy Condition\\
PBH & Primordial Black Hole \\
PGW & Primordial Gravitational Waves\\
QNM & Quasi Normal-Mode\\
SEP & Strong Equivalence Principle \\
TeVeS & Tensor-Vector-Scalar \\
WEP & Weak Equivalence Principle \\
\end{tabular}
\end{glossary}

\begin{abstract}[Abstract]
General Relativity (GR) remains the cornerstone of gravitational physics, providing remarkable success in describing a wide range of astrophysical and cosmological phenomena. However, several challenges underscore the urgent need to explore modified gravity theories. GR struggles to reconcile with quantum mechanics, fails to provide fundamental explanations for dark matter and dark energy, and faces limitations in describing extreme regimes such as black hole singularities and the very early universe. Furthermore, some observations at cosmological and astrophysical scales hint at potential deviations from GR, emphasizing the importance of theoretical extensions. 
This review provides an organized perspective on modified gravity theories by classifying them based on the principles of GR they preserve or violate. Specifically, we consider three broad categories: (1) metric theories that uphold local Lorentz invariance (LLI) and gauge invariance, (2) theories that break gauge invariance, LLI, or parity, and (3) beyond-metric theories that violate the Einstein's equivalence principle (EEP). This classification highlights the underlying assumptions of GR that these theories challenge or extend, providing a framework for understanding their motivations and implications.
The review also discusses the current and upcoming experimental and observational tests of GR, including those probing its foundational principles, such as LLI, gauge invariance, and EEP. For each class of modified theories, we examine their ability to address critical open questions in cosmology and black hole physics. These include their potential to explain the accelerated expansion of the current universe, the nature of dark matter, and deviations in black hole dynamics from GR predictions. This review aims to provide a structured understanding of modified gravity theories and their observational implications in the multimessenger era by focusing on the principles preserved or violated.
\end{abstract}

\textbf{Key points}
\\[1pt]
\begin{itemize}
\item Need for Modifications in GR at different curvature scales
\begin{enumerate}
\item Limitations of GR in extreme conditions (strong and weak curvature regimes).  
\item Motivations for exploring beyond GR: singularities, dark matter, and dark energy.  
\item Observational discrepancies and theoretical gaps in GR's framework. \\
\end{enumerate}

\item Testing Foundational Principles of GR at different curvature scales  
\begin{enumerate}
\item   Equivalence Principle tests across scales: weak, intermediate, and strong gravity regimes.  
\item Observational constraints on GR principles using laboratory, astrophysical, and cosmological systems. \\ 
\end{enumerate}

\item Metric theories of gravity preserving both LI and Gauge Invariance
\begin{enumerate}
\item Higher-derivative gravity theories (e.g., \( f(R) \), Quadratic and Lovelock gravity).  
\item Impacts of higher-curvature terms on gravitational dynamics.  
\item Examples of consistent modifications and their phenomenology. \\
\end{enumerate}

\item Theories of gravity breaking Lorentz or Gauge or parity invariance   
\begin{enumerate}
\item Theoretical motivations and construction of such models (e.g., Einstein-Aether, Horava-Lifshitz and Chern-Simons gravity).  
\item Implications for preferred frames, causality, and observational tests.  \\ 
\end{enumerate}

\item Beyond Metric Theories of Gravity 
\begin{enumerate}
\item Scalar-tensor, TeVeS theories
\item Challenges in experimental verification and compatibility with observations. \\
\end{enumerate} 

\item Tests of Modified Theories of Gravity Using Multimessenger Astronomy   \\

\item Implications of Modified Theories of Gravity in BH Physics and Late-Time Cosmology  
\end{itemize}

\section{Introduction}\label{sec:Intro}
On October 27, 1930, in his tribute to Albert Einstein, George Bernard Shaw remarked: "Ptolemy made a universe which lasted 1,400 years. Newton also made a universe which has lasted 300 years. Einstein has made a universe, and I cannot tell you how long that will last"~\citep{Pais-Book}. 
By comparing Einstein's work with Ptolemy and Newton, Shaw underscores the monumental shifts each brought to our comprehension of the cosmos. While Newtonian physics worked well at solar system scale, Einstein's insights into the fabric of spacetime fundamentally altered our understanding of the Universe, especially in extreme conditions like black holes (BHs) and the expanding Universe. Shaw acknowledges that, like previous theories, Einstein's Universe may be surpassed one day, but at the time of his tribute and after one century, General Relativity (GR) represents the most advanced and groundbreaking scientific understanding of the cosmos~\citep{MTW,1973-Hawking.Ellis-Book,2010-Padmanabhan-GravitationFoundationsFrontiers,Book-Schutz}.

Besides the above fascinating predictions, GR demonstrates striking concurrence with a wide array of precision tests of Solar System gravity, including gravitational redshift, gravitational lensing of light from distant background stars, anomalous perihelion precession of Mercury, Shapiro time-delay effect, and Lunar laser experiments. These tests remain valid within weak gravitational fields, such as near the Sun's surface~\citep{Will:2014kxa}.

Outside our solar system, current efforts are underway to substantiate predictions of GR by observing the alterations in the orbits of binary pulsars emitting gravitational waves (GWs) and BH mergers. Testing gravity through GW observations has emerged as a powerful method to probe the nature of gravity in extreme conditions and to explore possible deviations from GR~\citep{Sathyaprakash:2009xs,Barack:2018yly,LISA:2022kgy}. LIGO-VIRGO-KAGRA's detection of GWs from binary BH mergers and neutron star collisions has opened new avenues to test GR in regimes where traditional methods, such as solar system or cosmological tests, are less effective~\citep{LIGOScientific:2020tif,LIGOScientific:2021sio}.

The first three observing runs of Advanced LIGO and Virgo detectors have detected close to 100  coalescing binary merger events~\citep{LIGOScientific:2018mvr,LIGOScientific:2021sio,KAGRA:2021vkt,LIGOScientific:2021usb}. Also recently, the Event Horizon Telescope (EHT) collaboration provided 
the first direct visual confirmation of the existence of a Super-massive black hole in M87~\citep{EHT2019I} and compelling evidence was provided that the central object in M87 behaved as predicted by GR~\citep{EHT2019IV}. The size and shape of the shadow were consistent with predictions based on GR for a BH of $6.5 \times 10^9 M_{\odot}$~\citep{EHT2019VI}. Interestingly, the EHT collaboration could also image the super-massive BH in our galaxy~\citep{EHT2022I}, with a mass of around $4 \times 10^6 M_{\odot}$~\citep{EHT2022VI}.

These observations are well predicted by GR with statistical significance. Furthermore, the evidence strongly suggests that the object is an extremely compact one, consistent with the characteristics of a BH. However, the existence of infinite spacetime curvature in GR indicates that it cannot be a universal theory of spacetime~\citep{2022-Shanki.Joseph-GRG}. Furthermore, GR is neither part nor compatible with the quantum field theory that forms the basis of the standard model of particle physics. There is a broad consensus that these two standard frameworks are fundamentally incompatible. Quantum exchanges of virtual particles explain the strong, weak, and electromagnetic (EM) forces. In contrast, gravity is classically described as the curvature of spacetime caused by matter and energy. Given that quantum field theory cannot incorporate gravity and GR predicts singularities at the center of BHs, either framework is unlikely to be considered fundamental~\citep{Burgess:2003jk,Donoghue:2017pgk}.

On the largest observable scales, the biggest surprise from observational cosmology has been the acceleration of the Hubble expansion~\citep{SupernovaSearchTeam:1998fmf,SupernovaCosmologyProject:1998vns,WMAP:2003elm}. The late-time acceleration of the Universe can be explained by the presence of an exotic matter source referred to as dark energy~\citep{2003-Padmanabhan-PRep,2003-Peebles.Ratra-RMP,Copeland:2006wr,Huterer:2017buf,Kamionkowski:2022pkx}. However, it also provides an intriguing possibility to test gravity on cosmological scales and investigate modified theories of gravity~\citep{Nojiri:2006ri,Woodard:2006nt,Alexander:2009tp,Sotiriou:2008rp,Capozziello:2009dz,Maartens:2010ar,DeFelice:2010aj,Capozziello:2011et,Nojiri:2010wj,Joyce:2014kja,2022-Shanki.Joseph-GRG}.

The Lambda Cold Dark Matter (\( \Lambda \text{CDM} \)) model is the most widely accepted cosmological framework describing the Universe's large-scale structure and its evolution from the Big Bang to the present day~\citep{2000-Padmanabhan-TheoreticalAstrophysicsVolume, 2005-Mukhanov-PhysicalFoundationsCosmology, 2008-Weinberg-Cosmology}. However, this  is based on extrapolating local laws to the Universe as a whole~\citep{Coley:2019yov,Scott:2018adl}. \( \Lambda \text{CDM} \) is a six-parameter model that incorporates the following key features:  \\
\begin{enumerate}
\item Cosmological Constant (\( \Lambda \)) that accounts for $\sim 70\%$ of the total energy density of the Universe. $\Lambda$ is responsible for the current acceleration of the Universe.

\item CDM a non-relativistic, pressureless matter that makes up around $25\%$ of the total energy density of the Universe.  CDM is responsible for the formation of galaxies and larger structures via gravitational clustering.
\item Baryonic Matter (protons, neutrons, and electrons) and Radiation (neutrinos and photons) that constitutes around $5\%$ of the Universe.
\item Cosmic Microwave Background (CMB) radiation which is remnant of the early Universe. 
The \( \Lambda \text{CDM} \) model accurately matches the anisotropy spectrum of the CMB~\citep{WMAP:2003elm,Planck:2018jri}. 
\item Predicts a spatially flat Universe based on observations of the CMB and large-scale structures.
\item Hierarchical growth of structures, where small objects form first and merge to create larger systems, driven by dark matter. \\
\end{enumerate}
Despite its success in explaining a wide range of cosmological observations, \( \Lambda \text{CDM} \) has unresolved issues, like explaining the nature of dark energy, the unknown nature and composition of CDM, and  tensions with observations~\citep{2016-Riess.others-Astrophys.J.,2019-Riess.etal-Astrophys.J.,Schoneberg:2021qvd,2022-Perivolaropoulos.Skara-NewAstron.Rev.}. For instance, there is a persistent discrepancy between the value of the Hubble constant (\( H_0 \)) derived from CMB observations and local measurements and observations suggest a slower growth of large-scale structure than predicted by \( \Lambda \text{CDM} \)~\citep{2021-DiValentino.etal-Class.Quant.Grav.,2022-Perivolaropoulos.Skara-NewAstron.Rev.}. These tensions highlight potential gaps in the $\Lambda$CDM model and raise the possibility of modifications to dark matter properties or new forms of dark energy, or the need for modifications to GR, or new physics in the early Universe or even new particle physics models~\citep{2024-Wang-Rept.Prog.Phys.,2024-Giare-}.

Thus, astrophysics, cosmology, and gravity confront pressing questions~\cite{Rees:2022znv,Elizalde:2020kpn,2022-Melia-Review}: What drives the universe's accelerated expansion? What constitutes dark matter? How did the first cosmic structures emerge? How are supermassive BHs formed? What triggered magnetogenesis and baryon asymmetry? 
These questions are central to the \href{https://baas.aas.org/astro2020-science}{Decadal Survey on Astronomy and Astrophysics 2020}, which outlines ambitious projects spanning various electromagnetic (EM) bands,  gravitational waves (GWs), cosmic rays and neutrinos. By combining cutting-edge technology and multimessenger approaches, these initiatives will decode the secrets of black holes, neutron stars, and the universe's earliest moments, addressing fundamental problems in astrophysics, cosmology and gravity.

These upcoming missions will generate \emph{petabytes of data daily}, equivalent to digitizing billions of pages or filling millions of hard drives. While computational tools are advancing to handle such vast datasets~\cite{Moriwaki:2023sdh,Olvera:2021jlq}, the real challenge lies in uncovering the physics behind the data. \emph{What stories does this data tell about the universe? What new phenomena can it help us discover and explain?}  One compelling and essential research direction is testing gravity across both strong-field and cosmological scales, combining theoretical and observational approaches to develop modified gravity models with testable predictions. Such models could resolve some of the most pressing issues in fundamental physics.

This review focuses on modified gravity theories and their implications in BHs and cosmology. Although it requires a relativistic theory of gravity, there are differences in terms of curvature and the length scales. In order to understand the differences and applicability the review will take a slightly different approach --- defining quantifying tools and length scales in relativistic gravity.  Unlike other reviews in the field, which largely focus on theoretical models, this review discusses the key questions that any modified gravity theories must address for the next generation experiments: First, how to test the foundational principles of GR via multimessenger
astronomy, using both EM and gravitational wave observations. Second, examining the range of compact objects with similar curvature to analyze shared characteristics among these systems to identify the key feature of GR that can be common among all these systems. 

This review provides an organized perspective on modified gravity theories by classifying them based on the principles of GR they preserve or violate. Specifically, we consider three broad categories: (1) metric theories that uphold local Lorentz invariance (LLI) and gauge invariance, (2) theories that break gauge invariance, LLI, or parity, and (3) beyond-metric theories that violate the Einstein's equivalence principle (EEP). This classification highlights the underlying assumptions of GR that these theories challenge or extend, providing a framework for understanding their motivations and implications.

The review is organized as follows: Section \eqref{sec:GRreview} provides a concise overview of GR and its foundational principles. Section \eqref{sec:Why?} addresses the critical question: \emph{Why go beyond GR?} It uses Einstein's equations to identify potential limitations through two parameters --- the characteristic curvature scale \({\cal R}\) and the compactness parameter \(\Phi\). Section \eqref{sec:testing} explores various tests of GR and its principles, highlighting how these tests can indicate the need for modifications to GR. 
Section \eqref{sec:MGTheories-Classi} introduces a framework for classifying modified gravity theories. Section \eqref{sec:MGModels-ClassA} focuses on metric theories that preserve local Lorentz invariance (LLI) and gauge invariance, discussing their implications for Inflation and BHs. Section \eqref{sec:MGModels-ClassB} examines theories that violate gauge invariance, LLI, or parity, emphasizing their observational consequences. Section \eqref{sec:MGModels-ClassC} briefly covers beyond-metric theories that break the Einstein equivalence principle (EEP) and their potential to explain key cosmological phenomena. 
The review concludes with a forward-looking discussion of the prospects for these theories, informed by upcoming experiments and observations. The four appendices contain the details of some of the calculations for easy access. 

We use the $(-,+,+,+)$ signature for the 4-D spacetime metric. Lower-case Greek (Latin) alphabets denote the 4-D spacetime (3-space) coordinates. We set $\kappa^2 = 8 \pi G/c^4$ and Planck's constant ($\hbar$) to unity. An overdot corresponds to derivative w.r.t time, $t$. 

\section{A rapid review of General relativity: Foundations and Principles}
\label{sec:GRreview}

This section provides a rapid review of GR, focusing on principles required for discussing modified gravity theories in subsequent sections.

The \emph{principle of Relativity}, which is the foundation of Special Relativity, asserts that no particular frame of reference or state of \emph{uniform motion} is privileged. Physical theories must be represented by equations that are form invariant under Poincare transformations (translations, rotations, and boosts), implying that these theories must be constructed out of physical quantities which have well-defined transformation properties. 
Special Relativity is elegantly described within a four-dimensional spacetime framework, where the usual three spatial dimensions are unified with the temporal dimension. In GR, this principle was extended to a new principle, the \emph{principle of general covariance} --- there is no preferred coordinate system~\citep{Will:2018bme}.

Einstein proposed that gravity is the manifestation of the curvature of spacetime caused by the presence of matter. This idea is rooted in \emph{the Einstein's equivalence principle} (EEP), which states that: ``The outcome of any local non-gravitational experiment in a freely falling laboratory is independent of the velocity of the laboratory and its position in spacetime"~\citep{Pais-Book}.  EEP implies that non-gravitational phenomena remain unaffected by gravity when observed in a freely falling frame. A significant implication of EEP is that all entities, including light, adhere to the same laws. Consequently, EEP serves as the intermediary between gravity and other branches of physics. Unlike the gauge invariance of the EM field, EEP is not regarded as a fundamental symmetry but rather an empirical fact~\citep{2001-Damour-Talk,Jackson:2001ia}. 

Initially termed the equivalence hypothesis by Einstein, it was later elevated to the status of a principle due to its pivotal role in generalizing special relativity to encompass gravitation. Given the stringent constraints of modern tests, the fact that EEP is satisfied is a surprising and remarkable observation. GR has been thoroughly tested within the solar and stellar systems, but its validity beyond this scale still needs verification. Examining EEP is an excellent way to test GR. Therefore, any experimental evidence indicating a violation of the equivalence principle would also serve as evidence against GR. Additionally, local Lorentz invariance, a key component of GR, implies charge, parity, and time reversal (CPT) symmetry~\citep{1951-Schwinger-PR}. However, evaluating the Equivalence Principle on cosmic scales presents significant challenges. See the discussion in Sec.~\eqref{sec:testing}.

Gravity being a geometric phenomenon, a relativistic gravity theory describing it should be formulated in terms of differential geometry. Suppose spacetime is modeled as a pseudo-Riemannian manifold $\mathcal{M}$ with a metric $g_{\mu\nu}$. The metric that determines the infinitesimal distance between two spacetime points is described by the invariant quantity:
\begin{equation}\label{metric def}
ds^{2} = g_{\mu\nu}dx^{\mu}dx^{\nu}.    
\end{equation}
Its geometry resembles Minkowski spacetime in infinitesimally small regions, while its curvature becomes evident over finite regions. 
This is consistent with the EEP, which states that in infinitesimal regions $g_{\mu\nu} \to \eta_{\mu\nu}$ (Minkowski metric); however, due to the non-zero gravitational field, tidal forces emerge over larger regions. GR formalizes the gravitational action arising from spacetime geometry using curvature invariants. The Einstein-Hilbert action is:
\begin{equation}\label{E-H action}
S_{\rm E-H} = \frac{1}{2 \kappa^2}\int d^{4}x\sqrt{-g} \, \left(R - 2 \Lambda \right) + S_{(M)}
\end{equation}
where $g$ refers to the determinant of the metric tensor, $R$ is the Ricci scalar, $\Lambda$ is the cosmological constant, and $S_{(M)}$ corresponds to the action describing the matter dynamics on the spacetime manifold. As EEP demands, the matter action includes the \emph{universal minimal coupling} to the metric tensor~\citep{MTW}. Minimal coupling of the matter action refers to introducing a covariant integration measure $\mathrm{d}^4 x \rightarrow \mathrm{~d}^4 x \sqrt{-g}$ and promoting partial derivatives to covariant derivatives $\partial_\mu \rightarrow \nabla_\mu$. Varying the above action w.r.t the metric tensor leads to Einstein's equations: 
\begin{equation}
\label{Einstein's field equations}
G_{\mu\nu} + \Lambda g_{\mu\nu} = \kappa^2 \, T_{\mu\nu}^{(M)} \, ,  
\end{equation}
where $T_{\mu\nu}^{(M)}$ is the energy-momentum tensor of the matter field, given by: 
\begin{equation}
\label{def:StressTensor}
T_{\mu\nu}^{(M)} = - \frac{2}{\sqrt{-g}}\frac{\delta S_{\text{matter}}}{\delta g^{\mu\nu}}, 
\end{equation}
and $G_{\mu\nu}$ is the Einstein tensor given by
\begin{equation}
\label{def:EinsteinTensor}
G_{\mu\nu} = R_{\mu\nu} - \frac{1}{2}g_{\mu\nu}R \, .    
\end{equation}
$R_{\mu\nu}$ is the Ricci tensor of the geometry and the relation between Ricci tensor and Ricci scalar is given by $R = g^{\mu\nu}R_{\mu\nu}$. The Ricci tensor can be constructed out of the Riemann curvature tensor $R_{ \ \nu\rho\sigma}^{\mu}$ via the relation $R_{\mu\nu} = R_{ \ \mu\rho\nu}^{\rho}$, where,
\begin{equation}\label{Riemann curvature tensor def}
R_{ \ \beta\gamma\delta}^{\alpha} = \partial_{\gamma}\Gamma_{ \ \beta\delta}^{\alpha} - \partial_{\delta}\Gamma_{ \ \beta\gamma}^{\alpha} + \Gamma_{ \ \gamma\lambda}^{\alpha}\Gamma_{ \ \beta\delta}^{\lambda} - \Gamma_{ \ \delta\lambda}^{\alpha}\Gamma_{ \ \beta\gamma}^{\lambda}.    
\end{equation}
In the above equation, the right-hand side involves the Christoffel symbols, which are given by:
\begin{equation}\label{Christoffel symbols def}
\Gamma_{ \ \mu\nu}^{\rho} = \frac{1}{2}g^{\rho\lambda}[\partial_{\mu}g_{\lambda\nu} + \partial_{\nu}g_{\lambda\mu} - \partial_{\lambda}g_{\mu\nu}].    
\end{equation}
The Einstein's field equations \eqref{Einstein's field equations} are a set of $10$ non-linear, coupled partial differential equations. These equations dynamically determine the ten metric components and the corresponding matter fields for an initial matter field configuration. Since the Einstein field equations are quasi-linear, the same initial condition can have multiple solutions. Metrics, however, are not physically measurable quantities and multiple metrics could describe a single physical spacetime, each using a different set of coordinates. This implies that the action \eqref{Einstein's field equations} is diffeomorphism invariant. In  GR, the Bianchi identities lead automatically to the conservation of the energy–momentum tensor ( $\nabla^{\mu} T_{\mu\nu}^{(M)} = 0$). This is connected to the invariance of the theory by general diffeomorphisms $\nabla^{\mu} G_{\mu\nu} = 0$~\citep{MTW}.
The above relations imply that out of 10 metric components, only $6$ are the true dynamical variables of the theory\footnote{The Bianchi identity in GR is analogous to $\nabla_{\mu} F^{\mu\nu} = 0$ in Maxwell's Electrodynamics. Similarly the energy-momentum conservation in GR is analogous to the current conservation relation $\nabla_{\mu} J^{\mu} = 0$.}.

\section{What is the need to go beyond GR?}
\label{sec:Why?}

In the Introduction \eqref{sec:Intro}, we outlined both the achievements and limitations of GR. The current understanding strongly suggests that GR breaks down 
in the regime of  extremely strong gravitational fields. As a classical, geometric theory of spacetime, GR predicts infinite matter densities and curvatures in two specific scenarios. 
The Oppenheimer–Snyder equations, which describe the collapse of a dust cloud, predict the   BH formation with a singularity at its core~\citep{PhysRev.56.455}. Similarly, tracing the Friedmann equation --- which governs the evolution of a homogeneous and isotropic universe --- backward in time leads inevitably to a singularity, often referred to as the Big Bang. These singularities are widely regarded as unphysical and strongly point to the necessity of extending GR to a more comprehensive framework.

To assess the necessity of theories beyond GR and identify when such an extension is required, \emph{quantifiable tools} are essential. However, unlike other fundamental forces, the framework and foundations of GR make it challenging to define quantifiable tools similar to those in particle physics.
First, GR is a covariant theory with no inherently physical coordinate system. A particularly useful class of coordinates for Earth-based measurements is known as normal coordinates~\citep{MTW}, which serve as the closest analog to inertial coordinates in flat space. These coordinates define a locally inertial frame, and they can typically be constructed over regions comparable to the curvature scale of spacetime. Due to the equivalence principle, normal coordinates are always obtainable near any point in spacetime, preserving much of our flat-space intuition within this local frame.

Second, in GR, the relationship between length and energy scales is more complex. In particle physics, energy ($E$) and length scales $(\ell)$ are directly related by $\ell =\hbar c/E$, meaning a particle with rest energy $E$ cannot be localized within a region smaller than $\ell$. This relation does not apply in GR due to the non-linearity of Einstein's equations. Therefore, adding two gravitational fields and their sources—each obeying Einstein's equations—does not yield a straightforward combination of fields and sources~\citep{MTW}. As a result, GWs themselves contribute further to the gravitational field~\citep{1985-Braginsky-JETP,2009-Favata-ApJL,Chakraborty:2024ars}.

Interestingly, as we show, the Einstein's field equations themselves provide quantifiable tools to identify when such an extension to GR is required. Taking the trace of Einstein's equations \eqref{Einstein's field equations}, we get:
\begin{equation}
\label{eq:TraceEinstein}
R - 4 \Lambda = - \kappa^2 \, T^{(M)} \, , 
\end{equation}
where $T^{(M)} = g^{\mu\nu} T^{(M)}_{\mu\nu}$ represents the trace of the energy-momentum tensor. To explore quantifiable tools in this context, let us consider the scenario where the energy-momentum tensor includes only slow-moving objects. In this case, the trace is mainly determined by the energy density, as the pressure terms, being quadratic in velocity, are relatively small. Assuming $\Lambda = 0$, if the energy density is positive (as it is for ordinary matter), this indicates a negative scalar curvature for the resulting spacetime. The energy density of ordinary matter with total mass $M$ and size $L$ is given by $M c^2/L^3$, allowing us to define a \emph{characteristic curvature scale} ${\cal R}$ for such a matter:
\begin{equation}
\label{def:Curv}
R \equiv {\cal R}^2 \sim \frac{G}{c^4} \times \frac{M  c^2}{L^3} \sim \frac{G M}{c^2 L^3} 
\end{equation}
Physically, \({\cal R}^2\) represents the curvature of spacetime around an isolated compact object of mass $M$ and size $L$. We can also define the dimensionless compactness parameter $\Phi$ as follows~\citep{Buchdahl:1959zz}:
\begin{equation}
\label{def:Compactness}
\Phi = \frac{G M}{c^2 L} \, .
\end{equation}
The compactness parameter ($\Phi$) is the ratio of the gravitational potential experienced by a test particle in the presence of mass $M$ to its rest mass energy.  
\(\Phi \rightarrow 0\) corresponds to flat Minkowski spacetime, while \(\Phi \ll 1\) aligns with weak gravitational fields consistent with Newtonian gravity. By contrast, the strongest observable gravitational fields approach the limit \(\Phi \rightarrow \frac{1}{2}\), near the event horizon of a BH.
While the compactness parameter is useful for post-Newtonian expansions, it does not fundamentally describe gravitational fields in Einstein’s theory. The field equations in GR, or the Einstein–Hilbert action, rely on the Riemann tensor to quantify the curvature, not the gravitational potential. Hence, \({\cal R}\) as defined in Eq.~\eqref{def:Curv} provides insight into conditions where extensions of GR may be necessary.

To illustrate this, consider two cases --- BHs and cosmology. First, we consider a spherical symmetric non-rotating BH in an asymptotically flat spacetime, known as the Schwarzschild BH. For such a compact object of mass $M$, the characteristic length scale $L$ is $r_H = 2 GM/c^2$. The compactness parameter \(\Phi\) remains constant for BHs of any mass, but \({\cal R}\) varies with the BH mass. Although neutron stars are less massive than astrophysical BHs, their 
compactness is low, thus probing smaller curvature scales. As shown in Table \eqref{Table:BH}, \({\cal R}\) is considerably smaller for a supermassive BH at the center of a galaxy than for an Earth-sized BH. This implies that the same astrophysical object can have drastically different curvatures depending on its mass and length, which is a distinct feature of relativistic gravity.
Now, let us turn our attention to cosmology. Expressing ${\cal R}$ in terms of energy density, we get:
\begin{equation}
\label{def:R-Cosmo}
{\cal R} = \sqrt{\frac{G \, \rho}{c^4} } 
\end{equation}
For the flat Friedmann-Lema\^itre-Robertson-Walker (FLRW) line-element, the first Friedmann equation is~\citep{2000-Padmanabhan-TheoreticalAstrophysicsVolume,2008-Weinberg-Cosmology}
\begin{equation}
\label{eq:Friedmann}
H^2(t) = \kappa^2 \rho(t) \, , 
\end{equation}
where $H(t) = \dot{a}(t)/a(t)$ is the Hubble parameter, $a(t)$ is the scale factor and $\rho(t)$ is the energy density of the perfect fluid associated with the matter, radiation, and cosmological constant. In-terms of the Hubble parameter, ${\cal R}$ becomes:
\begin{equation}
{\cal R} = \frac{H(t)}{c} \, .
\end{equation}
Assuming the standard cosmology with a dark energy equation of state $w=-1$, the Hubble parameter (in-terms of redshift $z$) is given by~\citep{2000-Padmanabhan-TheoreticalAstrophysicsVolume,2008-Weinberg-Cosmology}:
\begin{equation}
H(z) = {H_0} \sqrt{\Omega_{m,0} (1 + z)^{3} + \Omega_{r,0} (1 + z)^4 + \Omega_\Lambda} \,
\label{eq:HubbleRelation}
\end{equation}
where $1+z=1/a(t)$. $\Omega_{\mathrm{m}}, \Omega_{\mathrm{r}}$ and $\Omega_{\Lambda}$ are the density parameters for the matter (cold dark matter and baryons), radiation and the cosmological constant, and $H_{0}$ represents the current value of the Hubble constant. As shown in Table \eqref{Table:FRW}, ${\cal R}$ increases for higher redshifts. In the early Universe, ${\cal R} \sim 1$ or higher.  Thus, different epochs of the Universe corresponds to different curvature scales and, in principle, enables us to test gravity at different curvatures.
\begin{table}
\parbox{.45\linewidth}{
\centering
\begin{tabular}{|c|c|c|}
\hline
& &  \\[-1.3em]
Black hole &  $r_H$  (m) & ${\cal R}(r_H)$\\
\hline 
& &  \\[-0.8em]
$10^6 M_{\odot}$ & $10^{9}$  & $10^{-9}$   \\
& &  \\[-1.0em]
\hline 
& &  \\[-0.8em]
$M_{\odot}$ & $10^{3}$  & $10^{-3}$   \\
& &  \\[-1.0em]
\hline 
& &  \\[-0.8em]
$10^{-3} M_{\odot}$  & $1$ & $1$  \\
& &  \\[-1.0em]
\hline 
& &  \\[-0.8em]
$10^{-5} M_{\odot}$  &  $10^{-2}$ &  $10^2$\\
& &  \\[-1.0em]
\hline 
& &  \\[-0.8em]
$10^{-10} M_{\odot}$ & $10^{-7}$ & $10^{7}$  \\
\hline
\end{tabular}
\caption{\label{Table:BH} The value of the characteristic curvature scale ${\cal R}$ for different BH masses starting from supermassive BHs $(10^6 M_{\odot})$ in the centre of the galaxy to micro BHs $(10^{-10} M_{\odot})$.}
}
\hfill
\parbox{.45\linewidth}{
\centering
\begin{tabular}{|c|c|c|}
\hline
&   \\[-1.3em]
$z$  &  ${\cal R}(z) (m^{-1})$\\
\hline 
&   \\[-0.8em]
$z = 0 $ &  $10^{-27}$   \\
& \\[-0.8em]
\hline 
&  \\[-0.8em]
$z = 10$ & $10^{-26}$   \\
&   \\[-0.8em]
\hline 
&   \\[-0.8em]
$100$ & $10^{-24}$  \\
\hline 
& \\[-0.8em]
$10^4$ & $10^{-21}$  \\
\hline 
\end{tabular}
\caption{\label{Table:FRW} The value of ${\cal R}$ at different redshifts, $z$.}
}
\end{table}
The second Friedmann equation and the conservation equation of perfect fluid~~\citep{2000-Padmanabhan-TheoreticalAstrophysicsVolume,2008-Weinberg-Cosmology} leads to:
\[
6\, H(t) \dot{H}(t) = \kappa^2 \dot{\rho}(t) \quad \Longrightarrow \quad 
\rho^2(t) = \frac{3 H^2(t)}{\kappa^2} 
\]
In the present epoch, the energy density of the Universe is dominated by DE leading to accelerated expansion~\citep{Planck:2019kim}. Rewriting the  above expression for the current epoch, we have: 
\[
\rho_{DE}^2(t_0) = \frac{3 H_0^2}{\kappa^2} \, .
\]
This is same as the critical density of the Universe~\citep{Planck:2019kim}. Since the accelerated expansion requires the DE to have negative pressure, one possibility is the Friedmann equation derived from GR needs modification on cosmological scales. 

From the above discussion, it is clear that ${\cal R}$ defines the characteristic curvature scale. Just as the linear-order binomial expansion of  $(1 + x)^n$ begins to break down when $x$ approaches $1$ or greater --- necessitating the inclusion of all higher-order terms --- higher-order corrections to gravity become essential when the characteristic curvature scale ${\cal R}$ approaches unity. At these scales, more than the simplistic leading-order approximation provided by GR is required. For instance, in regions of extremely high curvature, such as near black hole singularities or during the early moments of the universe, the contributions from higher-order curvature terms (like \(R^2\), \(R_{\mu\nu}R^{\mu\nu}\), or even non-local terms) become increasingly significant. This breakdown signals the limits of classical GR and the need for a more complete framework, such as quantum gravity or modified gravity theories, to fully describe spacetime dynamics. Much like how the entire binomial series must be summed for \(x \approx 1\), these higher-order corrections are imperative for maintaining consistency and predictive power in extreme regimes, ultimately offering insights into physics beyond Einstein's theory.

These quantities have been discussed by various authors in the literature~\citep{Psaltis:2008bb,Baker:2014zba,Yunes:2016jcc,Bailes:2021tot,Kalogera:2021bya}\footnote{In Refs.~\citep{Psaltis:2008bb,Baker:2014zba}, the parameters are represented by $\varepsilon$ and $\xi$.} However, to our knowledge, no author has presented an interpretation directly based on Einstein’s equations for the strong and weak-gravity regimes. The above approach cannot be directly applied to a radiation-dominated Universe since $T^{\rm (M)}$ vanishes in such cases. However, the interpretation holds for a point mass. Consequently, the parameters $\Phi$ and $\cal R$ create a parameter space that aids in quantifying the strength of gravitational fields across various gravity tests. 

\begin{figure}
\begin{center}
\includegraphics[width=0.7\textwidth]{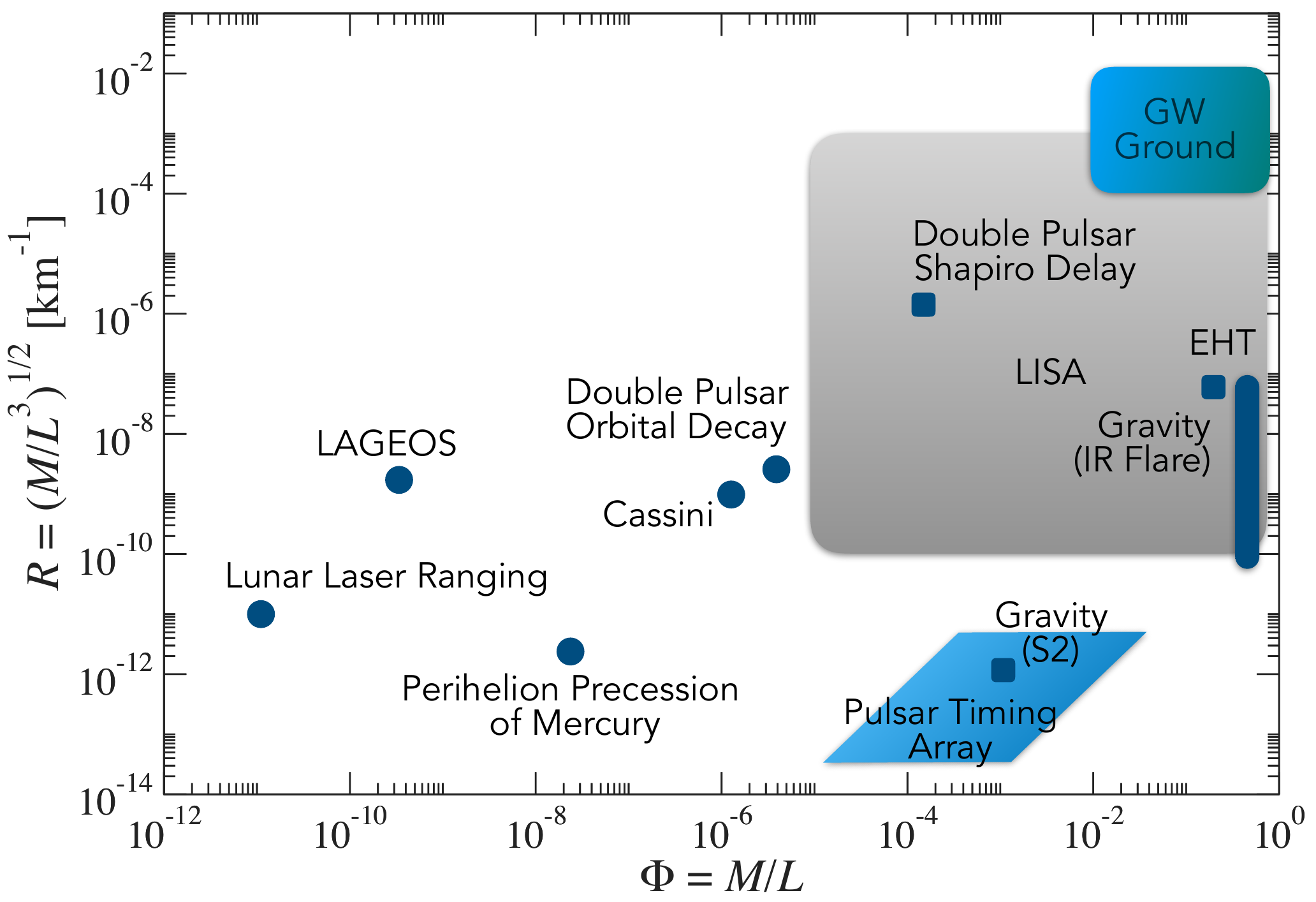}
\caption{\label{fig00} The figure shows the capabilities of past, current and future experiments to constrain GR in the curvature-surface potential plane. For a system of mass $M$ and size $L$ curvature scales as $M/L^3$ (in $c=G=1$ units) and the surface potential scales as $M/L.$ Ground-based gravitational wave observations can constrain GR on both the largest curvature and surface potential scales. Credit: \citep{Kalogera:2021bya}.}
\end{center}
\end{figure}

Figure~\eqref{fig00} illustrates this parameter space for \( \Phi \) and \( \cal R \), with the vertical axis representing the characteristic curvature length scale and the horizontal axis depicting the compactness parameter. Only a limited portion of this space is observable. For instance, regions where $\Phi > \frac{1}{2}$ correspond to distances smaller than the horizon radius, making them inaccessible. 

\subsection{Weak gravity regime}
\label{sec:Why-Weak}

All historical tests of GR have been conducted within our solar system. As a result, the strongest gravitational field accessible to these tests is that at the surface of the Sun, where the compactness parameter is:
\begin{equation}
\Phi_{\odot} \simeq \frac{GM_{\odot}}{R_{\odot}c^{2}} \simeq 2 \times 10^{-6}
\end{equation}
and the square-root of curvature is:
\begin{equation}
{\cal R}_{\odot}  = \sqrt{\frac{GM_{\odot}}{R_{\odot}^{3}c^{2}}} 
\simeq 10^{-14} \text{cm}^{-1}.
\end{equation}
Interestingly, the gravitational fields probed in tests involving double neutron stars are of similar magnitude, as the masses and separations of the neutron stars in these systems resemble those of Sun. However, these fields are significantly weaker than those found in the vicinity of neutron stars and stellar-mass BHs, which correspond to the compactness parameter of approximately $1/2$ and ${\cal R} \simeq 5 \times 10^{-6} \text{cm}^{-1}$. It is also elucidating to compare the extent to which current tests confirm the predictions of GR with the increase in the strength of the gravitational field when transitioning from the solar system to the vicinity of a compact object. Additionally, heightened theoretical research, including the development of Parameterized Post-Keplerian frameworks has bolstered efforts to quantify and test potential deviations from GR~\citep{Kramer:2021jcw}.

The ability to quantitatively test gravity in the weak gravity regime also depends on independent measurements of the mass generating the field. This is not always feasible, particularly in cosmology, where gravitational phenomena typically infer the presence of dark matter rather than test GR predictions. Dark matter is invoked in systems where acceleration falls below the so-called MOND acceleration scale, \( a_0 \simeq 10^{-8} \ \text{cm} \ \text{s}^{-2} \)~\citep{Milgrom1983,Sanders:2002pf, Bekenstein:2006bya}, comparable to \( H_{0} \). Systems requiring dark matter for their gravitational fields are characterized by~\citep{Psaltis:2008bb}
\begin{equation}
{\cal R}^2 \leq \frac{a_{0}^{2}}{\Phi} \simeq \frac{H_{0}^{2}}{\Phi}.
\end{equation}

Further, understanding deviations from GR across some regions of the parameter space requires an in-depth understanding of dark matter and dark energy, which is not yet fully developed. In low-curvature regimes, the effect of a non-zero cosmological constant becomes significant in gravitational experiments when:
\begin{equation}
{\cal R}^2 \leq \frac{1}{2} \left( \frac{H_0}{c}\right)^2 \Omega_\Lambda 
\end{equation}
where $\Omega_{\Lambda}$ represents the current dark energy density relative to critical density. Phenomena that investigate such low curvature values may yield quantitative tests of GR only under the assumption of a specific dark energy model (e.g., a cosmological constant).
Present constraints on deviations from GR in the Parameterized Post-Newtonian parameters are of the order of $\simeq 10^{-5}$~\citep{Will:2014kxa}. Thus, it is plausible that deviations within these constraints could become significant, approaching unity, when the compactness parameter increases by six orders of magnitude and ${\cal R}$ by fifteen. Therefore, naturally one may raise the following question ---- Is it conceivable that GR continues to accurately characterize phenomena occurring in the intense gravitational fields present near stellar-mass BHs and neutron stars?

\subsection{Strong gravity regime}
\label{sec:Why-Strong}
In the strong gravity regime, relevant to the evolution of the primordial Universe and the characterization of BHs and neutron stars, testing the predictions of GR has seen considerable advancements over the past decade~\citep{Stairs:2003eg,Will:2014kxa,Wex:2020ald,Freire:2024adf}. Previously, progress in this area was limited by two primary challenges. First, phenomena in strong gravitational fields often involve highly dynamic and complex events, such as explosive processes, making it difficult to identify observable properties that directly correlate with the gravitational field and allow for quantitative tests of gravitational theories. Second, the absence of a unified theoretical framework to systematically assess deviations from GR in strong-field conditions has hindered comprehensive exploration.

In recent years, however, significant technological improvements, such as advancements in gravitational wave detection and high-precision astronomical instruments, have expanded the observational reach and sensitivity in strong-field gravity.  These advances have made testing strong-field gravity a collaborative and rapidly evolving field, shedding new light on GR's validity under extreme conditions. For instance, the GW events GW150914~\citep{LIGOScientific:2016aoc} and GW230529~\citep{LIGOScientific:2024elc} probe regions where both curvature and potential are simultaneously high and dynamic, as shown in Fig.~\ref{fig00}. Pulsar timing arrays cover a large range of gravitational wave frequencies and total masses of supermassive BH binaries, allowing investigation over a finite parameter space area. As noted, smaller mass BHs exhibit higher \({\cal R}\), making coalescences of lighter astrophysical BHs ideal for testing GR’s strong-field predictions. Sub-solar-mass BH binaries, which are sources of high-frequency GWs~\citep{2021-Goryachev.etal-PRL,Domcke:2020yzq,2021-Aggarwal.etal-LRR,Kushwaha:2022twx}, would exhibit even greater curvature.

As shown in Fig.~\eqref{fig00}, experiments like Cassini and double pulsar orbital decay test moderately strong gravitational fields with low relative velocities w.r.t the speed of light. Meanwhile, the EHT and GRAVITY (S2) instruments test GR around super massive BHs~\citep{GRAVITY:2018ofz}, focusing on small curvature but high compactness regimes. Similarly, the Laser Interferometer Space Antenna (LISA) will cover an extensive range of curvature and compactness, from stellar-mass BHs (5–100 \( M_{\odot} \)), intermediate-mass BHs (\( 10^2 - 10^4 M_{\odot} \)), to supermassive BHs (\( 10^5 - 10^{10} M_{\odot} \))~\citep{LISA:2022kgy}. This broad range allows for testing GR across ten orders of magnitude in length scale and twenty orders of magnitude in curvature.

\subsection{Quantum effects and Gravity}
\label{sec:Why-QG}

Fig.~\eqref{fig00} does not include the regime of extreme gravitational fields where quantum effects are expected to become relevant. This scenario is anticipated when a gravitational test reaches distances from an object of mass $M$ close to its Compton wavelength, $\lambdabar_{C} \equiv {\hbar}/{(M c)}$. Thus, quantum effects are expected to dominate when
\begin{equation}
{\cal R} \geq \frac{\Phi}{L_{\rm P}^{2}},
\end{equation}
where $ L_{\rm P} \equiv (\hbar G/c^3)^{1/2} \approx 1.6 \times 10^{-33}$ cm is the Planck length. As noted, this region of parameter space is not shown in Fig.~\eqref{fig00}, as it is several orders of magnitude beyond the scales associated with astrophysical systems.

The robustness of classical gravitational theory hinges on minimal fluctuations at scales where gravitational effects become prominent, a concept Feynman initially explored~\citep{DeWitt:1957obj}. Feynman’s argument centers on a two-slit diffraction experiment with a mass indicator positioned behind the slits, forming a gravitational two-slit configuration. The dimensionless gravitational potential generated by a mass $M $ within the spatial volume of length $L$ is $\Phi = {G M}/{(L c^2)}$. Hence, the uncertainty is $\Delta \Phi = {G \Delta M}/{(L c^2)}$. 

We can compare this uncertainty to the quantum uncertainty in measuring $ \Phi$. In a spacetime region with spatial dimensions $ L$ and temporal dimensions \( L/c \), the uncertainty on the compactness parameter is:
\[
\Delta \Phi = \frac{L_{\rm P}}{L}
\]
By equating these expressions, the inferred mass uncertainty is:
\begin{equation}
\Delta M = \frac{c^2 L_{\mathrm{P}}}{G} \approx 10^{-5} \text{ g}
\label{def:MPlanck}
\end{equation}
when the observation period is shorter than \( L / c \). From this, Feynman concluded: “Either gravity must be quantized to avoid logical inconsistencies if the experiment is conducted with masses around \(10^{-5}\) grams, or quantum mechanics fails for masses near \(10^{-5}\) grams.”

Building on Feynman’s insight, recent proposals have suggested tabletop experiments using quantum information theory to investigate the quantum characteristics of gravity in a low energy laboratory settings~\citep{Bose:2017nin,Marletto:2017kzi}. Unlike other Planck units, such as length, energy or time, Planck mass lacks a definitive physical interpretation. Christodoulou and Rovelli proposed that these tabletop experiments could reveal the scale at which quantum superposition in curved spacetime becomes observable~\citep{2020-Christodoulou.Rovelli-Front}. 

From a theoretical standpoint, GR is a fundamentally classical framework. Power-counting arguments indicate that GR is not renormalizable within the conventional quantum field theory approach. This issue may be addressed by modifications in strong-field regimes; in fact, the theory can achieve renormalizability if quadratic curvature terms—i.e., high-energy or high-curvature corrections—are added to the Einstein-Hilbert action~\citep{Stelle1977-rd}. Moreover, these high-energy corrections could help prevent singularity formation, which is an inherent feature of classical GR as shown by the Hawking-Penrose singularity theorems \citep{Hawking:1970zqf}. Quantum gravity theories, including string theory and loop quantum gravity, suggest specific and potentially testable predictions about necessary modifications to GR at high energy scales. 

In summary, GR might have significant classical and quantum corrections at both extremely small and large scales. Throughout this review, we examine these classical and quantum modifications to GR, aiming to present frameworks for testing gravity across disparate curvature scales.


\section{Testing gravity at different curvature scales}
\label{sec:testing}

As discussed previously, gravity is well-tested on solar system scales but remains largely untested beyond these regions~\citep{Will:2014kxa,2022-Shanki.Joseph-GRG}. With the development of next-generation EM and gravitational wave experiments, two fundamental questions arise:
\begin{enumerate}
\item Does GR fall short in explaining the data? 
\item Does an alternative theory of gravity provide a better fit?
\end{enumerate}
While these questions may appear similar, they are distinct. As noted in Sec.~\eqref{sec:GRreview}, GR rests on several key assumptions: (i)  the existence of a symmetric metric and connection coefficients, (ii) all test bodies follow metric geodesics (corresponding to minimal coupling), and (iii) non-gravitational laws in local Lorentz frames are those of special relativity (corresponding to general covariance). Based on these principles, GR makes specific predictions, including the BH uniqueness theorem and appearance of BH image,  GWs being transverse, and decelerated expansion of the Universe for radiation and ordinary matter. 

While theories beyond GR may violate any or all of these assumptions, some alternative metric theories of gravity maintain these principles but yield different predictions, such as BH non-uniqueness, additional gravitational wave modes, or early-time and late-time accelerated expansion of the Universe. Where such predictions exist, the objective is to evaluate whether these alternative theories align better with observations than GR. This leads to a core question: \emph{How can we test gravity across varying curvature scales?}

While several approaches exist, this review follows a two-pronged strategy:
\begin{enumerate}
\item Test the foundational principles of GR via multimessenger
astronomy, using both EM and gravitational wave observations.
\item Examine the range of compact objects with similar curvature to analyze shared characteristics among these systems. 
\end{enumerate}

\subsection{Testing foundational principles of GR via multimessenger}

If the foundational principles of GR are found to be violated, it would suggest the need to explore alternative theoretical frameworks. Such a violation could indicate limits to the applicability of GR and motivate the consideration of new or modified theories of gravity that could account for deviations under specific conditions. The equivalence principle, in its various formulations, defines the geometric structure of the gravitational theory~\citep{Will:2014kxa,Will:2018bme}. \\
\begin{itemize}
\item  \emph{The Weak equivalence principle} (WEP) states that all the laws of motion for freely falling particles are the same as in an unaccelerated reference frame. 
\item \emph{The Einstein Equivalence Principle (EEP)} states that no local, non-gravitational experiment can differentiate between a frame at rest and one in free fall in a gravitational field. 
\item \emph{The Strong Equivalence Principle (SEP)} extends this, applying even to gravitational experiments.  \\
\end{itemize}
The EEP also includes \emph{Local Lorentz invariance} (LLI), and \emph{local position invariance} (LPI). EEP is satisfied if all these three sub-principles are satisfied~\citep{Will:2018bme}. As we discuss below, given its central role in gravitational theories, extensive research over the last century has rigorously tested the equivalence principle, primarily in the weak-field regime, setting highly precise upper bounds on potential violations --- reaching accuracies up to one part in $10^{15}$~\citep{Berge:2023sqt}.

Testing the equivalence principles involve a variety of experimental and observational tools across different physical systems and scales. For instance, torsion balance, atomic clock experiments, optical cavity resonators and muon lifetime experiments use terrestial setups to 
test the LLI and LPI~\citep{Turyshev:2007qy,Turyshev:2008dr,Tino:2020nla}. With extensive multimessenger data that will now be available in the \emph{weak gravity regime} --- such as observations near the event horizons of supermassive BHs~\citep{2020-Broderick-apj,2020-Johnson-Sci.Adv.}, large-scale structure data from multiple surveys~\citep{Bonvin:2020cxp}, and gravitational wave signals from intermediate-mass BH merger from space-based interferometers~\citep{Barack:2018yly} --- we have robust opportunities to rigorously test scenarios that extend beyond GR.

Multimessenger astronomy expands on traditional astronomy, which primarily relies on detecting EM radiation (photons) from cosmic events. This expanded approach combines conventional observations with detections of neutrinos, GWs, and cosmic rays~\citep{2019-Murase.Bartos-AnnRevNuclPartPhys,Guepin:2022qpl}. Neutrinos, due to their minimal interaction with matter, can emerge from dense astrophysical environments that are opaque to photons, offering unique insights into the interiors of stars and other obscured regions. 
The first observed multimessenger event was Supernova 1987A, where a neutrino burst was detected hours before the supernova appeared in the EM spectrum~\citep{1989-Arnett.etal-ARAA}. Meanwhile, GWs across various frequency ranges provide complementary information on diverse processes throughout the Universe~\citep{2021-Aggarwal.etal-LRR,NANOGrav:2020gpb,Kalogera:2021bya,Bailes:2021tot}.

Modern astronomical instruments allow for the detection of astrophysical photons across a vast frequency range, from low-frequency radio waves around $10$~MHz (for instance, LOFAR)~\citep{2013-LOFAR-AA} to very high-energy gamma rays about $10^{28}~$Hz (for instance, LHAASO)~\citep{LHAASO:2019qtb}. These instruments capture the intensity which is related to the  energy of the incoming EM signals~\citep{Wei:2021vvn}. When the photons strike a detector, they are absorbed, leading to a shift in energy levels. This energy change which decreases with distance as \(1/r^2\), produces the observed signal. Hence, EM flux goes as $1/r^2$.

GW detectors, however, do not absorb the energy of these waves, but, 
measure the distortions in spacetime caused by passing GWs, Hence, 
GW detectors measure the amplitude and not the energy. Consequently, the sensitivity of GW detectors to distant sources fall off more slowly, at $1/r$. Even though the amplitude of the generated GW is small, the GW detectors can detect signals at a far-off distance. This is the crucial reason why the next-generation GW detectors can detect binary BH events up to a redshift of $100$~\citep{Kalogera:2021bya,Evans:2023-CE}.
The rest of this subsection briefly outlines the experimental status of the foundational principles of GR and discusses the possibilities with next generation experiments.

%

\subsubsection{Weak equivalence principle}

Experimental tests of the WEP focus on comparing the accelerations ($a_1$ and $a_2$) of two test masses, $m_1$ and $m_2$, in the same gravitational field. This comparison is often expressed using the E\"otv\"os parameter $\eta$~\citep{Tino:2020nla}: 
\begin{equation}
\eta = 2 \frac{|a_1 - a_2|}{a_1 + a_2} 
\label{def:EotvosP}
\end{equation}
In E\"otv\"os’s experiment, two masses of different compositions were placed at opposite ends of a suspended rod. The Earth's gravitational pull and centrifugal forces from Earth’s rotation would produce a torque on the balance if the masses had different inertial-to-gravitational mass ratios, indicating a potential WEP violation~\citep{labtest1}.

Advancements in torsion balance technology have continually increased experimental precision. The most stringent limits to date come from the Eöt-Wash group, which has placed an upper bound on WEP violation at $2 \times 10^{-13}$~\citep{roll1964equivalence,su1994new,adelberger2001new,2007-Schlamminger.etal-PRL}. See left panel of Figure 2 for the bound on $\eta$ as a test of WEP.
Despite these advancements, systematic errors from environmental gravitational gradients, thermal fluctuations, and magnetic fields still present challenges, though they can be minimized through innovative experimental techniques. For comprehensive reviews of torsion pendulum tests of gravity, including WEP experiments, see~\citep{Will:2018bme, Will:2014kxa}.


Space-based experiments enable significantly longer free-fall times than ground-based tests, with comparable driving accelerations~\citep{Turyshev:2007qy,Turyshev:2008dr}. This setup allows space-based experiments to potentially exceed the sensitivity of torsion balance experiments, which, while benefiting from long integration times, experience much smaller driving accelerations. Crucially, weightlessness in space, along with advanced drag-free technology that cancels non-gravitational external forces, creates a stable environment for precise measurement of tiny accelerations and forces. In space, free-fall experiments can be conducted in compact setups, which simplifies control over external disturbances.  Another advantage of space-based tests is the ability to rotate the entire apparatus to adjust the detection frequency of potential WEP violations. 

MICROSCOPE (Micro-Satellite \'a tra\^in\'ee Compens\'ee pour l’Observation du Principe d’Equivalence) used a drag-free satellite that shielded its payload from non-gravitational forces, reducing noises as low as \(10^{-12} \, \text{m/s}^2 \, \text{Hz}^{-1/2}\)~\citep{Berge:2015dya,Berge:2023sqt}. 
 In 2022, MICROSCOPE confirmed the equivalence principle to within \(\eta = 10^{-14}\), setting the tightest constraint yet on WEP violations~\citep{MICROSCOPE:2022doy}. Although not explicitly designed to test the equivalence principle, LISA Pathfinder developed key technologies for precise force measurements, crucial for any WEP test. LISA Pathfinder demonstrated a drag-free environment, achieving extremely low residual differential accelerations (\(10^{-15} \, \text{m/s}^2 \, \text{Hz}^{-1/2}\)), paving the way for gravitational wave observatories like LISA and influencing WEP missions~\citep{Armano:2016bkm}.

In the coming decade, three space-based missions --- Galileo Galilei mission~\citep{2018-GGMission-PRD}, Space-Time Explorer and Quantum Equivalence Principle Space Test (STE-QUEST)~\citep{STE-QUEST:2022eww} and Atomic Clock Ensemble in Space (ACES)~\citep{ACESMission:2024uqz} --- are planned to test the WEP up to $\eta = 10^{-17}$. GG plans to use a spinning satellite to amplify any WEP violations by adjusting the modulation frequency, helping distinguish potential signals from noise, hence testing the WEP at \(10^{-17}\) accuracy. STE-QUEST proposes to combine atomic interferometry with satellite technologies, aiming for \(10^{-17}\) precision. ACES plans to test time dilation and effects of gravity on atomic clocks, indirectly probing principles related to WEP.

Astrophysical observations offer unique opportunities to test WEP over vast distances and extreme environments. For instance, deviations in pulsar timing signal an EP violation by revealing differences in the paths or velocities of signals influenced by gravitational fields~\citep{Wex:2020ald,Freire:2024adf}. On the other extreme, GRBs emit high-energy EM waves, and any frequency-dependent time delay in the arrival times of signals from a distant GRB, beyond what can be explained by travel distances, could suggest a violation of the EEP. This technique has constrained EP violations down to \(10^{-8}\) for specific events~\citep{Wu:2017yjl}. 

Binary Neutron star mergers allow us to test EEP by looking at the simultaneous arrival of GWs and EM signals from the same astrophysical event. A difference in arrival times that could not be attributed to other factors would suggest an EP violation. For instance, the detection of GW170817 and its EM counterpart~\citep{LIGOScientific:2017vwq} allowed us to compare travel times over hundreds of millions of light-years and put stringent constraints on various modified gravity models~\citep{Kreisch:2017uet}. 


%
\begin{figure*}[t!]
    \centering
    \begin{subfigure}[t]{0.55\textwidth}
        \centering
        \includegraphics[width=0.99\textwidth]{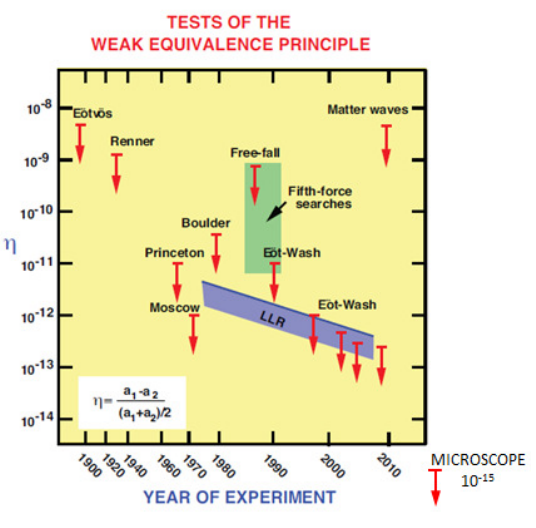}
    \end{subfigure}%
    ~ 
    \begin{subfigure}[t]{0.45\textwidth}
        \centering
        \includegraphics[width=0.99\textwidth]{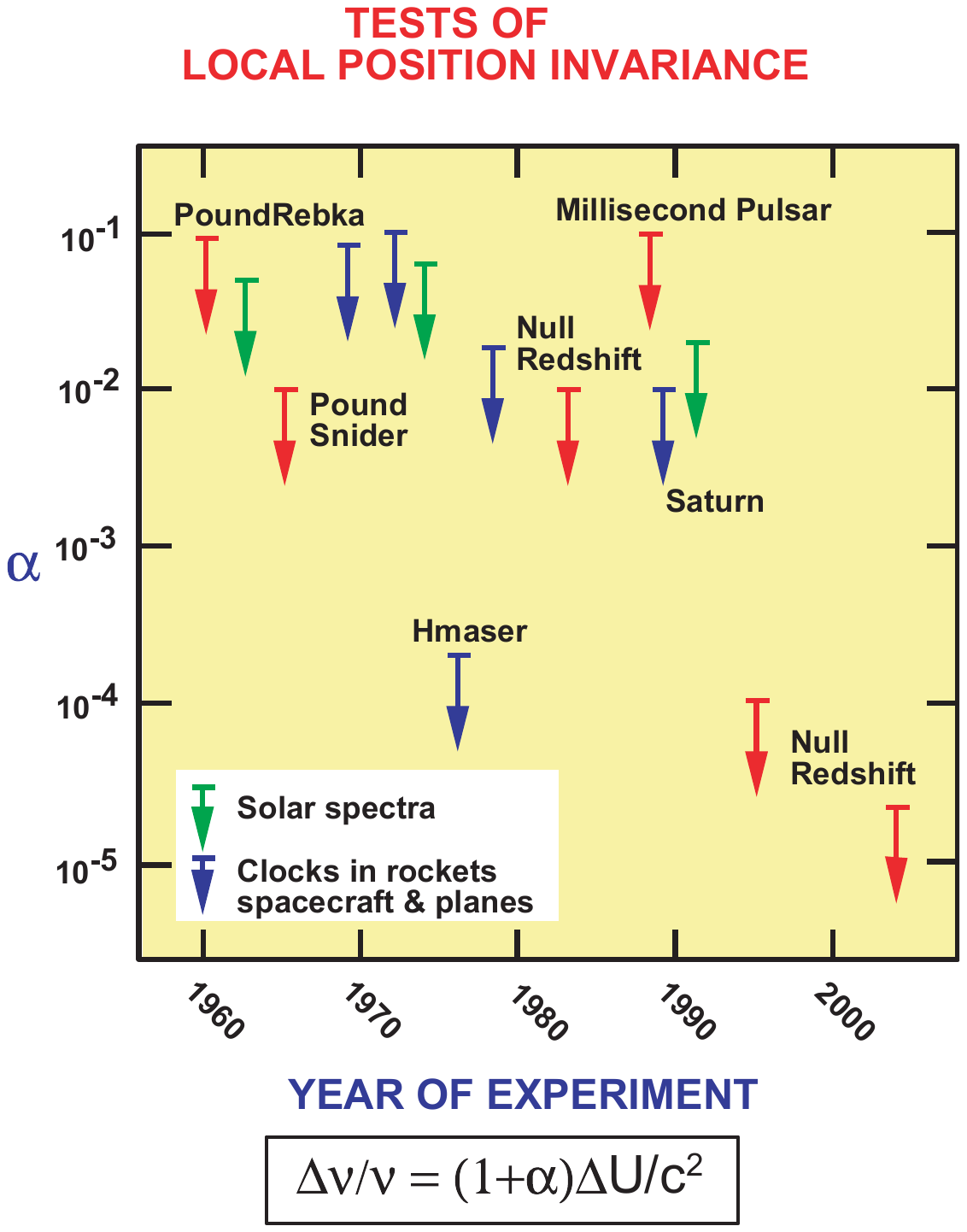}
    \end{subfigure}
    \caption{\label{fig:EEPTest} The above figure shows different tests of WEP (left) and Local position invariance (right) and different 
bounds on the parameter that quantifies it. Credit: \citep{Berge:2015dya,Will:2014kxa}}
\end{figure*}

\subsubsection{Einstein's Equivalence principle (EEP)}

As mentioned above, the EEP encompasses three principles --- WEP, LLI and LPI. The LLI stipulates that experimental outcomes in a freely falling inertial frame are independent of the orientation and velocity of the laboratory. Historically, local Lorentz invariance was examined through the Michelson-Morley experiment, which tested the hypothesis that EM waves propagate through a medium, resulting in the well-known null result~\citep{2005-Mattingly-LRR}. Although the Michelson-Morley experiment confirmed the isotropy of the speed of light, subsequent investigations, such as the Kennedy-Thorndike experiment conducted in the 1930s~\citep{PhysRev.42.400}, sought to detect any dependence on laboratory velocity. Through these and numerous other experiments, local Lorentz invariance has become a fundamental assumption rigorously tested in contemporary physics.

High-sensitivity measurements using superconducting gravimeters have also been employed to test LLI by examining the behavior of gravitational interactions under precise laboratory conditions. This allows us to set strict limits on the potential Lorentz-violating parameters, particularly concerning gravitational-matter couplings. These gravimeters have provided some of the best terrestrial constraints on LLI, with improvements in sensitivity across multiple operators related to possible violations~\citep{Flowers:2016ctv,Zhang:2023hey}. Experiments involving cold atoms and precision interferometry are developing sensitivity to subtle effects of LLI violations. Efforts are underway to achieve sub-millimeter-level measurements of gravitational time dilation~\citep{Delva:2017znr}. 

Astrophysical and cosmological observations also provide constraints on LLI~\citep{Hees:2016lyw}. For instance, analyzing CMB data for polarization patterns has yielded constraints on LLI. Any anisotropies in the polarization of the CMB could suggest Lorentz violation~\citep{Komatsu:2022nvu}. Recent analyses of CMB data from experiments like the Planck satellite have shown no significant anomalies, further supporting the invariance of Lorentz symmetry on cosmological scales~\citep{Gubitosi:2009eu,Staicova:2023vln}. Observations of GRBs have been extensively analyzed to test for any time delays in the arrival of photons of different energies, which would imply a violation of LLI. A recent study using data from the Swift satellite evaluated the time-lag of photons from GRBs at different redshifts, allowing for the constraining of possible Lorentz-violating effects with unprecedented precision~\citep{PhysRevD.108.123023}. 

A fundamental aspect of Einstein’s equivalence principle is LPI, which posits that measurement results should remain independent of spatial or temporal location. LPI, as part of special relativity, holds in flat, free space. When gravitational effects are weak, LPI can still be valid if proper adjustments are made, approximating a locally flat space. However, if gravity is partially mediated by a long-range vector field, a universal preferred rest frame might exist, possibly aligning with the CMB frame (this will be discussed further in the next chapter).  

A prevalent approach to test LPI is using the graviational redshift.  This involves introducing a parameter to the GR redshift term in the formula for the frequency shift of an EM wave:
\begin{equation}
\frac{\delta\nu}{\nu} = (1 + \alpha)\Delta U,
\end{equation}
where $\Delta U$ represents the difference in gravitational potential between two locations at which the frequency shift $\delta\nu$ is measured. Theories extending beyond GR that predict a nonzero value for $\alpha$ typically assume that clock rates may vary with spatial position, thereby violating the Local Position Invariance. Within the framework of Effective Field Theory (EFT), redshift tests are sensitive to various forms of Lorentz violation. In ~\cite{PhysRevD.83.016013}, the modification of gravitational and matter interactions due to Lorentz and CPT violation results in deviations from the standard GR redshift prediction.
Also, interpreting certain pulsar spectra suggests possible EEP violations~\citep{Wex:2020ald,Freire:2024adf}. The detection of Lorentz invariance violation in gravitational interactions --- manifesting as preferred frame effects ---can result in violation of angular momentum conservation, while tests of parameters that leads to violation of total momentum conservation have provided stringent limits on the two Parameterized Post-Newtonian parameters~\citep{1968-Nordtvedt-PRb,Wex:2020ald,Freire:2024adf}. See right panel of Figure 2 for the bound on the $\alpha$ parameter as a test of LPI.

Thus, any experimental evidence for EP violation would also challenge GR. Additionally, one of EP’s key components, local Lorentz invariance, implies CPT symmetry~\citep{1951-Schwinger-PR}, which has been thoroughly tested on Earth~\citep{2013-Liberati-CQG}. However, confirming the validity of the Einstein Equivalence Principle (EEP) on cosmic scales poses significantly greater challenges.

\begin{table}
\parbox{0.25\linewidth}{
\centering
\begin{tabular}{|c|c|c|}
\hline
& &  \\[-1.3em]
Object  & ${\cal R} (m^{-1})$ & $\Phi$ \\
\hline
& &  \\[-0.8em]
Earth & $10^{-11}$  & $10^{-8}$ \\
& & \\[-0.8em]
\hline 
 & & \\[-0.8em]
Sun &  $10^{-14}$  & $10^{-5}$ \\
& & \\[-0.8em]
\hline 
& & \\[-0.8em]
WD &  $10^{-8}$  & $10^{-3}$ \\
 & & \\[-0.8em]
\hline 
 & & \\[-0.8em]
NS &  $10^{-7}$  & $0.1$ \\
 & & \\[-0.8em]
\hline 
 & & \\[-0.8em]
BH & $10^{-5}$ & $1$ \\
\hline
\end{tabular}
\caption{\label{Table:Compactobjects} $\Phi$ and ${\cal R}$ for $\sim$ solar scale compact objects.}
}
\parbox{0.05\linewidth}{
\centering
$~~~$ 
}
\parbox{0.70\linewidth}{
\centering
\begin{align}
\nonumber
\begin{array}{|l|c|c|c|l|}
\hline \text{Source} & \text{Rate density (Gpc$^{-3}$ Yr$^{-1}$)} & \text{Luminosity (erg/s)}   & \text{Duration (s)} \\ 
\hline & &  & \\[-0.8em] 
\text{GW} & 0.002 - 18 & 10^{54} - 10^{56} & < 1 \\
\hline & & & \\[-0.8em] 
\text { Long GRB } & 0.1-1 & 10^{51}-10^{52} & 10-100 \\
\hline & &  &\\[-0.8em] 
\text { Short GRB } & 10-100 & 10^{51}-10^{52} & 0.1-1 \\
\hline & & & \\[-0.8em] 
\text { Low-luminosity GRB } & 100-1,000 & 10^{46}-10^{47} & 1,000-10,000  \\
\hline & &  & \\[-0.8em] 
\text { GRB afterglow } & \sim 10 & <10^{46}-10^{51} & >1-10,000  \\
\hline & &  & \\[-0.8em] 
\text{ FRB} & 339^{+1097}_{-330} & 10^{41}-10^{44} & 10^{-3} - 1 \\
\hline 
\end{array}
\end{align}
\caption{\label{Table:Transients-energy} List of high energy transients. For more details, see ~\cite{2019-Murase.Bartos-AnnRevNuclPartPhys,2020-Luo.Zhang-MNRS,2024-Jana.etal-Arx}.
}
}
\end{table}

\subsection{Testing fundamental physics with compact objects and transients}

Compact objects include dense stellar remnants such as white dwarfs, neutron stars, and BHs, along with supermassive BHs that are found at the centers of galaxies.
These objects are characterized by their strong gravitational fields (see Table \ref{Table:Compactobjects} containing 
$\Phi$ and ${\cal R}$ for some of these objects) GR predicts distinctive properties for each type, including unique spacetime geometries and horizon structures. Interestingly, the formation of these compact objects and/or the presence of these compact objects can lead to high-energy astrophysical transients --- brief, intense bursts of radiation across various wavelengths and messengers like cosmic rays, GWs or  neutrinos~\citep{2019-Murase.Bartos-AnnRevNuclPartPhys,Guepin:2022qpl}. 
Table \ref{Table:Transients-energy} contains some of the well-known astrophysical events and the corresponding luminosity. 

Since the compact objects have relatively larger ${\cal R}$  and the transients are highly energetic, cataclysmic events, it is unclear whether GR can be an accurate description of gravity in this regime. Also, since the Einstein's equation are non-linear, adding two gravitational fields and their sources --- each obeying Einstein's equation --- does not yield a straightforward combination of fields and sources. Hence, these are the best places to study deviations from GR. 
In this section, we briefly discuss some of the well observed high-energy transients that offer unique window to test gravity in various extreme environments.

\subsubsection{Gravitational waves from compact binary coalescence}

GW transients are produced by massive accelerating compact object mergers, such as BH-BH, neutron star-neutron star, or neutron star-BH coalescence~\citep{Sathyaprakash:2009xs}. Since all masses have the same gravitational sign, they tend to clump together and produce large coherent bulk motions that generate energetic, coherent 
GWs~\citep{2007-Hendry.Woan-AG}.  Typical GWs from a compact binary coalescence with
strain amplitude $h \sim 10^{-22}$ carry energy of the order of $10^{27}$~ergs~\citep{Book-Schutz}.  

A stellar mass binary BH collision can lead to a peak GW luminosity of 
$10^{56}$ ergs/s, which is three orders lower than the absolute upper limit on luminosity for any source in GR~\citep{2017-LIGOScientific.AnnalPhy}.  To understand this, consider a spherical region of radius \( R \) filled with light of total energy \( Mc^2 \). If this energy is released instantaneously, it escapes the sphere within a time \( R/c \), yielding an average luminosity of \( Mc^3/R \). However, gravitational constraints limit how small \( R \) can be. If \( R \) is less than the Schwarzschild radius \( R_S = 2GM/c^2 \), the light cannot escape, as it would be trapped within the event horizon of a BH.  Thus, the maximum possible luminosity, independent of mass, is~\citep{Hogan:1999hz}:
\[
L_{\text{GR}} = \frac{c^5}{2G} = \frac{m_{\text{Planck}}^2}{2} \approx 1.81 \times 10^{59} \, \text{erg/s},
\]
where \( m_{\text{Planck}} = \sqrt{\hbar c / G} \approx 10^{-5} \, \text{g} \) is the Planck mass defined in Eq.~\eqref{def:MPlanck}. While this limit is derived using classical GR, it does not depend on quantum mechanics since \( \hbar \) cancels out when expressed in luminosity units. In principle, this can be different for other theories of gravity and hence, GW transients are a good test bed for modified gravity theories.

\subsubsection{Gamma ray bursts (GRBs)}

GRBs are among the most powerful explosions observed, releasing intense gamma-ray radiation lasting from milliseconds to several minutes~\citep{2004-Piran-RevModPhys}. For instance, GRB 221009A, dubbed the brightest of all time, released intense gamma-ray emissions around $10^{54}~$ ergs/s~\citep{Frederiks:2023bxg}.  They are typically categorized into short GRBs (lasting $< 2$ seconds ) and long GRBs (lasting $ > 2$ seconds)~\citep{Oates:2023ghw}. Short GRBs are thought to originate from mergers of compact objects like neutron stars~\citep{LIGOScientific:2017vwq}, while long GRBs are associated with the collapse of massive stars in supernova events~\citep{Woosley:2006fn} 

The Large High Altitude Air Shower Observatory Project (LHAASO) has recently~\citep{LHAASO:2024lub} detected an exceptionally large number of very high-energy photon events originating from the luminous GRB 221009A at its earliest observable epoch, thereby marking the first identification of a TeV GRB afterglow. These distinctive characteristics present a rare opportunity to investigate Lorentz violation within the photon sector. By employing both cross-correlation function and Machine Learning techniques, they searched for the potential  Lorentz violation-induced delays in the arrival times of high-energy photons. Both methodologies yield consistent constraints on the LIV energy scale, denoted as $E_{QG}$. Specifically, for a linear modification of the photon dispersion relation, they derive a lower bound of $E_{QG,1} > 1.0 \times 10^{20} \text{GeV} $ (or $E_{QG,1} > 1.1 \times 10^{20} \, \text{GeV}$ for superluminal  Lorentz violation). This result is comparable to the most stringent lower limit on $E_{QG,1}$ previously obtained from the GeV emission of GRB 090510~\citep{PhysRevD.87.122001}. For the quadratic case, they establish a limit of $E_{QG,2} > 6.9 \times 10^{11} \, \text{GeV} $ (or $E_{QG,2} > 7.0 \times 10^{11} \, \text{GeV}$ for superluminal  Lorentz violation), representing the most stringent time-of-flight constraint to date, thus improving upon previous bounds by factors of 5 and 7, respectively~\citep{PhysRevD.87.122001}.

\subsubsection{Supernova}

Supernovae are explosive deaths of stars, leading to the release of massive amounts of energy in optical and ultraviolet light~\citep{Burrows:2012ew}. Superluminous supernovae are a particularly energetic class, up to 10-100 times brighter than standard supernovae. Typical supernovae occur in white dwarfs in binary systems or massive stars, while superluminous supernovae may involve unusual processes, such as interactions with dense circumstellar material or powerful magnetic fields in rapidly rotating neutron stars (magnetars).

Type Ia supernovae (SNe Ia) have played a pivotal role in uncovering the accelerated expansion of the Universe~\citep{SupernovaSearchTeam:1998fmf,SupernovaCosmologyProject:1998vns}.
 These stellar explosions act as ``standard candles," objects with a consistent intrinsic brightness, which allows astronomers to measure cosmic distances with remarkable precision~\citep{Goobar:2011iv}. Observations of SNe Ia across different redshifts are used to test and refine models of cosmic evolution, including potential modifications to GR~\citep{Leibundgut:2018mfn}. Combining SNe Ia data with other cosmological measurements, such as baryon acoustic oscillations and GWs, enhances the robustness of our understanding of late-time acceleration of the Universe~\citep{Bernal:2020vbb}.

\subsubsection{Fast Radio Bursts (FRBs)}

FRBs are millisecond-long bursts of radio waves that are highly energetic, lasting just a few milliseconds~\citep{CHIMEFRB:2021srp,Petroff:2021wug}. Their origin remains  largely enigmatic, and they are observed primarily in radio wavelengths~\citep{Lorimer:2024ysl}. The repeating FRB 121102 was the first FRB observed to repeat, challenging theories that FRBs originate solely from cataclysmic events~\citep{Spitler:2016dmz}. These bursts are particularly intriguing because they could serve as unique tools for studying extreme astrophysical conditions and probing the distribution of matter across the Universe~\citep{Bhandari:2021thi}. 

Recently, it has been suggested that FRBs can provide indirect evidence for high-frequency GWs (HFGWs)~\citep{Kushwaha:2023poh,Kushwaha:2022twx}. Primordial BHs, exotic compact objects, and phenomena from the early Universe are predicted to produce HFGWs in the MHz to GHz range~\citep{2008-Akutsu.etal-PRL,2008-Nishizawa.etal-PRD,2017-Chou.etal-PRD,2021-Aggarwal.etal-LRR}. HFGWs could reach frequencies as high as 14 GHz~\citep{2020-Ito.etal-EPJC}.

\subsubsection{Exotic Compact objects (ECOs)}

Modified gravity theories predict the existence of ECOs, which might resemble BHs or neutron stars but with notable differences~\citep{Cardoso:2016oxy}. Examples include gravastars, boson stars, and BH mimickers, which can arise in theories with additional fields or modifications to the gravitational interaction at small scales. ECOs have been proposed to address conceptual challenges associated with BHs, such as the pathological nature of their internal structure and the information loss paradox~\citep{Cardoso:2019rvt,Bezares:2024btu}. Penrose's theorem establishes that under general conditions, an apparent horizon inevitably conceals a curvature singularity, where Einstein's theory ceases to be valid~\citep{1965-Penrose-PRL}. ECOs are theorized as potential alternatives that avoid these issues. ECOs have properties that allow them to avoid event horizons or singularities, and they could exhibit unique signatures in GW and EM observations that could help differentiate them from traditional BHs or neutron stars~\citep{Cardoso:2019rvt,Bezares:2024btu}. 

\section{Modified theories of gravity}
\label{sec:MGTheories-Classi}

Having discussed the current constraints on the fundamental principles of GR and testing GR with compact objects and transients, we now look at the theoretical frameworks of modified theories of gravity. GR has been remarkably successful in describing
gravitational phenomena across a wide range of various curvature scales, however, unresolved challenges --- such as the nature of dark matter, dark energy, and the behavior of spacetime near singularities --- have driven the search for alternative theories of gravity. While GR is self-consistent within the classical framework and provides accurate predictions in terrestrial, solar, and stellar systems, it may only serve as an approximation of a more complete theory, particularly in extreme regimes like those near singularities or at cosmological scales,

Singularities in GR, such as those predicted by the Hawking-Penrose singularity theorems, arise under specific conditions, including the validity of energy conditions like the NEC~\citep{1965-Penrose-PRL,Hawking:1970zqf,1973-Hawking.Ellis-Book,Senovilla:2014gza}. The violation of the Null Energy Condition (NEC) has profound implications for the structure of spacetime, particularly in the context of BHs. While the NEC is traditionally considered a fundamental requirement for forming singularities, its violation opens the door to exotic solutions that exhibit non-singular BH configurations. In such scenarios, the matter and energy content of the spacetime, as described by the stress-energy tensor, allows for modifications of the usual gravitational collapse, potentially avoiding the formation of a curvature singularity at the center of the BH. These solutions often involve exotic matter or fields with negative energy density, which can lead to alternative scenarios such as regular BHs, bounce cosmologies, and traversable wormholes. Notably, models like the loop quantum gravity-inspired BHs and the regular BHs in higher-dimensional theories demonstrate that, by violating the NEC, one can obtain BH solutions that possess finite density at the core, thereby preventing the formation of an infinite curvature singularity. This opens exciting possibilities for resolving long-standing paradoxes in BH physics, such as the information paradox and the fate of matter that falls into a BH. For detailed discussion, see~\cite{rubakov2014null, creminelli2006starting, baldi2005inflation, visser1995scale, alexandre2021tunnelling, cai2024primordial, hochberg1998null, dubovsky2006null}.

While the foundational principles of GR have been tested to great accuracy in weak-gravity regimes, no unique theoretical necessity mandates Einstein's field equations~\citep{Will:2018bme}. A viable self-consistent theory of gravity can be built for any action satisfying four criteria~\citep{MTW,Will:2018bme}: \\
\begin{itemize}
\item it must yield Minkowski spacetime in the absence of matter and a cosmological constant,
\item it should be formulated solely from the Riemann curvature tensor and metric,
\item it must respect the symmetries and conservation laws of the stress-energy tensor,
\item it should reduce to Poisson's equation in the Newtonian limit. \\
\end{itemize}
\begin{figure}
\center{
\includegraphics[scale=1.0]{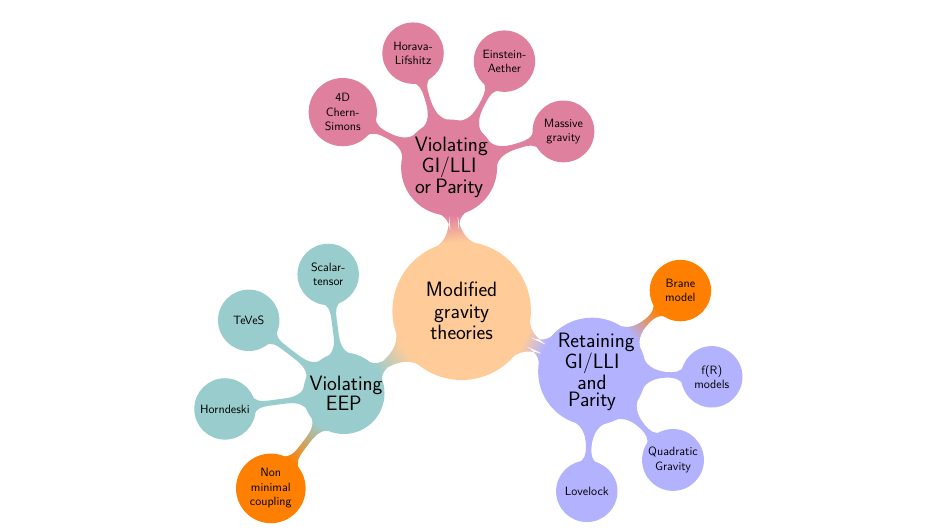}
}
\caption{Classification of modified gravity theories}
\label{fig:Mindmap}
\end{figure}

As mentioned in Sec.~\eqref{sec:Why?}, ${\cal R}$ defines the characteristic curvature scale. Just as the linear-order binomial expansion of  $(1 + x)^n$ begins to break down when $x$ approaches $1$ or greater --- necessitating the inclusion of all higher-order terms --- higher-order corrections to gravity become essential when the characteristic curvature scale ${\cal R}$ approaches unity. At these scales, more than the simplistic leading-order approximation provided by GR is required. For instance, in regions of extremely high curvature, such as near black hole singularities or during the early moments of the universe, the contributions from higher-order curvature terms (like \(R^2\), \(R_{\mu\nu}R^{\mu\nu}\), or even non-local terms) become increasingly significant. This breakdown signals the limits of classical GR and the need for a more complete framework, such as quantum gravity or modified gravity theories, to fully describe spacetime dynamics. Much like how the entire binomial series must be summed for \(x \approx 1\), these higher-order corrections are imperative for maintaining consistency and predictive power in extreme regimes, ultimately offering insights into physics beyond Einstein's theory.

Several modified gravity models have been proposed to provide crucial insights and testable predictions in the strong gravity and at cosmological scales~\citep{Woodard:2006nt,Nojiri:2006ri,Sotiriou:2008rp,Capozziello:2009dz,Alexander:2009tp,Nojiri:2010wj,DeFelice:2010aj,Capozziello:2011et,Clifton:2011jh,Hinterbichler:2011tt,deRham:2014zqa,Joyce:2014kja,2022-Shanki.Joseph-GRG}. There are several ways to classify these modified gravity theories, as shown in Fig. \eqref{fig:Mindmap},  we classify them into the following three categories:  \\
\begin{enumerate}
\item \emph{Metric theories of gravity preserving LLI and gauge invariance}: These theories modify the Einstein-Hilbert action to address specific issues close to the Big Bang singularity or the behavior of gravity at large scales to explain the current acceleration of the Universe. 
They are also used to improve the behavior of the theory at high energies, making it renormalizable under specific conditions (for instance, quadratic gravity)~\citep{Stelle1977-rd,Nenmeli:2021orl}. However, some of these theories may contain higher-derivative terms and may introduce ghosts (unphysical degrees of freedom), which challenge the consistency of these models at Planck scales.

\item \emph{Theories of gravity that violate gauge invariance, LLI or Parity:} 
In the late 1990s and early 2000s, a novel approach emerged that centered around the systematic development of a comprehensive dynamical EFT framework for evaluating the principles of relativity, as detailed in several seminal publications \citep{PhysRevD.55.6760, PhysRevD.58.116002, PhysRevD.66.056005}. The foundational assumption of this approach is that local Lorentz symmetry is upheld to a high degree of accuracy, implying that any deviations from this symmetry are minimal. Such deviations are then characterized by the interaction of a \emph{matter field} with a generic background tensor field, with components referred to as \emph{coefficients for Lorentz violation}. An EFT framework incorporating GR and the Standard Model of particle physics enables the systematic categorization of all the potential terms that violate symmetries such as CPT and Lorentz invariance. This framework has proven adaptable to numerous proposed quantum gravity models, related theories, and various testing methodologies. Numerous experiments have been evaluated within this framework \citep{PhysRevD.88.096006, PhysRevD.80.015020, RevModPhys.83.11}, facilitating comparative analysis across a broad range of tests, from astrophysical observations to Earth-based laboratory experiments. 

\item \emph{Theories of gravity that violate EEP:} As discussed earlier, EEP is a cornerstone of GR. Theories of gravity that violate the EEP arise from attempts to address fundamental questions that GR cannot resolve, such as the unification of gravity with quantum mechanics or the nature of dark matter and dark energy. In many of these theories, gravity is mediated by both a metric tensor and a scalar field. Since the scalar field also evolves in the 4-D spacetime, this leads to violations of LPI or LLI. In some of these modes, the additional 
(scalar) fields couple directly to matter, leading to non-minimal coupling and, hence, variations in the trajectories of test particles. \\
\end{enumerate}

Interestingly, the field equations in the above class of modified theories of gravity can also be expressed as $G_{\mu\nu} = \kappa^2 T_{\mu\nu}^{\text{eff}}$. In this case, the effective energy-momentum tensor depends on the energy-momentum tensor of matter and geometric quantities. As a result, in modified theories of gravity, the violation of the NEC plays a crucial role in shaping the physical consequences and predictions of these frameworks. Many extensions of GR, such as theories involving scalar-tensor gravity, $f(R)$ gravity, and others, allow for the possibility of NEC violations under certain conditions. These violations are often a consequence of the introduction of non-trivial fields or modified interactions between geometry and matter. For example, in $f(R)$ gravity, where the gravitational action is generalized to include arbitrary functions of the Ricci scalar, the field equations may permit energy-momentum tensors that do not satisfy the traditional NEC, enabling phenomena such as the existence of wormholes, dark energy, or even non-singular BHs. Similarly, in scalar-tensor theories, a scalar field can lead to exotic matter configurations that violate the NEC, facilitating the construction of solutions that allow for traversable wormholes or avoiding singularities in BH spacetimes. The violation of the NEC in these modified gravity theories often provides new avenues for resolving outstanding issues in cosmology and astrophysics, such as the cosmic no-hair conjecture, the structure of singularities, and the nature of dark matter and energy.

In the rest of this chapter, we will discuss some popular models and their observational implications in the multimessenger astronomy era. Due to paucity of space, we will not cover Brane world models and non-minimal coupling models of gravity explicitly. We refer to the readers to excellent reviews~\citep{Maartens:2010ar,Cipriano:2024jng,DiMicco:2019ngk,Cheong:2021vdb}

\section{Metric theories of gravity preserving Lorentz invariance and gauge invariance}
\label{sec:MGModels-ClassA}

The modified gravity theories in this category contain quadratic or higher-order curvature terms in the action, like $R^2, R_{\mu\nu} R^{\mu\nu}$ and $R_{\mu\alpha\nu\beta} R^{\mu\alpha\nu\beta}$. One of the key motivations of these theories of gravity is to alleviate the divergences in the quantum theory~\citep{Barth:1983hb,Belenchia:2016bvb} and modify the cosmological expansion of the early and late Universe. These models critically address the limitations of the $\Lambda$CDM Model and the absence of a definitive theory of quantum gravity. These models aim to develop a semi-classical framework that extends the successes of GR by incorporating higher-order curvature invariants and minimally or non-minimally coupled scalar fields, reflecting their origins in gauge theories and general bundle structures~\citep{Nojiri:2006ri, Nojiri:2010wj, DeFelice:2010aj, Sotiriou:2008rp, Capozziello:2009dz}.

This approach is founded on the premise that GR is gauge theory and can be organized within a general bundle structure \citep{Capozziello:2009dz}. Moreover, in unification schemes such as Superstrings, Supergravity, or Grand Unified Theories, effective actions consider non-minimal couplings to geometry or higher-order terms of the curvature invariants. These contributions arise from one-loop or higher-loop corrections in strong gravitational fields, in the QG (Quantum Gravity) regime, which is not yet available \citep{Nojiri:2006ri}. Specifically, this approach was adopted to address the quantization on curved spacetimes, resulting in corrective terms in the Einstein-Hilbert action due to quantum scalar fields interacting with background geometry and gravitational self-interactions \citep{Birrell:1982ix}. Furthermore, it has been recognized that such corrective terms are essential to obtain the effective action of QG at scales close to the Planck scale \citep{Vilkovisky:1992pb}. These approaches do not constitute full QG but are necessary as interim steps, leading towards it. Incorporating higher-order terms in curvature invariants (such as $R^{2}, R_{\mu\nu}R^{\mu\nu}, R_{\mu\nu\alpha\beta}R^{\mu\nu\alpha\beta}, R\Box R, \text{or} \ R\Box^{k}R$) and non-minimally coupled terms between scalar fields and geometry (such as $\phi^{2}R$) is essential when accounting for quantum corrections. These adjustments are crucial components of the effective Lagrangian of a gravity theory, as evidenced in the effective Lagrangian of strings or in Kaluza-Klein theories when implementing dimensional reduction~\citep{Gasperini:1992ym}.

However, the field equations derived from these modified gravity theories are typically of fourth order, contrasting with the second-order equations of GR. The higher-order derivatives in these theories have sparked mixed reactions due to their implications and challenges~\citep{Simon:1990ic,Mueller-Hoissen:1990jsv, Belenchia:2016bvb}. One of the key advantages of these models is that they serve as a low-energy approximation to a yet-unknown quantum theory of gravity, offering insights into how gravity might behave at extreme scales. Also, these models provide a broader unifying theory by incorporating higher-order terms, potentially bridging gaps between GR and other extended modified gravity models. Lastly,  these modifications can resolve issues in GR and cosmology, like taming the singularity, testing the validity of Birkhoff and no hair theorems beyond GR, explaining the late-time acceleration without introducing any exotic matter~\citep{DeFelice:2010aj,2007-Starobinsky-JETPLett,2007-Hu.Sawicki-PRD,Johnson:2019vwi}.

However, these models face significant criticisms~\citep{Simon:1990ic, Belenchia:2016bvb}. It is argued that some solutions predicted by these theories may lack physical relevance, as they could lead to unphysical gravitational fields or instabilities~\citep{Simon:1990ic, Belenchia:2016bvb}. Determining whether solutions correspond to physical reality or mathematical artifacts remains an ongoing challenge. 
Some models still need a well-posed initial value formulation, making their predictive power questionable. Also, incorporating quantum effects and reconciling these models with quantum field theory need deeper exploration. 

Despite the associated challenges, these modified gravity theories remain instrumental as alternatives to GR and as frameworks for investigating fundamental properties of spacetime. They provide critical platforms to test and explore key concepts, including the resolution of singularities, mechanisms driving early and late-time cosmic acceleration, the structure of the Hamiltonian in gravitational systems, and the compatibility of such models with energy conditions and causality principles. 

In the following subsections, we delve into specific modified gravity models, examining their implications for strong gravity regimes, cosmological dynamics, and their alignment with observational data. These explorations highlight the potential and limitations of such theories in addressing unresolved issues within the standard gravitational paradigm.

\subsection{$f(R)$ theories of gravity}
\label{sec:f(R)}

Initially proposed by Buchdahl in 1970, $f(R)$ gravity theory is one of the most straightforward extensions of GR~\citep{1970-Buchdahl-MNRAS}. It is formulated by relaxing the assumption that the Einstein-Hilbert action for the gravitational field is strictly linear in the Ricci curvature scalar, i.e. $f(R) = {R}/{(16\pi G)}$. This formulation embodies a class of theories defined by arbitrary functions of the Ricci scalar $R$ and is thus seen as the simplest example of an extended theory of gravity~\citep{Woodard:2006nt, Nojiri:2006ri, Sotiriou:2008rp, Capozziello:2009dz, Nojiri:2010wj, DeFelice:2010aj, Capozziello:2011et}. $f(R)$ theories of gravity have garnered significant interest in cosmology due to their natural ability to address the limitations of the Standard Cosmological Model based on GR. In addition to their fundamental physics implications, these theories demonstrate inflationary behaviors that align with the Cosmic Microwave Background Radiation (CMB) observations, making them realistic cosmological models \citep{Starobinsky:1980te, La:1989za}. Furthermore, through conformal transformations, it can be shown that the higher-order and non-minimally coupled terms can be equated to Einstein gravity along with one or more minimally coupled scalar fields \citep{Teyssandier:1983zz, Maeda:1988ab,  Adams:1990pn, Wands:1993uu, Capozziello:1998dq}.

In $f(R)$ gravity, the Einstein-Hilbert action \eqref{E-H action} is generalised as
\begin{equation}\label{f(R) action with matter}
S = \frac{1}{2\kappa^{2}}\int d^{4}x \sqrt{-g} f(R) + \int d^{4}x \mathcal{L}_{M}(g_{\mu\nu}, \psi_{M}).    
\end{equation}
Varying the above action w.r.t the metric $g_{\mu\nu}$ leads to the following equations of motion:
\begin{equation}\label{f(R) field equations}
f_{,R}(R)R_{\mu\nu} - \frac{1}{2}g_{\mu\nu}f(R) + g_{\mu\nu}\Box f_{,R}(R) - \nabla_{\mu}\nabla_{\nu}f_{,R}(R) = \kappa^{2}T_{\mu\nu}^{(M)},  
\end{equation}
where the energy-momentum tensor of the matter fields is given by Eq.~\eqref{def:StressTensor}. A detailed derivation can be found in Appendix \eqref{App:f(R)}. 
Like in GR, the stress-tensor satisfies the continuity equation:
\begin{equation}
\nabla^{\mu}T_{\mu\nu}^{(M)} = 0,    
\end{equation}
which follows from the covariant derivative of the metric field equations (\ref{f(R) field equations}). Taking the trace of the field equations, (\ref{f(R) field equations}), we get
\begin{equation}
3\Box f_{,R}(R) + f_{,R}(R)R - 2f(R) = \kappa^{2}T^{(M)} \, ,    
\end{equation}
where $T^{(M)} = g^{\mu\nu}T_{\mu\nu}^{(M)}$, and with the identifications
\begin{equation}\label{identification of scalar degree}
\phi_{S} \equiv f_{,R}(R), \ \frac{dV_{\text{eff}}}{d\phi_S} \equiv \frac{2f(R) - f_{,R}(R)R + T^{(M)}}{3},
\end{equation}
a Klein-Gordon equation for the effective $\phi_{S}$ scalar field can be obtained as
\begin{equation}
\Box\phi_S - \frac{dV_{\text{eff}}}{d\phi_S} = 0.    
\end{equation}
This implies that there exists a propagating scalar degree of freedom $\phi_S$, which is sometimes called Chameleon~\citep{Brax:2008hh}. It is interesting to note that when $f(R) = R$, $f_{,R}(R) = 1$ leading to vanishing of $\Box f_{,R}(R)$. Since the scalar field $\phi$ 
vanishes, GR does not have any propagating scalar degree of freedom. 
Instead, we obtain $R = \kappa^{2}T$. It is worth mentioning that by solving 
(\ref{identification of scalar degree}) \textit{w.r.t} $\phi_S$, we can in principle express the Ricci curvature $R$ in terms of $\phi_S$, $R = R(\phi_S)$.

Cosmology stands as a foundational area driving the application of $f(R)$ gravity. As elaborated upon below, $f(R)$ gravity may serve as a geometric framework to elucidate the enigmatic facets of the Universe \citep{Woodard:2006nt, Nojiri:2006ri, Sotiriou:2008rp, Capozziello:2009dz,DeFelice:2010aj, Capozziello:2011et}. The crux of this interpretation hinges on the capacity to amalgamate the additional degrees of freedom of $f(R)$ gravity into an effective curvature stress-energy tensor \citep{Capozziello:2012uv}, consequently engendering manifestations of dark energy.
The Starobinsky model provides an elegant explanation for the early Universe's rapid expansion~\citep{Starobinsky:1980te}. This model modifies the Einstein-Hilbert action 
\eqref{E-H action} to include quadratic corrections in the Ricci scalar that hold particular importance in the early Universe.

\subsubsection{Applications of f(R) in early Universe}
\label{sec:Starobinsky}

The Starobinsky model is
\begin{equation}
\label{eq:Starobinskyaction}
f(R) = R +  \frac{R^{2}}{6 M^2},    
\end{equation}
where $M$ is a dimensionful constant of mass dimension.  \( R^2 \) term is usually interpreted as a quantum correction to GR, arising in semi-classical gravity when quantizing fields in curved spacetime~\citep{Birrell:1982ix, Vilkovisky:1992pb}. The higher-order curvature term (\( R^2 \)) contributes to avoiding or smoothing singularities.

Substituting the above action in the field equations \eqref{f(R) field equations}, we get:
\begin{equation}
\begin{split}
G_{\mu\nu} & + \frac{1}{6 M^2}\Big[2R\left(R_{\mu\nu} - \frac{1}{4}g_{\mu\nu}R\right) + 2(g_{\mu\nu}\Box R - \nabla_{\mu}\nabla_{\nu}R)\Big] = \kappa^{2}T_{\mu\nu}^{(M)}.
\end{split}
\end{equation}
The trace of the above equation leads to:
\begin{equation}
\begin{split}
\Box R & - M^2 (R + \kappa^{2}T^{(M)}) = 0.
\end{split}   
\end{equation}
The above equation can be seen as an effective Klein-Gordon equation for the effective scalar field degree of freedom $R$ (sometimes called scalaron)~\citep{DeFelice:2010aj}. In other words, the \( R^2 \) term modifies the dynamics of GR and introduces an effective scalar degree of freedom.

The Starobinsky model is often reformulated in the Einstein frame using conformal (Weyl) transformations to understand the implications better. Appendix \eqref{App:ConformalTransf} contains the details that maps $f(R)$ gravity to a scalar-tensor theory. The conformal transformation introduces an effective scalar field \( \phi \), referred to as the inflaton, whose potential ($V(\phi)$) governs the inflationary dynamics~\citep{Kolb:1990vq,Bassett:2005xm,Nojiri:2017ncd, Odintsov:2023weg}:
\begin{equation}
V(\phi) = \frac{3M^2 M_{\text{Pl}}^2}{4} \left( 1 - e^{-\sqrt{\frac{2}{3}} \frac{\phi}{M_{\text{Pl}}}} \right)^2.
\end{equation}
This potential is exponential-like and nearly flat for large \( \phi \), supporting the slow-roll conditions required for inflation~\citep{Bassett:2005xm}. The model naturally provides the required 50-60 e-folds of inflation, solving the flatness and horizon problems. 

Using the standard slow-roll conditions for the scalar perturbation equations~\citep{Mukhanov:1990me, Malik:2008im}, 
the Starobinsky model predicts the scalar spectral index to be $\approx 0.965$, which is consistent with observations from the Planck satellite~\citep{Planck:2018jri}. From the Planck observations, the inflationary scale is determined by \( M \), which is related to the amplitude of the primordial power spectrum. Observations suggest \( M \sim 10^{13} \) GeV. The model predicts a tiny tensor-to-scalar ratio, \( r \sim 0.004 \), which aligns with the lack of strong detections of primordial GWs~\citep{Planck:2018jri}. Also, it predicts minimal non-Gaussianity in the primordial density fluctuations, consistent with current data~\citep{Planck:2019kim}.

The model's predictions remain a benchmark for inflationary theory. Current and upcoming missions such as the Simons Observatory, CMB-S4,~\citep{Ghigna:2020wat} and space-based gravitational wave detectors~\citep{LISA:2022kgy} aim to test its predictions more rigorously by searching for subtle imprints of inflation in the CMB and large-scale structure.

\subsubsection{Applications of f(R) in late-time Universe}

As mentioned in the Introduction, understanding the late-time accelerated expansion of the Universe is one of the most pressing challenges in modern cosmology. GR alone cannot account for this acceleration when considering ordinary matter and radiation. This has led to the hypothesis of dark energy (DE), a mysterious form of energy that permeates space and drives the current accelerated expansion of the Universe~\citep{2003-Padmanabhan-PRep,2003-Peebles.Ratra-RMP, Copeland:2006wr,Huterer:2017buf, Kamionkowski:2022pkx}. 
The primary candidate for dark energy is the cosmological constant (\( \Lambda \)), which represents a constant vacuum energy density. While \( \Lambda \) is consistent with many observations, its value inferred from cosmology is many orders of magnitude smaller than the vacuum energy density predicted by quantum field theory. This discrepancy, known as the \emph{cosmological constant problem}, has motivated the search for alternative explanations. Dynamic dark energy models, like quintessence, k-essence, phantom energy, and chameleon fields introduce scalar fields that vary over time, allowing the dark energy density to evolve with the Universe. These models aim to address the limitations of a static cosmological constant while providing a framework to describe late-time acceleration. An alternative to dark energy is to modify GR~\citep{Joyce:2014kja}. Among these, \( f(R) \) models are particularly appealing due to their mathematical simplicity. 

Let us consider the FLRW line element: 
\begin{equation}
\label{eq:FRWmetric}
ds^2 = -dt^2 + a^2(t) \, d\Sigma^2 \quad {\rm where} \quad 
d\Sigma^2 = \frac{dr^2}{1- k r^2} + r^2 d\Omega^2 \, , 
\end{equation}
where $a(t)$ is the scale factor, and $k$ is the spatial curvature of the Universe, which can be $1, 0,$ or $-1$. For the above metric, the modified field equations \eqref{f(R) field equations} reduce to the following:
\begin{equation}
\begin{split}
H^{2} + \frac{k}{a^{2}} & = \frac{\kappa^2}{3} \rho_{\rm total} \, ,
\qquad
\dot{H} - \frac{k}{a^{2}}  = - \frac{\kappa^2}{2} (\rho_{\rm total} + P_{\rm total}).
\end{split}
\label{Modified Friedmann equation}
\end{equation}
where 
\begin{equation}
\begin{split}
\rho_{total} = \frac{1}{f_{,R}}\rho_{M} + \frac{\rho_{\rm curv}}{\kappa^2},  & \quad P_{total} = \frac{1}{f_{,R}}P_{M} + \frac{P_{\rm curv}}{\kappa^2}, \\
\rho_{\rm curv} = - \frac{1}{f_{,R}}\Big[ \frac{f - Rf_{,R}}{2} + 3Hf_{,RR}\dot{R}\Big], & \quad 
P_{\rm curv} = \frac{1}{f_{,R}}\left(\frac{f - Rf_{,R}}{2} + f_{,RR}\ddot{R} + f_{,RRR}\dot{R}^{2} + 6f_{,RR}\dot{R}\right)\, ,
\end{split}
\end{equation}
$\rho_M (P_M)$ refers to the density (pressure) of the matter content of the Universe,
$\rho_{\rm curv} (P_{\rm curv})$ refer to the density (pressure) contributions from the non-linear terms in the gravity action,  $H(t) \equiv \dot{a}(t)/a(t)$ is the Hubble parameter. Note that we have ignored the radiation contribution. From the \textit{r.h.s} of Eq.~\eqref{Modified Friedmann equation}, the acceleration is achieved if $\rho_{\text{tot}} + 3p_{\text{tot}} < 0$. This leads to the following condition: 
\begin{equation}
\rho_{\text{curv}} + 3p_{\text{curv}} \equiv \frac{3}{f_{,R}(R)}\Big[\dot{R}^{2}f_{,RRR}(R) + H(t)\dot{R}f_{,RR}(R) + \ddot{R}f_{,RR}(R) - \frac{1}{3}[f(R) - Rf_{,R}(R)]\Big] < - \rho_M \, .
\end{equation}
Given a form of $f(R)$, the above modified Friedmann equations \eqref{Modified Friedmann equation} and the conservation of the matter sector, should, in principle, determine the late-time dynamics. However, $f(R)$ must satisfy some conditions. For instance, from the above equation, we see that $f_{,R}(R) \neq 0$; otherwise, this will lead to divergence of $\rho_{\rm curv} (P_{\rm cur})$. Similarly, to avoid any ghost instabilities, $f(R)$ must satisfy the following conditions~\citep{DeFelice:2010aj}:
\begin{equation}
f_{, R}>0 \text { and } f_{, R R}>0 \text { for } \quad R \geq R_0 \quad(>0),
\label{eq:fRconstraints}
\end{equation}
where $R_0$ is the present value of $R$. Moreover, from the observational perspectives, a viable $f(R)$ model reproducing the matter-dominated era, satisfying the local gravity constraints and consistent with the equivalence principle, should satisfy the following condition:
$$
f(R) \rightarrow R-2 \Lambda, \text { for } \quad R \geq R_0
$$
where $\Lambda$ is a constant and to depict a late-time stable de Sitter solution~\citep{2007-Hu.Sawicki-PRD}, the $f(R)$ model also needs to satisfy
$$
0<\left(\frac{R f_{,R R}}{f_{, R}}\right)_r<1 \quad \text { at } \quad 
r \equiv -\frac{R f_{, R}}{f}=-2 .
$$
Many $f(R)$ models have been proposed that lead to the late time acceleration of the Universe~\citep{2007-Hu.Sawicki-PRD,2007-Starobinsky-JETPLett,2007-Amendola.etal-PRD,2007-Appleby.Battye-PLB,2008-Tsujikawa-PRD}. $f(R)$ gravity has the advantage of describing both inflationary dynamics and the dark energy era in a unified framework \citep{nojiri2003modified, nojiri2006modified, nojiri2007unifying, odintsov2019unification, oikonomou2021rescaled, oikonomou2021unifying}. Among these, the most popular models are the Hu-Sawicki~\citep{2007-Hu.Sawicki-PRD} and Starobinsky model~\citep{2007-Starobinsky-JETPLett}.  Recently, \cite{Johnson:2019vwi} considered a generic $f(R)$ satisfying the above constraints \eqref{eq:fRconstraints} and using the late-time expansion history realizations constructed by \cite{Shafieloo2006}  in the range $0<z<1.2$. Table \eqref{tab:fRModels} gives the best fit with root mean square error (RMSE) for the constructed realizations of $f_{,R}(z)$ in the range $0<z<1.2$ for the two popular $f(R)$ that explain the late-time acceleration. For an arbitrary $f(R)$ model that leads to late-time accelerated expansion, it was explicitly shown that the density perturbations and its time-derivative evolves differently compared to the $\Lambda$CDM model at lower redshifts~\citep{Johnson:2019vwi}. This has potential observable consequences in the future Euclid mission~\citep{2018-Amendola.others-LRR}.

\begin{table}[!h]
\begin{tabular}
{| p{1cm} | p{8cm}| p{8cm} | } \hline 
& ~~~~~~~~~~~~~~~~~~~~~~~~{\bf $f_{,R}(R)$} & {\centering ~~~~~~~~~~~~~~~~~~~~~~~~~~\bf Best fit} \\ 
\hline
\centering 1 & 
$$
1 - 2\lambda n \dfrac{R}{R_0}\left[ 1+ \left(\dfrac{R}{R_0}\right)^2 \right]^{-(n+1)}
$$
\centerline{Starobinsky \citep{2007-Starobinsky-JETPLett}}
 &
\begin{equation*}
n=3.676, \quad
\lambda = 1.312 \times 10^6, \quad R_0 = H_0^2
\end{equation*}

\centerline{
$RMSE = 6.8 \times 10^{-4}$}
 \\ 
\hline
\centering 2 & \begin{equation*}
\label{eq:Husawifit}
1 -  n \dfrac{c_1}{c2} \dfrac{\left(\dfrac{R}{R_0}\right)^{n-1}}{\left[\left(\dfrac{R}{R_0}\right)^n - 1 \right]^2}
\end{equation*}
\centerline{Hu \& Sawicki \citep{2007-Hu.Sawicki-PRD}}&

\begin{equation*}
n=7.176, \quad c_1/c_2=8.67 \times 10^5, \quad R_0=H_0^2
\end{equation*}

\centerline{$
RMSE = 6.6 \times 10^{-4}
$} \\ 
\hline
\end{tabular}
\caption{Best fit for the two popular $f(R)$ models that can explain the late-time acceleration. More details, see Ref. \citep{Johnson:2019vwi}.}
\label{tab:fRModels}
\end{table}
%

\subsubsection{Implication for BHs}

According to GR, three measurable quantities (mass, charge, and angular momentum) fully characterize isolated BHs in equilibrium~\citep{Heusler:1996jaf, Chrusciel:2012jk, Bambi:2015kza, Mann:2021mnc}. In other words, any \emph{deformations of the BH horizon} will finally result in a BH configuration with the above three quantities~\citep{Heusler:1996jaf}. Therefore, any object's material properties are unmeasurable as it collapses into or gets absorbed by a BH~\citep{1998-Bekenstein-Arx}. 

When two Black holes (BHs) merge to form another BH, the remnant BH's event horizon is highly distorted. It radiates GWs until it settles down to an equilibrium configuration~\citep{Vishu}. GWs emitted, referred to as quasi-normal modes (QNMs), are a superposition of damped sinusoids and depend only on the parameters characterizing the BH, namely, its mass and spin (astrophysical BHs are not likely to be electrically charged)~\citep{Nollert:1999ji, Kokkotas:1999bd, Konoplya:2011qq, Chen:2021cts}.  QNMs are the fingerprints of the final BH. The simplicity of the spectrum allows one to identify the Kerr solution~\citep{Barack:2018yly}.
\begin{figure}[!htb]
\begin{center}
 \includegraphics[width=13cm]{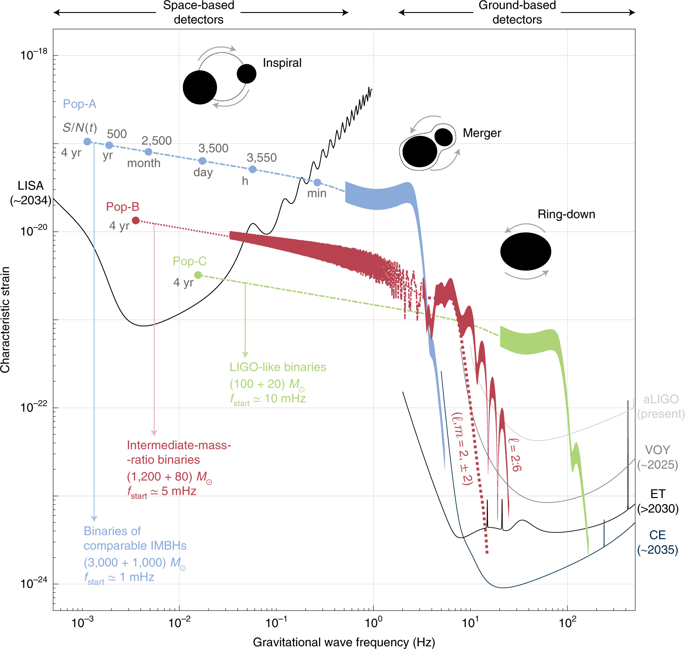}
\end{center}
\caption{\label{fig:IMBHQNM} This plots the characteristic strain $h$ as a function of observable frequency for (i) several ground-based detectors, the currently operational advanced LIGO (aLIGO), advanced Virgo (AdV), and future proposed Einstein Telescope (ET), Cosmic Explorer (CE), and (ii) Space-based detectors like LISA. The three (green, red, and blue) curves correspond to the three different scenarios that the interferometers will detect. Source: \cite{2019-Jani.etal-NatureAstro}}
\end{figure}
Figure \ref{fig:IMBHQNM} gives three possible binary BH collision scenarios~\citep{2019-Jani.etal-NatureAstro}. It is evident that in the next generation of GW detectors (e.g., the Cosmic Explorer~\cite{Evans:2023-CE}), the signal-to-noise ratio in the quasi-normal mode regime alone could be as large as ${\rm SNR} > 50$. Specifically, using a signal-to-noise ratio threshold with a lower cutoff of 100, CE and ET are projected to detect $\sim 100$ events annually~\citep{Branchesi:2023, Evans:2023-CE}. Since the ringdown phase signal is a characteristic of the final BH, and due to a large signal-to-noise ratio in Cosmic Explorer and Einstein Telescopes, BH binary systems can confirm or contradict GR.

As mentioned above, if GR is the correct theory of gravity, the QNM frequencies will depend only on the final BH's mass and angular momentum. However, as mentioned earlier, singularities indicate that GR cannot be a universal theory of space-time and needs strong gravity modifications. Unlike GR, whose field equations contain only up to second-order derivatives, the equations of $f(R)$ gravity models include higher derivatives. Unlike lower-derivative corrections, the higher-order derivative corrections, with many degrees of freedom,  cannot be treated as a perturbation to the original theory, especially in the early Universe. However small they may appear in the present epoch, higher-derivative terms make the new theory drastically different from GR~\citep{Barth:1983hb, Simon:1990ic, Woodard:2006nt, Woodard:2015zca}. See also the detailed discussion in Appendix~\ref{App:HDTheories}.

This has two immediate implications. First, it can lead to new BH solutions. Recently, \cite{Xavier:2020ulw} obtained an infinite number of exact static, Ricci-flat spherically symmetric vacuum solutions for a class of $f(R)$ theories of gravity that includes all higher-powers of $R$. Specifically, they analytically derived two exact vacuum BH solutions for the same class of $f(R)$ theories. The two BH solutions have event horizons at the same radial distance; however, their asymptotic features differ. This explicitly showed that these theories do not support Birkhoff theorem~\citep{Schmidt:2012wj}.
Later, \cite{Sunny:2023dgz} explicitly constructed multiple slow-rotating BH solutions, up to second order in rotational parameter, for the same class of $f(R)$ models. Here again, they analytically showed that multiple vacuum solutions satisfy the field equations up to the second order in the rotational parameter. Specifically, they showed that the multiple vacuum solutions depend on arbitrary constants, which depend on the coupling parameters of the model. Interestingly, they showed the violation of the no-hair theorem for $f(R)$ theories directly from a metric formalism without conformal transformation.

Second, the dynamics of the higher derivative theory is much more complicated than GR. Hence, the gravitational wave propagation in these theories will also be more complicated than in GR. In particular, in the ringdown phase (see Figure 3) it can lead to three possibilities~\citep{Barack:2018yly}: First, modified theories can have a spectrum of even and odd modes while preserving isospectrality. Second, they break isospectrality; they do not emit GWs with equal energy in the two polarization states. Third, they mix the two GW polarization modes, so there is no clear distinction between them. The breaking of isospectrality has been systematically investigated for an arbitrary $f(R)$ gravity~\citep{Bhattacharyya:2017tyc,Bhattacharyya:2018qbe}. It was shown that the longitudinal mode couples to one of the polarization modes. This work has culminated in a definite proposal of a test parameter that can differentiate between modified gravity and GR~\citep{Shankaranarayanan:2019yjx}.

\subsection{Quadratic gravity}

Expanding the Einstein-Hilbert action to include  $f(R)$ theories represents a minimal modification that introduces higher-order curvature invariants. Beyond \( f(R) \), other curvature invariants can be incorporated into the action, such as terms constructed from the Ricci or Riemann tensors, e.g., \( R^2 \), \( R_{\mu\nu}R^{\mu\nu} \), or \( R_{\mu\nu\sigma\lambda}R^{\mu\nu\sigma\lambda} \). 
There is no unique principle dictating the preference for one invariant over another, allowing for combinations such as \( R_{\mu\nu\sigma\lambda}R^{\mu\sigma}R^{\nu\lambda} \) or terms involving the Weyl tensor. Since these terms are inherently higher-derivative and non-linear, the resulting equations of motion become significantly more complex. We consider invariants constructed from the Riemann tensor and the metric up to the second order to keep things transparent. The generalized gravitational Lagrangian is given by~\citep{Stelle1977-rd}: 
\begin{equation}
\label{eq:GenIIorderGraviaction}
    \mathcal{L}=\frac{1}{2\kappa^2}\,R+ \alpha R^2+\beta R_{\mu\nu}R^{\mu\nu}+\gamma R_{\mu \nu \sigma \lambda}R^{\mu \nu \sigma \lambda}
\end{equation}
where \( \alpha \), \( \beta \), and \( \gamma \) are constants. When \( \gamma = 0 \), this framework is commonly referred to as \emph{quadratic gravity}. As we will elaborate in Sec.~\eqref{sec:lovelock}, the following action represents the most general second-order action constructed from the Riemann tensor:
\begin{equation}\label{ch2eqnQGMatr}
    S=\int d^4x \, \sqrt{-g}\left[\frac{1}{2\kappa^2}\, R-\alpha R_{\mu\nu}R^{\mu\nu} +\beta R^2 \right]+S_{Matter}(g_{\mu \nu},\psi) \, ,
\end{equation}
where $S_{Matter}$ is the matter action. Like in \( f(R) \) theories, these parameters $(\alpha,\beta)$ can be constrained through experimental observations~\citep{Nenmeli:2021orl,2021-Kim_Picker-EuPhyJ,Das:2022hjp,Chowdhury:2022ktf}. Including an \( R_{\mu\nu\sigma\lambda}R^{\mu\nu\sigma\lambda} \) term does not alter the equations of motion, as it can be absorbed via the Gauss-Bonnet term~\citep{1938-Lanczos-AnnalMath,Yale:2011usf}\footnote{See the detailed discussion in Sec.~\eqref{sec:lovelock}}. Note that no particular relation between $\alpha$ and $\beta$ can make the equations of motion second order~\citep{2005-Nojiri_Odintsov-PLB, Nenmeli:2021orl}. Hence, the  action~\eqref{ch2eqnQGMatr} is called fourth-order gravity since it leads to fourth-order equations~\citep{1990-Schimming_Schmidt}. 

This extended action forms the basis of Quadratic Gravity, which addresses the incompleteness of GR in regions of strong gravity~\citep{Stelle1977-rd}. Quadratic gravity introduces higher-order curvature terms into the gravitational action, resulting in field equations of fourth order. They were initially studied by Utiyama and DeWitt~\citep{1962-Utiyama.DeWitt-JMathPhy} and formalized by Weinberg and Deser~\citep{2018-Salvio-FrontinPhy}, this theory gained prominence when Stelle demonstrated its renormalizability~\citep{Stelle1977-rd}. Quadratic gravity extends GR by adding new propagating degrees of freedom: a massive scalar and a massive spin-2 graviton, with masses determined by \( \alpha \) and \( \beta \). To ensure physical consistency (e.g., avoiding negative energy solutions), constraints like \( \alpha > 0 \) and \( 3\beta > \alpha \) are required.

The classical instability of quadratic gravity, mainly when \( \alpha \neq 0 \), has raised concerns about its practical applicability. Nonetheless, these modifications lead to intriguing possibilities from a quantum perspective, such as richer phenomenology in strong gravity regimes and potential quantum corrections. These aspects will be explored further in the rest of this subsection.

\subsubsection{Generalized uncertainty principle and quadratic gravity}

Even without a definitive ultraviolet completion, quantum gravitational phenomena well below a specified cutoff scale can be captured using an effective quantum field theory of gravity, which introduces corrections to GR~\citep{Duff:1974ud, Hamber:1995cq, Donoghue:1997hx, Donoghue:2017pgk}. 
Models of quantum gravity that seek to reconcile GR and Quantum Mechanics predict the existence of a minimum measurable length --- a fundamental limit on positional uncertainty, typically around the Planck length. In other words, quantum gravity suggests the existence of a minimum measurable length and/or a maximum measurable particle momentum~\citep{1994-Garay-IJMPD,2012-Hossenfelder-LRR}. Quantum gravity Phenomenology investigates the experimental implications of this fundamental scale~\citep{Amelino-Camelia2013-xs, Scardigli2014-wm,2022-Addazi.etal-PPNP}. This scale is often modeled by modifying the traditional Heisenberg uncertainty principle to a Generalized Uncertainty Principle (GUP):
 \begin{equation}
      \Delta_{x} \geq \frac{\hbar}{2} (\frac{1}{\Delta_{p}}+\beta\Delta_{p}) 
  \end{equation}
where $\beta$ is a parameter characterizing the GUP. While, theoretically $\beta$ is assumed to take values near the Planck scale, this needs to be fixed by observations.  

The GUP extends the standard Heisenberg Uncertainty Principle to incorporate effects from quantum gravity. This framework predicts minimal measurable length, maximal momentum, and non-commutative spacetime geometry. The GUP is derived by deforming the classical position-momentum Poisson algebra, which results in modified uncertainty relations. \cite{1995-Kempf-PRD} proposed one of the most prominent GUP models to account for a minimal length scale, often linked to Planck-scale physics. (See also, \cite{1995-Maggiore-PRD, 1994-Kempf-JMatPhy}). A minimal length is interpreted as a lower bound on the uncertainty in position measurements, a fundamental feature of the GUP.  
To introduce the GUP, consider the position operator \(\hat{x}\) and momentum operator \(\hat{p}\). Their commutation relation is modified from the standard one to reflect the effects of quantum gravitational corrections, leading to expressions consistent with a minimal length formulation:
\begin{equation}
\left[ \hat{x},\hat{p}\right]=i\hbar \left[1+\zeta \hat{x}^2 + \xi \hat{p}^2  \right]
\end{equation}
lead the uncertainty relation with $\delta= \zeta \left< \hat{x} \right>^2 + \xi \left<  \hat{p} \right>^2$ ($\zeta$ and $\xi$ are constants) as:
\begin{equation}
\Delta x~\Delta p \geq \frac{\hbar}{2}\left[1+\zeta (\Delta x)^2 + \xi (\Delta p)^2+\delta \right],
\end{equation}
leads to a non-zero minimal uncertainty in both position $\Delta x_{0}$ (for $\zeta>0$) and momentum $\Delta p_{0}$ (for $\xi>0$). Minimal uncertainty in position or momentum implies that the corresponding operators are no longer self-adjoint but remain symmetric. Consequently, position eigenstates \(|x \rangle\) or momentum eigenstates \(|p \rangle\) do not exist within the framework of the modified Heisenberg algebra. This modification prevents the algebra from being represented on traditional position or momentum wavefunctions, such as \(\langle x|\psi \rangle\) or \(\langle p|\psi \rangle\), in a Hilbert space.

By setting \(\zeta = 0\) (eliminating minimal uncertainty in position), the algebra can still be expressed in a convenient representation of momentum-space wavefunctions. Under these conditions, the associative Heisenberg algebra generated by \(\hat{x}\) and \(\hat{p}\), along with their modified commutation and uncertainty relations, holds for \(\zeta = 0\) and \(\xi > 0\). This allows for simplified analysis in momentum space while preserving the effects of minimal position uncertainty:
\begin{align*}
\left[ \hat{x},\hat{p}\right]=i\hbar \left[1+ \xi \hat{p}^2 \right]     \quad \Longrightarrow \quad 
\Delta x \,\Delta p  \geq \frac{\hbar}{2}\left[1+ \xi (\Delta p )^2 + \xi \langle p\rangle^2 \right]
\end{align*}
gives the minimal position uncertainty in position as
\begin{equation}
\Delta x_{min}(\langle p\rangle)=\hbar\sqrt{\xi}\sqrt{1+\xi\langle p\rangle^2}.
\end{equation}
Thus, the minimal uncertainty in position can be read as $
\Delta x_{min}(\langle p\rangle)=\hbar\sqrt{\xi}$. \cite{2018-Todorinov-AnnalPhy} expanded this algebra to the full Minkowski spacetime:
\begin{equation}
[\hat{x}^{\mu},\hat{p}^{\nu}]=i\hbar\,\eta^{\mu \nu}[1-\gamma ~\hat{p}^{\sigma}\,\hat{p}_{\sigma}]-2i\gamma \hat{p}^{\mu}\hat{p}^{\nu}
\end{equation}
where $\gamma = \gamma_0 \kappa^2$. 
It is important to note that \(\hat{x}^\mu\) and \(\hat{p}^\nu\) represent the physical position and momentum operators, even though they are not canonically conjugate in the standard sense. To define a conjugate algebra, one introduces auxiliary 4-vectors \(\hat{x}^\mu_0\) and \(\hat{p}^\nu_0\), where \(\hat{x}^\mu_0\) resembles the standard position operator. The physical momentum is then expressed as \(\hat{p}^\mu = \hat{p}^\nu_0 / (1 + \gamma \, \hat{p}^\sigma_0 \hat{p}_{\sigma 0})\). Expanding to first order in \(\gamma\), the physical momentum takes the form:
\begin{equation}
\hat{p}^{\mu}\simeq\hat{p}^{\nu}_{0}\left(1-\gamma ~\hat{p}^{\sigma\,0}\,\hat{p}_{\sigma \,0}\right)
\end{equation}
which in term modifies the Klein-Gordon equation~\citep{2020-Todorinov-AnnalPhy,2021-Todorinov-AnnalPhy}. It has been shown that for energies below the Planck energy, the effective action for the modified Klein-Gordon equation does not have Ostrogradsky ghosts. The corresponding operators in the position eigenspace are represented as;
\begin{align}\label{GUPtrans}
x	&	\rightarrow ~~x;				\quad
\partial_{\nu}\rightarrow\partial_{\nu}\left(1-\gamma~\Box\right).
\end{align}
\cite{Nenmeli:2021orl} extended the analysis to spin-2 field using Gupta-Feynman formalism~\citep{1968-Gupta-PR,Feynman2002-hb}. 
Specifically, they showed that the GUP corrected gravity action reduces to 
the quadratic gravity action \eqref{ch2eqnQGMatr} when $\alpha=2\beta=\gamma_0/\kappa^{2}$. For details see Appendix~\ref{App:GuptaFeynman}.

Stelle gravity --- unlike $f(R)$ --- has extra massive spin-2 and spin-0 modes~\citep{Stelle1978-du}. Intriguingly, for this class of Stelle theories, the masses of the spin-0 and spin-2 modes  \emph{coincide} $(1/\sqrt{2 \gamma_0})$. Also, this class of Stelle theories does not have malicious ghosts~\citep{Antoniadis1986-hj}. In the rest of this section, we now look at the implications of this in the early Universe and gravitational waves.

\subsubsection{Implications for early Universe }

The equations of motion for generic Stelle gravity action~\eqref{ch2eqnQGMatr} (for $\alpha$ and $\beta$ unconstrained) are
\begin{eqnarray}
\label{app:StelleEOM01}
 \frac{1}{\kappa^2} G_{\mu\nu}&+& 2(\alpha-2\beta) \nabla_{\mu} \nabla_{\nu} R 
-2 \alpha \Box R_{\mu\nu} -(\alpha- 4 \beta)g_{\mu\nu}\Box R   
- 4 \alpha R^{\rho\sigma}R_{\mu\rho\nu\sigma} + 
4\beta RR_{\mu\nu}+ g_{\mu\nu}(\alpha R^{\rho\sigma}R_{\rho\sigma}-\beta R^{2})  = T_{\mu\nu}\,.
\end{eqnarray}
Substituting the flat FLRW line-element \eqref{eq:FRWmetric} in the above equations, we get: 
 \begin{eqnarray}
 \label{eq:TimeFriedmanGenStelle}
H^{2}-(12\beta-4\alpha)\kappa^{2}\left(\dot{H}^{2} 
-2 H \, \ddot{H} - 6H^{2}\dot{H}\right) = 0 & & ~~\\
\label{eq:SpaceFriedmanGenStelle}
 3H^{2}+2\dot{H}+ [6\beta-2\alpha] \kappa^{2} \left[ 18H^{2}\dot{H} + 9{\dot{H}}^{2}+12 H \ddot{H}+2\dddot{H}\right]= 0 & & ~~
 \end{eqnarray}
Using the fact that  $\alpha$ and $\beta$ have the exact same dimensions and are of the same order, we see that the two members of this family of differential equations and their solutions will differ only by a numerical constant.
Therefore, the  {qualitative dynamics predicted by a Stelle action is independent of the chosen parameter values}. As a result, we can conclude that the qualitative dynamics of \textit{any} Stelle theory can be studied by observing the dynamics of a particular case. In particular, $\alpha = 0$ 
corresponds to the Starobinsky model of inflation action discussed earlier in Sec.~\eqref{sec:Starobinsky}. In particular, setting $\beta = 1/(12 M \kappa^2)$ is the well-known Starobinsky action~\eqref{eq:Starobinskyaction} . As discussed earlier, the Starobinsky action leads to inflation with exit. From the above equations, it is clear that any other Stelle theory (and GUP modified gravity in particular) will also lead to inflation with exit.  

Repeating the analysis of \cite{Mukhanov:1990me},  \cite{Das:2022hjp} 
obtained  the following expressions for the power spectra of the scalar and tensor perturbations, respectively:
\begin{eqnarray}
\mathcal{P}(k)_{\mathcal{R}}= A_{\mathcal{R}}\left(\frac{k}{k^*}\right)^{n_{\mathcal{R}}-1} &;&
\mathcal{P}(k)_{T} = A_{T}\left(\frac{k}{k^*}\right)^{n_T}\,,
\end{eqnarray}
where $k_*$ is the wave number of the fluctuations at the horizon crossing. ${P}(k)_{\mathcal{R}}$, $A_{\mathcal{R}}$ and $n_{\mathcal{R}}$ are the power spectrum, the amplitude, and the spectral tilt of the scalar perturbations respectively. Analogously, ${P}(k)_{T}$, $A_{T}$, and $n_T$ are the power spectrum, the amplitude, and the spectral tilt of the tensor perturbations. During slow-roll inflation~\citep{Mukhanov:1990me,1995-Lidsey.etal-RMP}, in the leading order, they obtained the scalar (${P}(k)_{\mathcal{R}}$) and tensor (${P}(k)_{T}$) power-spectrum amplitudes to be: 
\begin{equation}
\label{eq:TheoreticalAs}
\!\!\!\!\! A_{\mathcal{R}}
=\frac{N^2}{18\pi\gamma_0 };~
A_{T} 
=\frac{1}{\pi\gamma_0 }  \, .
\end{equation}

\begin{figure*}[t!]
    \centering
    \begin{subfigure}[t]{0.45\textwidth}
        \centering
\includegraphics[width=0.99\textwidth]{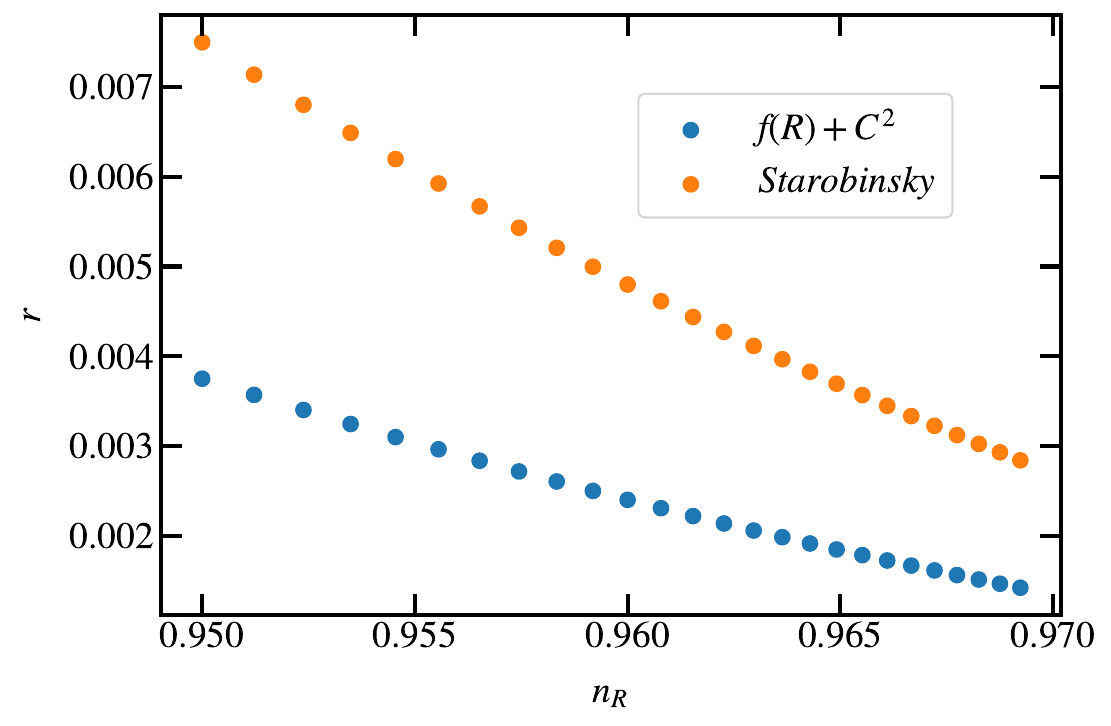}
        \end{subfigure}
 ~ 
    \begin{subfigure}[t]{0.45\textwidth}
        \centering
        \includegraphics[width=0.99\textwidth]{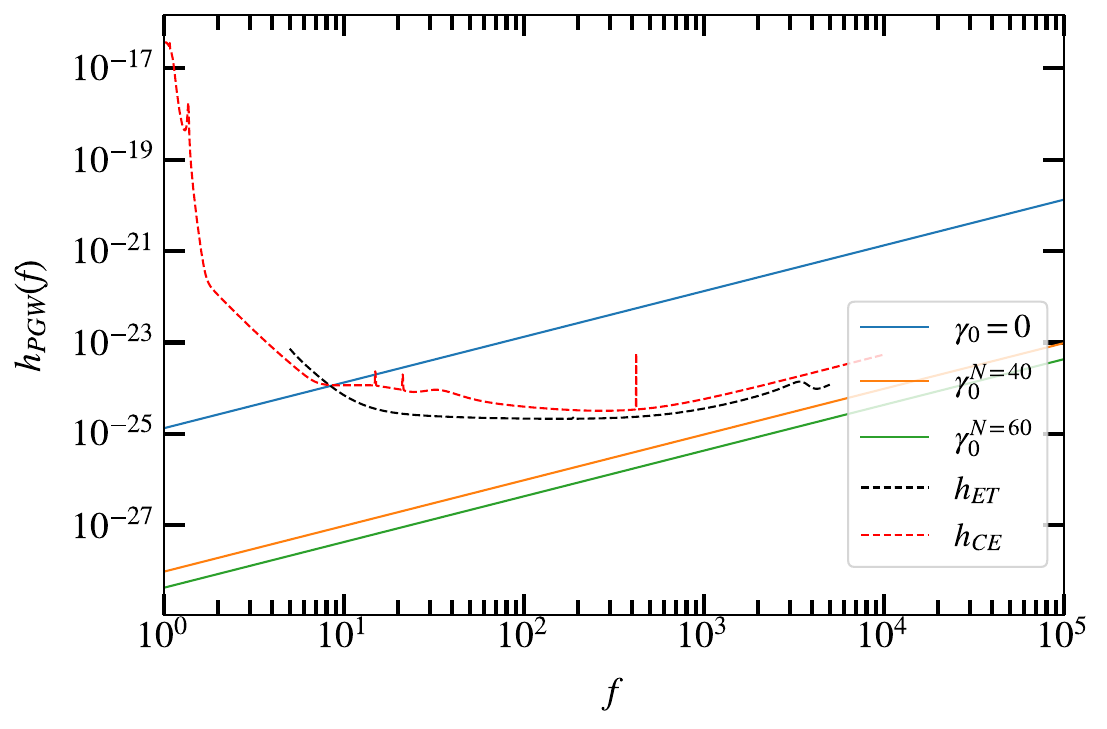}
    \end{subfigure}
    \caption{\label{fig:GUPInflation} The left plot represents the spectral tilt of the scalar perturbations $n_{\mathcal{R}}$ and the  tensor to scalar ratio $r= A_{T}/A_{\mathcal{R}}$ from $N=40$  to $N=70$ number of e-folds. The right plot is the Logarithmic graph of the characteristic strain of the gravitational waves produced during inflation $h_{GW}$ as a function of frequency $f$ and the GUP parameter $\gamma_0$, compared to the characteristic strain of ET $h_{ET}$ and CE $h_{CE}$. Source \cite{Das:2022hjp}}
    \end{figure*}

Using the Planck observations~\citep{Planck:2018jri,Bonga:2015xna}:
\begin{equation}\label{eq:ExperimentalAs}
    A_{\mathcal{R}} = 2.474 \pm 0.116 \times 10^{-9}\, ,
\end{equation}
\cite{Das:2022hjp} obtained bounds on the GUP parameter $\gamma_0$. Specifically, comparing the theoretical results of Eq. \eqref{eq:TheoreticalAs} with the above expression leads to the following value for $\gamma_0$ for two different e-folds ($N=40$ and $N=60$) of inflation:
\begin{equation}
\label{eq:gamma0values}
\gamma_{0}^{N=40} \approx 3.430\times 10^{10} \, , ~~\gamma_{0}^{N=60}\approx 7.719\times 10^{10}.
\end{equation}
Thus, the values for the intermediate scale are consistent with the bounds obtained earlier using completely different approaches~\citep{Nenmeli:2021orl,2018-Todorinov-AnnalPhy}. For the above range of $\gamma_0$, the authors showed that the tensor perturbations in the quadratic gravity model are suppressed (see Fig.~\eqref{fig:GUPInflation}).

For this range of $\gamma_0$, the authors evaluated the power spectrum of the primordial gravitational waves (PGWs) that can be generated during inflation. To quantify the detectability of PGWs in the upcoming GW detectors, they evaluated the energy density of PGWs~\citep{Smith:2005mm}: 
\begin{equation}\label{eq:GWEnergyDensity}
  \Omega_{\text{GW}}(k) = \mathcal P_{\rm PGW}(k) \, (k^2/{(12 H_0^{2})})\, ,
\end{equation}
where the scale factor at present is set to unity, 
 and $H_0=100 \,h\, \text{km}\,\text{s}^{-1}\,\text{Mpc}^{-1}$.  They plotted the characteristic strain of PGWs for three different values of $\gamma_0$ (corresponding to different e-foldings of inflation) along with the design sensitivity of ET and CE~\citep{Branchesi:2023, Evans:2023-CE}. From the right plot in figure~\eqref{fig:GUPInflation}, we infer that the characteristic strain of PGWs for the standard inflation (blue curve) is well within the ET and CE observational capability. However, the characteristic strain of PGWs for the GUP modified inflation (orange and green curves) are below the ET and CE observational capability. Hence, it is not possible to confirm the detection of GUP corrected inflation with confidence. However, the non-detectability of the PGWs will directly constrain the value of $\gamma_0$. Thus, ET and CE can severely constrain the value of the GUP parameter.

\subsubsection{Effect of massive spin-2 modes on GWs}

From the above discussion, we infer that quadratic gravity suppresses the power spectrum of the PGWs generated during inflation. This leads to the interesting question: What is the cause of the suppression? What are the other consequences of this suppression? 

To understand this, let us consider the equations of motion of quadratic gravity~\eqref{app:StelleEOM01} and set $ \alpha = {\gamma_0}/{\kappa^2}$ and $\beta = {\gamma_0}/{2 \kappa^2}$. 
Linearizing the field equation~\eqref{app:StelleEOM01} about the Minkowski space-time ($\eta_{\mu\nu}$), we get, 
\begin{equation}
g_{\mu\nu}=\eta_{\mu\nu}+ \epsilon h_{\mu\nu}~,~ g^{\mu\nu}=\eta^{\mu\nu}- \epsilon h^{\mu\nu}~\mbox{and}~  h=\eta^{\mu \nu} h_{\mu \nu}, 
\end{equation}
where $h_{\mu\nu}$ is the metric perturbation, and $\epsilon$ is a book-keeping parameter. This leads to the following linearized equations:
\begin{equation}\label{eq:linEOM_flat}
\mathcal{\delta G}_{\mu \nu}\equiv \left(1- 2 \gamma \bar\Box\right)\left(\delta R_{\mu \nu}-\frac{1}{2} \eta_{\mu \nu} \delta R\right)=0~,
\end{equation} 
where $\bar\Box \equiv \eta^{\mu\nu}\partial_{\mu}\partial_{\nu}$ and
\begin{eqnarray}
\delta R_{\mu \nu}&=& \frac{1}{2}\left( \partial_\mu \partial_\rho h^\rho_\nu + \partial_\nu \partial_\rho h^\rho_\mu -\partial_\mu \partial_\nu h - \bar{\Box} h_{\mu \nu}\right),\label{eq:ricci}\\
\delta R &=& \eta^{\mu \nu}R^{(1)}_{\mu \nu} = \left(\partial_\mu \partial_\rho h^{\rho\mu}-\bar{\Box} h\right) \, . \label{eq:ricciscalar}
\end{eqnarray}
To separate the contribution of the different spin-components from the linearized field equation~\eqref{eq:linEOM_flat}, we use the following ansatz for the metric perturbations~\citep{Chowdhury:2022ktf}:
\begin{equation}\label{eq:ansatz}
\!\!\! h_{\mu\nu}=\left[\psi_{\mu \nu}- \frac{\eta_{\mu\nu}}{2} \psi \right]+ \left[C_1+ \frac{C_2}{4}\right] R^{(1)} \eta_{\mu\nu} - C_2 \hat{R}^{(1)}_{\mu\nu}~,
\end{equation}
where $\psi= \eta^{\alpha \beta} \psi_{\alpha \beta}$ and $C_1, C_2$ are arbitrary constants, and $\hat{R}^{(1)}_{\mu\nu}=R^{(1)}_{\mu\nu}-\frac{1}{4} \eta^{\mu\nu} R^{(1)}$
is the traceless part of $ R^{(1)}_{\mu\nu}$. In GR, the constants $C_1$ and $C_2$ vanish because the linearized field equations only contain a massless spin-2 (graviton) mode. Thus, $\psi_{\mu \nu}$ reduces to the familiar trace-reversed metric perturbations in GR. In the case of the Starobinsky model~\eqref{sec:Starobinsky}, the field equations contain additional contribution from the massive spin-0 mode $(R^{(1)})$ only, hence $C_1=-6\gamma_0$ and $C_2=0$~\citep{Bhattacharyya2018EPJC}.
However, in the case of Stelle gravity, there is an additional massive tensor field and hence, $C_2 \neq 0$. We get $C_1 = - 2 \gamma_0$ for the GUP-inspired quadratic gravity model. Substituting Eq.~\eqref{eq:ansatz} in Eqs.~(\ref{eq:ricci},\ref{eq:ricciscalar}) and using the gauge conditions, one gets: 
\begin{equation}\label{eq:GW_flat}
2 \, \mathcal{\delta G}_{\mu \nu}=-\bar{\Box}\psi_{\mu\nu} ~.
\end{equation}
where $C_1=-2\gamma$ and $C_2=4\gamma$. Using Eqs.~(\ref{eq:linEOM_flat},~\ref{eq:GW_flat}), the propagation equation for the graviton, the massive spin-2 and spin-0 modes are, 
\begin{eqnarray}
\bar{\Box}\psi_{\mu\nu}= 0~,\label{eq:M0S2_flat}\\
\bar{\Box}\hat{R}^{(1)}_{\mu \nu} -  \hat{R}^{(1)}_{\mu \nu}/{2 \gamma} = 0~, \label{eq:MS2_flat}\\
\bar{\Box}R^{(1)} -  R^{(1)}/{2 \gamma}=0~
\label{eq:lintrcEOM_flat}
\end{eqnarray}
Thus, the metric perturbations of the quadratic gravity are described by a massless graviton ($\psi_{\mu\nu}$); a massive spin-2 ($\hat{R}^{(1)}_{\mu \nu}$) and a massive spin-0 (${R}^{(1)}$). Using the above separation of modes,  \cite{Chowdhury:2022ktf} obtained the energy and momentum of the GWs in the above quadratic gravity model. Using the \cite{isaacson1968PR1} averaging approach, these authors obtained the effective stress-energy tensor of the emitted gravitational waves in Ricci-flat background (see also \cite{isaacson1968PR}):
\begin{equation}
\label{eq:teff_full_trcls}
\begin{split}
\!\!\! t_{\mu\nu}^{\rm GW} = &\frac{1}{\kappa^2}\Bigl[\left<\frac{1}{4}\bar\nabla_\mu \psi^{\rho \sigma}\bar\nabla_\nu \psi_{\rho \sigma}
-  \gamma \left(\bar\nabla_\mu \psi^{\rho \sigma} \bar\nabla_\nu\hat{R}^{(1)}_{\rho \sigma} + \bar\nabla_\nu \psi^{\rho \sigma} \bar\nabla_\mu\hat{R}^{(1)}_{\rho \sigma}\right)\right.\\
&\left.-\gamma^2 \left(4 \bar\nabla_\mu \hat{R}^{(1)\rho \sigma} \bar\nabla_\nu \hat{R}^{(1)}_{\rho \sigma}- \bar\nabla_\mu R^{(1)}\bar\nabla_\nu R^{(1)} \right)\right>+\left<\mathcal{A}_{\mu\nu}+\gamma \mathcal{B}_{\mu\nu}+\gamma^2 \mathcal{C}_{\mu\nu}+\gamma^3 \mathcal{D}_{\mu\nu}\right>\Bigr],
\end{split}
\end{equation}
where $\mathcal{A}_{\mu\nu},\mathcal{B}_{\mu\nu},\mathcal{C}_{\mu\nu},\mathcal{D}_{\mu\nu}$ are related to the background Riemann tensor. The 
 first term within the triangular bracket gives the dominant contribution of the graviton mode as in GR~\citep{isaacson1968PR1}. However, the \emph{crucial difference is the dominant contribution} of the massive spin-2 mode. The massive spin-2 mode contribution is proportional to $\gamma$, whereas the massive spin-0 mode contribution is proportional to $\gamma^2$. Thus, the above expression explicitly shows that the massive spin-2 mode carries more energy than the massive spin-0 mode. This is consistent with the fact that PGWs will be suppressed. 
However, this study demonstrates that the $f(R)$ theories \emph{overlook crucial information} concerning the massive spin-2 mode, and the next-generation GW detectors can, in principle, constrain GUP parameters from observations.

\subsection{Lovelock Theories}
\label{sec:lovelock}
The previous sections suggest that adding higher-order curvature terms to the gravitational action generally results in higher-derivative equations of motion. This raises an important question: Can a second-order action constructed from the Riemann tensor, as in \eqref{eq:GenIIorderGraviaction}, avoid higher-order equations of motion, regardless of the values of \(\alpha\), \(\beta\), and \(\gamma\)? In 1938, Lanczos identified a specific combination of curvature invariants, now known as the \emph{Gauss-Bonnet term}, which leads to second-order equations of motion~\citep{1938-Lanczos-AnnalMath}. This term is:
\begin{equation}
\label{def:GBterm}
\mathcal{G} = R^2 - 4R_{\mu\nu}R^{\mu\nu} + R_{\mu\nu\sigma\lambda}R^{\mu\nu\sigma\lambda} \, .
\end{equation}
Setting \(\beta = -4\alpha\) and \(\gamma = \alpha\) in the Lagrangian \eqref{eq:GenIIorderGraviaction}, and varying the action leads to the Lanczos tensor
\begin{equation}\label{Lancoz tensor}
\begin{split}
A_{ \ \nu}^{\mu} = - \frac{1}{2}\alpha_{0}\delta_{\nu}^{\mu} + \alpha_{1}\left(R_{\nu}^{\mu}
- \frac{1}{2}\delta_{\nu}^{\mu}R\right)
+ \alpha_{2}\left(2R^{\mu\alpha\rho\sigma}R_{\nu\alpha\rho\sigma} - 4R^{\rho\sigma}
 R_{ \ \rho\nu\sigma}^{\mu} - 4R^{\mu\rho}R_{\nu\rho} + 2RR_{ \ \nu}^{\mu} - \frac{1}{2}
 \delta_{\nu}^{\mu}\mathcal{G}\right).
\end{split}
\end{equation}
The Lanczos tensor is analogous to the Einstein tensor and exhibits unique properties. First, in 4-D spacetime, the Gauss-Bonnet term does not contribute to the field equations because it reduces to a total derivative. Its integral over a closed manifold relates to the Euler characteristic, a topological invariant~\citep{1971-Lovelock-JMP,1981-Bunch-IOP, Yale:2011usf}. Consequently, the most general classical action in four dimensions that is linear in second-order curvature invariants is given by \eqref{ch2eqnQGMatr}. Second, in dimensions greater than four, the Gauss-Bonnet term contributes dynamically to the field equations, leading to the generalization of general relativity (GR) known as \emph{Einstein-Gauss-Bonnet gravity}. Unlike in 4-D, GR is no longer unique in higher dimensions, as extensions incorporating higher-order curvature terms can still yield second-order equations of motion. Hence, Lanczos-Lovelock theories are considered a natural generalization of GR in higher dimensions~\citep{1972-Lovelock-JMP,2013-Paddy_Dawood-PhyRept}.

This raises another intriguing question: Can higher-order curvature terms constructed from the Riemann tensor produce quasi-linear, second-order equations of motion? Lovelock extended Lanczos's findings, showing that higher-order curvature terms could be constructed to satisfy this condition~\citep{1971-Lovelock-JMP, Yale:2011usf}.
\begin{equation}
\mathcal{L}_{N} = \alpha_{0}L_{0} + \alpha_{1} L_{1} + \alpha_{2} L_{2} + \alpha L_{3} + \cdots + \alpha_{N}L_{N},
\end{equation}
whose zeroth-, first-, and second-order terms are cosmological constant, the Ricci scalar, and Gauss-Bonnet respectively, $\alpha_i$'s are coupling constants, and
\begin{eqnarray}
\label{eq:Lovelockseries}
L_{0} &=& \sqrt{|g|}\Lambda, \ L_{1} = \sqrt{|g|}R, \ L_{2} = \sqrt{|g|}(R^{2} - 4R_{\mu\nu}R^{\mu\nu} + R_{\mu\nu\rho\sigma}R^{\mu\nu\rho\sigma}) \\
L_{3} &=& R^{3}+3 R R^{\mu \nu \alpha \beta} R_{\alpha \beta \mu \nu}-12 R R^{\mu \nu} R_{\mu \nu}+24 R^{\mu \nu \alpha \beta} R_{\alpha \mu} R_{\beta \nu}  +16 R^{\mu \nu} R_{\nu \alpha} R_{\mu}^{\alpha} +24 R^{\mu \nu \alpha \beta} R_{\alpha \beta \nu \rho} R_{\mu}^{\rho}+8 R^{\mu \nu}{ }_{\alpha \rho} R_{\nu \sigma}^{\alpha \beta} R_{\mu \beta}^{\rho \sigma}+2 R_{\alpha \beta \rho \sigma} R^{\mu \nu \alpha \beta} R^{\rho \sigma}{ }_{\mu \nu} \nonumber 
\end{eqnarray}
The term \( L_3 \) serves as a boundary term in 5-D and 6-D spacetimes but becomes dynamically relevant in 7-D and higher dimensions. More generally, \( n^{\text{th}} \)-order curvature terms are non-dynamical when \( n < D/2 \). These terms are topological in nature for \( n = D/2 \) (in even dimensions). This can be formalized using the delta notation for the \( n^{\text{th}} \)-order Lovelock terms~\citep{Yale:2011usf}:
\begin{equation}
L_{N} = \sqrt{|g|}\delta_{\eta_{1}\ldots\eta_{2n}}^{\sigma_{1}\ldots\sigma_{2n}}R_{\sigma_{1}\sigma_{2}}^{ \ \ \eta_{1}\eta_{2}}\ldots R_{\sigma_{2n - 1}\sigma_{2n}}^{ \ \ \eta_{2n - 1}\eta_{2n}}.
\end{equation}
In this expression, all terms with \( n > D/2 \) vanish due to repeated antisymmetrized indices. The Lanczos-Lovelock theories provide a powerful extension of GR, as they include quasi-linear second-order differential equations and reduce to GR in four dimensions. There is no inherent theoretical reason to omit these terms in higher-dimensional spacetimes. However, their inclusion introduces modifications that may pose specific phenomenological challenges, necessitating careful consideration in applications beyond standard GR.

\subsubsection{Black holes in Lovelock Gravity}

Although the equations of motion for these gravity theories are complex compared to GR, an appealing feature of Lovelock gravity is that they admit explicit static spherically symmetric black hole solutions. In the case of vacuum, these solutions are treated as deformations of the \emph{Schwarzschild geometry}~\citep{Garraffo:2008hu}. 
It is easy to see this using the \emph{symmetric criticality principle,} which states that for a group-invariant Lagrangian, the equations obtained by the restriction of the Euler-Lagrange equations to group-invariant fields are equivalent to the Euler-Lagrange equations of a canonically defined, symmetry-reduced Lagrangian~\citep{Palais:1979rca, Fels:2001rv}.

The most general $D$-dimensional static spherically symmetric metric in Schwarzschild coordinates is~\cite{Deser:2003up}:
\begin{equation}
ds^2 = -a(r) \, b^2(r) \, dt^2 + \frac{dr^2}{a(r)} + r^2 \, d\Omega_{D-2}^{2} \,,
\label{eq:DdSchw}
\end{equation}
where \( d\Omega_{D-2}^{2} \) is the metric on the unit $(D-2)$-dimensional sphere \( S^{D-2} \). Substituting the above metric in Eq.~\eqref{eq:Lovelockseries}, we get: 
\begin{align}
\label{eq:Lovelockseries-SS} 
{L}_{1} = &
- \frac{r \left(b^{\prime} \left(3 r a^{\prime}+2 (D-2) a\right)+2 r a b^{\prime\prime}\right)+b \left(r^2 a^{\prime\prime}
+(D-2) \left(2 r a^{\prime}+(D-3) (a-1)\right)\right)}{r^2 b}, \\
{L}_{2} = {}& \frac{(D-3) (D-2)}{r^4 b} \Big(
2 r \big\{ b^{\prime} \left[ r (5 a-3) a^{\prime}+2 (D-4) (a-1) a \right] +2 r (a-1) a b^{\prime\prime} \big\} \Big. \nonumber \\
& 
+b \big\{ a \left[ 2 r^2 a^{\prime\prime}-2 (D-4) \left(-2 r a^{\prime}+D-5\right) \right]
+2 r \left[ a^{\prime} \left(r a^{\prime}-2(D-4) \right) -r a^{\prime\prime} \right]  +(D-5) (D-4) (a^2+1)  \big\} \Big) \,, \nonumber 
\end{align}
where a prime indicates derivative with respect to the coordinate $r$. 
For 5-D, the above expressions reduce to the following simpler form:
\begin{equation}
I \to \int_{0}^{\infty} dr \, r^3 \, b \, 
\Big( -2 \Lambda + \alpha_{1}  {\cal L}_{1} + \alpha_{2}  {\cal L}_{2} \Big)  \,,
\label{eq:Lovelockseries-ss01}
\end{equation}
where we have ignored the angular integrations.
Variation of the above-reduced action w.r.t $ a(r)$ and $b(r)$ leads to two uncoupled differential equations~\citep{Deser:2007za}. Solving these equations leads to the well-known Boulware-Deser solution~\citep{Boulware:1985wk}:
\begin{eqnarray}
b(r) = 1 \,, \qquad  a_{\pm}(r) = 1+ \frac{\alpha_{1} \, r^2}{4 \alpha_{2}} 
\left( 1 \pm \sqrt{1 + \frac{4 \Lambda \alpha_{2}}{3 \alpha_{1}^{2}} + \frac{m}{r^4} } \right) \,,
\label{eq:LovelockSeries-5dim}
\end{eqnarray}
for generic values of the parameters \( \alpha_{1}, \alpha_{2} \) and \( \Lambda \), and an integration constant $m$ that is related to the mass of the black hole(s). The two solutions ($\pm$) result from the fact that field equations are quadratic in $a(r)$. The above analysis can be generalized to arbitrary dimensions and higher-order curvature terms, leading to 
\begin{equation}
b(r) = 1 \, , \quad 
\sum_{k=1}^{[(D-1)/2]} \, \omega_{k} \, \left( \frac{(1-a)}{r^2} \right)^{k} =
{\cal M} \, r^{1-D} + \frac{2 \Lambda}{(D-1)(D-2)} \,,
\label{eq:LovelockSeries-Ddim}
\end{equation}
where ${\cal M}$ is an integration constant. The modifications to black hole solutions in this framework reveal a deeper, more intricate structure of spacetime and gravitational dynamics, with far-reaching consequences for our understanding of gravitational phenomena, especially in higher-dimensional or quantum contexts. For instance, higher-order curvature terms alter the thermodynamic properties of black holes. The usual relations between the entropy and the horizon area, encapsulated by the Bekenstein-Hawking entropy, may require modification to incorporate contributions from the higher-order terms.
Specifically, the entropy of these black holes is not proportional to the horizon area ($A_H)$~\cite{Myers:1988ze}: 
\begin{equation}
 S = \frac{A_H}{4} + \alpha_2 \frac{(D-2)(D-3)(A_{D-2})^{\frac{2}{D-2}}}{2}A_H^{\frac{D-4}{D-2}}, \label{WaldEn}
\end{equation}
where $A_{D -2}$ is the area of the $(D- 2)$ sphere. Thus, the black holes of Lovelock gravity do not generally obey the Bekenstein-Hawking area law. 
Interestingly, the correction term is controlled in general by the curvature of the event horizon~\citep{Clunan:2004tb}. For smaller mass BHs, the correction term will be significant and can dominate. Hence, these higher-order curvature modifications allow a better understanding of critical phenomena associated with black hole thermodynamics, such as the behavior of black hole formation and evaporation near the critical point. Additionally, the equations of state and the specific heat of black holes in Lovelock gravity can differ significantly from their GR counterparts, particularly in non-trivial spatial curvature. For details see \cite{louko1997hamiltonian,  dehghani2005thermodynamics, cai2010black,  kastor2010smarr, kastor2011mass, hennigar2017thermodynamics, petrov2021conserved, hull2021thermodynamics, bai2023topology,  hull2023exotic}.

\subsubsection{4-D Gauss-Bonnet gravity}

As discussed above, the Gauss-Bonnet term is purely topological in 4-D and does not contribute dynamically. This has been the main reason to discard such contribution in realistic cosmology.  In order to circumvent the stringent requirements of Lovelock's theory, and in an attempt to introduce the Gauss-Bonnet term in 4D gravity directly, \cite{glavan2020einstein} employed a limiting procedure that reinterprets the contributions of the Gauss-Bonnet term in a 4-D context without requiring compactified extra dimensions. Specifically, they proposed rescaling the coupling constant $\alpha_2$ such that: 
\begin{equation}
\alpha_{2} \rightarrow \frac{\alpha_{2}}{D - 4}.
\end{equation}
This quantity is evidently divergent in the limit $ D \to 4 $. However, Glavan and Lin made a non-trivial suggestion that if this re-scaling were introduced into the Lancoz tensor, the terms containing this quantity as a factor might remain finite and non-zero. Specifically, they postulated that the divergence introduced into $\alpha_{2}$ might be sufficient to cancel the fact that additional terms in Eq. \eqref{Lancoz tensor} tend to zero as $D \to 4$. If this were the case, the Gauss-Bonnet term could then have a direct effect in the four-dimensional theory of gravity. The motivation for this radical new approach stems from the trace of the Lancoz tensor (\ref{Lancoz tensor}), which in $D$ dimensions gives
\begin{equation}
A_{\ \mu}^{\mu} = - \frac{1}{2} D \alpha_{0} - \frac{1}{2} (D - 2) \alpha_{1} R - \frac{1}{2} (D - 4) \alpha_{2} \mathcal{G}.
\end{equation}
This expression explicitly shows the vanishing of the term from the Einstein tensor in $D = 2$ and the vanishing of the Gauss-Bonnet contribution in $D = 4$, both of which are due to a pre-factor of the form $(D - n)$ (note that $R$ and $\mathcal{G}$ can be non-zero only if $D > 1$ and $D > 3$, respectively). By applying the re-scaling described above, the factor that typically causes the contribution from the Gauss-Bonnet term to vanish is entirely removed, leaving a term that can, in general, be non-zero in the limit $D \to 4$.
Thus technique by Glavan and Lin is seen as a way to retain higher-order corrections in a purely 4D framework. 

Despite its appeal, the 4D Gauss-Bonnet gravity theory has faced criticism for relying on a non-standard limiting process, raising questions about its mathematical rigor and physical interpretation~\citep{fernandes20224d}. This has spurred interest in systematically deriving a non-topological version of the Gauss-Bonnet term in four dimensions. Techniques such as the counterterm regularization approach, dimensional derivative regularization, and Kaluza-Klein reduction have been proposed to achieve this goal. For a comprehensive review, see \cite{fernandes20224d}.
While these approaches provide valuable insights into the dynamical origin of 4D Gauss-Bonnet gravity, they remain incomplete as they do not fully incorporate the effects of quantum corrections. One-loop quantum effects, particularly from matter fields interacting with gravitons, are critical in the strong-gravity regime where Gauss-Bonnet terms become relevant. 
Although one-loop matter corrections to the graviton have been previously explored \citep{Barth:1983hb,Simon:1990ic, Buchbinder:2017lnd}, such an analysis is notably absent in the literature and is essential for the phenomenological study of four-dimensional Gauss-Bonnet gravity.

Recent work by \cite{Mandal:2024kic} offers a more robust mathematical framework to address these issues. Using dimensional regularization, a common tool in quantum field theory to handle divergences, they linearized the Einstein-Hilbert action coupled to massless scalar and fermion fields. By calculating one-loop self-energy corrections to the graviton and introducing counterterms, the authors demonstrated that the Einstein-Gauss-Bonnet gravity theory can be made non-topological in four dimensions when quantum matter effects are considered. This result establishes a more solid foundation for incorporating the simplest Lovelock gravity formulation into a four-dimensional framework that accounts for quantum phenomena.

Given this, 4-D Gauss-Bonnet gravity provides a richer setting for studying early universe cosmology, and other high-curvature phenomena. For instance, it offers new insights into potentially observable  astrophysical settings. To go about this, let us start by considering the following  Einstein-Gauss-Bonnet action:
\begin{equation}
S = \frac{1}{16\pi G}\int d^{D}x\sqrt{-g}\Big[ - 2\Lambda + R + \hat{\alpha}\mathcal{G}\Big],
\end{equation}
where $\hat{\alpha}$ is the rescale $\alpha$ coupling constant. Metric variation of the above action leads to the following field equations
\begin{equation}
G_{\mu\nu} + \Lambda g_{\mu\nu} = \hat{\alpha}H_{\mu\nu},
\end{equation}
where
\begin{equation}
H_{\mu\nu} = -2 \left(RR_{\mu\nu} - 2R_{\mu\alpha\nu\beta}R^{\alpha\beta} + R_{\mu\alpha\beta\sigma}
R_{\nu}^{ \ \alpha\beta\sigma} - 2R_{\mu\alpha}R_{\nu}^{ \ \alpha} - \frac{1}{4}g_{\mu\nu}\mathcal{G}
\right).
\end{equation}
As a result, we find
\begin{equation}
\hat{\alpha}g^{\mu\nu}H_{\mu\nu} = \frac{\alpha}{D - 4}\frac{1}{2}(D - 4)\mathcal{G} = \frac{\alpha}{2}\mathcal{G}.
\end{equation}
Glavan and Lin suggested that this non-vanishing contribution may not be exclusive to the trace of the field equations, but could be manifest in the full theory.
Substituting the $D-$dimensional FLRW spacetime line element \eqref{eq:FRWmetric} in the above field equations leads to 
%
%
the following modified Friedmann equations:
\begin{equation}
\begin{split}
H^{2} & + \frac{k}{a^{2}} + \alpha\left(H^{2} + \frac{k}{a^{2}}\right)^{2} = \frac{8\pi G\rho}{3} + \frac{\Lambda}{3}\\
\dot{H} & = - \frac{4\pi G(\rho + P)}{1 + 2\alpha\left(H^{2} + \frac{k}{a^{2}}\right)} + \frac{k}{a^{2}},
\end{split}
\end{equation}
where $\rho (P)$ correspond to density (pressure) of the perfect fluid. Interestingly, these Friedmann equations take exactly the same form as those derived in holographic cosmology \citep{apostolopoulos2009cosmology, bilic2016randall}, from the generalized uncertainty principle \citep{lidsey2013holographic}, through the consideration of quantum entropic corrections \citep{cai2008corrected}, and from gravity with a conformal anomaly \citep{lidsey2009thermodynamics}.

The Friedmann equations in the 4D Gauss-Bonnet framework include non-trivial corrections, prompting investigations into their observable effects on the CMB and the PGW spectrum. In particular, \cite{glavan2020einstein}  analyzed perturbations of the spatial part of the FLRW metric, expressed as:
\begin{equation}
g_{ij} = a^{2}(\delta_{ij} + h_{ij}),
\end{equation}
where $\partial_{i} h_{ij} = 0$ and $h_{ii} = 0$ described the PGWs. In the 4-D limit, they obtained:
\begin{equation}
\ddot{\gamma}_{ij} + \left(3 + \frac{4\alpha\dot{H}}{1 + 2\alpha H^{2}}\right)H\dot{\gamma}_{ij}
- c_{s}^{2}\frac{\partial^{2}\gamma_{ij}}{a^{2}} = 0,
\end{equation}
where
\begin{equation}
c_{s}^{2} = 1 + \frac{4\alpha\dot{H}}{1 + 2\alpha H^{2}}.
\end{equation}
The Gauss-Bonnet term modifies both the Hubble friction term and the effective sound speed of GWs. These changes could have significant implications for observational cosmology, especially in the early Universe. The altered friction and propagation speed of tensor modes may lead to detectable signatures in the PGW spectrum, which can be probed through next-generation gravitational wave detectors or imprints on the CMB polarization. These features could provide a unique window into high-energy physics and early-universe dynamics.

\section{Metric theories of gravity that break gauge invariance, LLI or parity}
\label{sec:MGModels-ClassB}

Metric theories of gravity, such as GR, generally assume foundational symmetries like gauge invariance, LLI and parity to describe gravitational interactions. Lorentz invariance ensures that physical properties remain unchanged regardless of a system’s orientation or uniform motion. However, beyond-standard-model physics often considers scenarios where small violations of these symmetries may occur. For example, a Lorentz tensor field with a nonzero vacuum expectation value introduces a preferred spacetime direction, breaking Lorentz invariance. Such violations are systematically studied using EFT, which provides a model-independent framework to explore observable effects.

The EFT framework allows for a systematic description of observable effects arising from Lorentz symmetry breaking.  Notably, Lorentz symmetry underpins the CPT theorem, so its violation naturally impacts CPT invariance\footnote{However, the converse of this statement is not true: it is possible to violate Lorentz invariance while keeping CPT intact}. While these violations often preserve mass equivalence between particles and antiparticles, they can produce distinct Lorentz-violating dispersion relations and other CPT-violating effects. Lorentz invariance in GR leads to problems with ultraviolet behavior and renormalizability. Lorentz-violating theories, like Horava-Lifshitz gravity~\citep{Horava:2009uw,Herrero-Valea:2023zex}, introduce anisotropic scaling between space and time at high energies to improve ultraviolet behavior.

EFT also encompasses potential violations of broader spacetime symmetries, such as CPT symmetry and diffeomorphism invariance, which are fundamental to gravitational interactions in curved spacetime. This framework supports comparisons of diverse precision experiments, including short-range gravitational tests~\citep{bailey2015short, PhysRevD.91.102007, PhysRevLett.117.071102}, gravimeter measurements \citep{PhysRevLett.100.031101, Flowers:2016ctv}, tests of the WEP \citep{PhysRevLett.123.231102}, tests of the weak equivalence principle (WEP), and redshift experiments~\citep{PhysRevLett.106.151102}. By expressing results in terms of symmetry-violating coefficients, EFT enables direct comparisons across experimental domains, such as correlating high-energy cosmic ray observations with laboratory tests.

In GR, gauge invariance reflects the freedom to choose coordinate systems without altering physical predictions. However, certain contexts, such as the existence of the CMB rest frame --- where motion relative to the CMB photons is zero --- seemingly challenge the universality of special relativity (see, for instance, \cite{Rameez:2019wdt}). While relativistic effects like Doppler shifts make the CMB appear different across frames, the physical laws governing its interactions remain consistent. Theoretical frameworks such as unimodular gravity restrict allowed coordinate transformations (e.g., to volume-preserving ones), effectively breaking gauge invariance~\citep{Unruh:1988in}. 
However, without gauge invariance, interpreting the physical meaning of the theory becomes more complex~\citep{Will:2018bme}.

This section explores the consequences of violating Lorentz and gauge invariance in metric gravity theories, focusing on their implications for fundamental physics and cosmological phenomena.

\subsection{Gravity theories that break LLI}

LLI is a cornerstone of both GR and quantum field theory. In traditional metric theories, gravity is described by a metric tensor $g_{\mu\nu}$ that governs the spacetime geometry, and the theory is invariant under local Lorentz transformations. However, in some modified gravity theories, this symmetry is explicitly broken. The breaking of LLI can occur at various scales, often motivated by fundamental physics beyond the Standard Model, such as string theory, quantum gravity, or the need to explain cosmological observations that challenge standard cosmological models~\citep{Safronova:2017xyt}.

\subsubsection{Einstein-aether theory}

One way to break LLI is by introducing a time-like vector $u_{\alpha}$ with a fixed unit norm, given by
\begin{equation}
g_{\alpha\beta} u^{\alpha} u^{\beta} = -1,
\end{equation}
and coupling it to gravity. This approach is employed in Einstein-aether theory~\citep{Jacobson:2000xp}.  To preserve general covariance, the vector $u_{\alpha}$ —- which is referred to as the aether 
-- must be dynamical. The most general covariant action 
(without matter coupling) quadratic in derivatives of the metric 
$g_{\alpha\beta}$ and the aether $u_{\alpha}$, up to a total divergence is:
\begin{equation}
\begin{split}
S_{AE} &= \frac{1}{16\pi G_{AE}} \int \sqrt{-g} \, d^{4}x \Big[ R - c_1 (\nabla_{\mu} u_{\nu}) (\nabla^{\mu} u^{\nu}) - c_2 (\nabla_{\mu} u^{\mu})^2 - c_3 (\nabla_{\mu} u_{\nu}) (\nabla^{\nu} u^{\mu}) + c_4 (u^{\alpha} \nabla_{\alpha} u_{\mu}) (u^{\beta} \nabla_{\beta} u^{\mu}) + \lambda (u_{\alpha} u^{\alpha} + 1) \Big],
\end{split}
\end{equation}
where $c_{1}, c_{2}, c_{3}, c_{4}$ are dimensionless constants, and the Lagrange multiplier 
$\lambda$ enforces the unit constraint. The relationship between the bare gravitational constant 
$G_{AE}$ and the Newtonian gravitational constant $G_{N}$ can be determined by taking the Newtonian limit, yielding~\citep{Foster:2005dk}:
\begin{equation}
G_{AE} = \left(1 - \frac{c_1 + c_4}{2}\right) G_{N}.
\end{equation}
By linearizing over a Minkowski background and assuming a constant aether, it can be shown that the theory possesses five massless degrees of freedom: one spin-2 mode, one spin-1 mode, and one 
spin-0 mode. The squared speeds of these modes are determined by a combination of the coupling constants in the theory~\citep{Foster:2005dk}. The absence of Cherenkov radiation in ultra-high-energy cosmic ray observations provides constraints on these speeds, which must be superluminal. Whether superluminal speeds lead to violations of causality remains a subtle issue. Furthermore, the theoretical requirements for stability and positive energy of the modes impose additional constraints on the coupling constants~\citep{Chen:2021bvg, deRham:2021bll}.

\subsubsection{Horava-Lifshitz gravity}

Horava-Lifshitz gravity (HL gravity)  presents an alternative to GR by proposing a framework for quantum gravity that employs anisotropic scaling between space and time~\citep{Horava:2009uw}. 
The primary aim of HL gravity is to address the 
issue of renormalizability of gravity at high energies by exploiting the Lifshitz-type scaling, making the theory well-suited for the ultraviolet regime. The core idea of HL gravity lies in the introduction of anisotropic scaling of spacetime coordinates, characterized by transformations:
\begin{equation}
t \to b^z t, \quad x_i \to b x_i,
\end{equation}
where $t$ is the time coordinate, $x_i$ are the spatial coordinates, and $b$ is a scaling factor. The parameter \(z\), known as the dynamical critical exponent, quantifies the anisotropy between time and space. 
To understand this, let us consider the HL gravity action~\citep{Horava:2009uw}: 
\begin{equation}
S = \int d^4x \sqrt{-g} \left( \frac{2}{\kappa^2} \left( K_{ij} K^{ij} - \lambda K^2 \right) + \frac{\mu_4}{\Lambda^4} R^2 + \cdots \right)
\end{equation}
where $K_{ij}$ is the extrinsic curvature of the boundary, $K$ is its trace, and $R$ is the Ricci scalar. The constants $\lambda$ and $\mu_4$ regulate the higher-order corrections, and 
the parameter $\Lambda$ sets the scale of the transition between classical and quantum gravity. As we can see, HL gravity modifies the Einstein-Hilbert action by incorporating higher-order spatial derivatives, leading to significant changes in black hole solutions and cosmological dynamics at small scales. These higher-order terms ensure the theory's renormalizability, allowing it to remain well-defined in the ultraviolet regime, making HL gravity a compelling candidate for quantum gravity. Unlike GR, where gravity becomes non-renormalizable at short distances, the modifications introduced in HL gravity prevent divergences, ensuring that quantum corrections are finite and manageable. These terms act as an ultraviolet completion for gravity, offering potential insights into quantum gravitational phenomena such as spacetime discreteness and Planck-scale physics (see, for instance, \cite{visser2009lorentz, contillo2013renormalization, vernieri2015power, barvinsky2016renormalization, barvinsky2017hovrava, barvinsky2019towards}).

A key departure from GR in HL gravity lies in its introduction of anisotropic scaling between space and time. This scaling inherently leads to a degree of Lorentz violation, particularly at high energies. The dynamical critical exponent \( z \), which governs this anisotropic scaling, differentiates HL gravity from Einstein's relativistic framework. In the IR limit, where the theory converges with GR, \( z \) approaches 1, restoring Lorentz invariance. However, in the ultraviolet regime, \( z > 1 \), breaking the symmetry between time and space. This asymmetry results in non-relativistic behavior at high energies, a signature feature of HL gravity. The violation of Lorentz invariance in the ultraviolet limit has profound implications for the underlying spacetime structure and the fundamental symmetries governing gravitational interactions.

HL gravity introduces modifications to the dispersion relations for gravitons and possibly other fields, stemming from the anisotropic scaling between time and space. In GR, dispersion relations for gravitons remain relativistically invariant. However, in HL gravity, higher-order spatial derivatives alter these relations, especially in the ultraviolet regime. For example, the graviton's dispersion relation takes the form:  
\[
\omega^2 = \vec{k}^2 + \mu^2 (\vec{k}^2)^2 + \ldots,
\]
where \( \omega \) is the frequency, \( \vec{k} \) is the wavevector, and \( \mu \) governs the higher-order corrections. These deviations from the standard GR relation (\( \omega^2 = \vec{k}^2 \)) highlight the effects of Lorentz violation, leading to anisotropic or frequency-dependent propagation of gravitational waves. As a result, high-frequency waves may travel at different speeds, diverging from the speed of light, which could cause measurable time delays in astrophysical events like black hole mergers. Observations from LIGO and Virgo offer opportunities to test such deviations, potentially providing evidence for Lorentz violation at quantum gravity scales. See, for instance, \cite{pospelov2012lorentz, coates2019revisiting}.  

In cosmology, HL gravity impacts the dynamics of the early universe and inflation. The higher-order terms in its action modify the evolution of the scale factor \( a(t) \), leading to changes in the Hubble parameter and the universe's expansion history. These adjustments could create a new inflation mechanism distinct from GR. Additionally, HL gravity provides a possible resolution to the Big Bang singularity. By smoothing out infinite density and temperature through higher-order corrections, the theory allows for a continuous transition from a quantum gravitational phase to a classical spacetime. This implies that the early universe might have emerged from a non-singular state. Detailed studies explore how these modifications influence inflationary dynamics and singularity resolution. See for instance, \cite{brandenberger2009matter,  dutta2009observational, leon2009phase, mukohyama2009scale, mukohyama2010hovrava, carloni2010analysis, gao2010cosmological,bertolami2011hovrava, appignani2010cosmological, obregon2012quantum, saridakis2011aspects,    wang2009thermodynamics, pitelli2012quantum, misonoh2017stability, nilsson2019hovrava}.

In HL gravity, the introduction of higher-order spatial derivatives alters classical black hole solutions. For instance, the \cite{Kehagias:2009is} BH solution derived in the limit where the theory approaches detailed balance is a testable framework for studying deviations from GR in strong gravity regimes and has implications for astrophysics, quantum gravity, and early-universe cosmology. This spacetime is significant as it represents a static, spherically symmetric vacuum solution that serves as a counterpart to the Schwarzschild black hole in GR. However, due to the modified gravitational dynamics of HL gravity, this BH solution includes additional features not present in GR black holes. The Kehagias-Sfetsos  metric can be expressed as:
\[
   ds^2 = -f(r)dt^2 + f(r)^{-1}dr^2 + r^2(d\theta^2 + \sin^2\theta \, d\phi^2),
   \]
   where the lapse function \( f(r) \) is given by:
   \[
   f(r) = 1 + \Omega r^2 - \sqrt{r(\Omega^2 r^3 + 4\Omega M)} \, ,
   \]
$M$ is the mass of the black hole, and \( \Omega \) governs deviations from 
GR.  Depending on the value of \( \Omega \), there can be one, two, or no horizons.  The solution smoothes some of the singularities present in GR, reflecting the ultraviolet completion goals of Horava-Lifshitz gravity.  Also, this black hole modifies classical thermodynamic quantities such as temperature, entropy, and the specific heat due to the presence of higher-order corrections in HL gravity. These deviations have implications for black hole evaporation and stability~\citep{cai2009thermodynamics, cai2009topological, cai2010horizon, chen2009strong, mukohyama2009scale, mukohyama2010hovrava,   myung2009thermodynamics,  aliev2010slowly,   koutsoumbas2010black,  myung2010entropy, myung2010thermodynamics, majhi2010hawking, peng2010hawking, biswas2011black, blas2011hovrava, saridakis2011aspects, zhou2011black, barausse2013slowly,   atamurotov2013shadow, xu2020black}. 

\subsubsection{Other gravity theories}

Currently, there are no definitive observations confirming Lorentz violation --- only indirect hints suggest its possibility \citep{ bertolami2000spontaneous, bertolami2000proposed, jenkins2004spontaneous,  kostelecky2004gravity, 
alfaro2005quantum, bojowald2005loop, gabadadze2005lorentz, heinicke2005einstein,
collins2006lorentz, berezhiani2007spontaneous,cheng2006spontaneous,
bluhm2008spontaneous,
blas2009lorentz,  sotiriou2009quantum, sotiriou2009phenomenologically,  visser2009lorentz, seifert2009vector, kostelecky2009gravity,
armendariz2010effective, moffat2010lorentz, brax2012lorentz,
gielen2012spontaneously,  pospelov2012lorentz, 
barausse2013black, bluhm2014observational, liberati2015lorentz, bailey2015short,       
lin2017no, sakstein2017baryogenesis,   eichhorn2020lorentz,     illuminati2021spontaneous}. While there are several theoretical frameworks that can lead to Lorentz violation, the generic consequences are as follows: 

The violation of Lorentz symmetry implies the existence of preferred directions or frames in spacetime, leading to anisotropic effects in physical phenomena. This anisotropy could manifest as direction-dependent propagation speeds or deviations from the isotropy observed in the cosmic microwave background (CMB), with significant implications for cosmological models, particularly those concerning inflation or early-universe dynamics~\citep{Martin:2000xs,Niemeyer:2000eh,Shankaranarayanan:2002ax,Shankaranarayanan:2005cs}. 

In quantum gravity models, such as those incorporating \emph{spontaneous Lorentz violation or doubly special relativity}, deviations from Lorentz invariance may emerge at high energies. These deviations could lead to measurable effects, including alterations in the behavior of black holes, the propagation of gravitational waves, or singularity formation. High-energy astrophysical phenomena like gamma-ray bursts and black hole evaporation might exhibit unique features stemming from these deviations, offering a potential observational window into Lorentz-violating physics~\citep{kostelecky2009gravity, pospelov2012lorentz,liberati2015lorentz, illuminati2021spontaneous}.

Additionally, Lorentz violation could induce anisotropic spacetime curvature, affecting gravitational lensing~\citep{kostelecky2004gravity,Filho:2024isd}. For example, light bending around massive objects might depend on the direction relative to the preferred frame, leading to unconventional lensing effects such as non-standard deflection angles or unique multiple-image configurations. These phenomena could be detectable through precision studies of gravitational lensing, providing further insights into possible Lorentz violations. Additionally, even if LV does occur, applying effective field theory (EFT) to describe its low-energy effects may not always be valid. Despite these uncertainties, constraints derived from the straightforward considerations outlined here remain valuable. They leverage advancements in observational precision to constrain plausible scenarios potentially linked to Planck-scale physics. This approach provides critical guidance for quantum gravity research~\citep{liberati2015lorentz}.

\subsection{Gravity theories that break gauge Invariance}

As mentioned earlier, in GR, gauge invariance is tied to diffeomorphism invariance --- the principle that physical laws remain unchanged under arbitrary coordinate transformations. This invariance ensures that GR respects the principle of general covariance, meaning the laws of physics are the same regardless of the observer's frame of reference. Also, this invariance guarantees that the gravitational interaction is mediated by the metric field in a manner consistent with the geometric structure of spacetime.

Since gauge invariance is intimately related to the conservation of the stress tensor, breaking this symmetry can lead to anomalies in energy-momentum conservation, potentially affecting the dynamics of matter and radiation~\citep{giddings1991spontaneous, bluhm2008spontaneous,    bahr2009breaking,  charmousis2009strong,    anber2010breaking, alberte2012massive, bluhm2015explicit, lin2016effective,  kostelecky2018lorentz,  momeni2020massive, bluhm2021gravity, bluhm2023explicit, bailey2024explicit}. This violation of gauge invariance can, in particular, 
significantly alter GW physics. Instead of propagating as purely transverse traceless modes, GWs could develop additional degrees of freedom or new polarization states, which could be detectable through GW astronomy. Observations of GW spectra and polarization patterns might offer direct evidence for such symmetry violations~\citep{LIGOScientific:2021sio}.

Further, if gauge invariance is violated, the gravitational field might not be described purely by the metric tensor. For example, modifications might introduce additional fields or higher-dimensional terms in the gravitational action, such as \textit{scalar-tensor fields} or \textit{vector fields}. This could lead to \textit{modified dynamics} of spacetime, which may alter the propagation of gravitational waves or the response of matter to the gravitational field. Moreover, non-local interactions can arise \citep{deser1994gauge, carone2020aspects}. In other words, the behavior of gravity may not only depend on the properties of matter and energy at a point but also on distant parts of the spacetime. Non-local terms in the gravitational action could modify the standard form of the Einstein-Hilbert equations, potentially leading to unexpected physical behaviors like deviations from the inverse-square law of gravity or altered propagation of gravitational waves.

Gauge invariance often constrains the number of independent physical degrees of freedom in a theory. For example, in GR, the metric tensor is symmetric, but its physical degrees of freedom are constrained by the diffeomorphism invariance, which removes redundant components associated with coordinate freedom. If gauge invariance is absent, new spurious or unphysical degrees of freedom could emerge. These additional degrees of freedom might manifest as extra scalar, vector, or tensor fields that are not associated with any physical matter or energy but are artifacts of the loss of gauge symmetry. These unphysical modes can lead to inconsistencies in the theory, such as violations of causality or the appearance of ghost fields, which can introduce instabilities in the theory.

One notable example of a theory breaking gauge invariance is \emph{massive gravity}~\citep{de2012ghost, deRham:2014zqa}.  By endowing the graviton with a small, non-zero mass, massive gravity alters the behavior of GWs, causing them to travel slower than light. This modification impacts the large-scale behavior of gravity, potentially explaining phenomena like the universe's accelerated expansion without invoking dark energy~\citep{Mandal:2024nhv}. In the rest of this section, we will focus on linearized massive gravity and possible way to generate the mass from spontaneous symmetry breaking (SSB) in the matter sector~\citep{bernstein1974spontaneous,kibble2015spontaneous,
beekman2019introduction}..

\subsubsection{Linearized massive gravity theory}
Linearized massive gravity represents a perturbative approach to the study of a gravitons with a small but nonzero mass, as opposed to the massless graviton assumed in GR. To keep things transparent we will consider 
linearized massive gravity about Minkowski flat spacetime as described by the Fierz-Pauli action~\citep{fierz1939relativistic, Hassan:2011vm,  de2012ghost, deRham:2014zqa, Gambuti:2021meo}:
\begin{equation}\label{eq. 0.1}
S_{FP} = \int d^{D}x \Big[ \mathcal{L}_{m = 0} - \frac{1}{2}m^{2}(h_{\mu\nu}h^{\mu\nu}
 - h^{2})\Big] 
\end{equation}
where $m$ is the graviton mass, $h_{\mu\nu}$ is a rank-2, symmetric tensor field defined about a D-dimensional Minkowski space-time $\eta_{\mu\nu}$, $h = \eta^{\mu\nu}h_{\mu\nu}$ is the trace of the metric perturbation, and $\mathcal{L}_{m = 0}$ is the Lagrangian density corresponding to the linearized  Einstein-Hilbert action:
\begin{equation}
\label{eq. 0.1a}
\mathcal{L}_{m = 0}  = - \frac{1}{2}\partial_{\lambda}h_{\mu\nu}\partial^{\lambda}
h^{\mu\nu} + \partial_{\mu}h_{\nu\lambda}\partial^{\nu}h^{\mu\lambda} - \partial_{\mu}
h^{\mu\nu}\partial_{\nu}h + \frac{1}{2}\partial_{\lambda}h\partial^{\lambda}h \, .
\end{equation}
The above Lagrangian is invariant under the following gauge transformation:
\begin{equation}\label{eq. 0.2}
\delta h_{\mu\nu} = \partial_{\mu}\xi_{\nu} + \partial_{\nu}\xi_{\mu},
\end{equation}
where, $\xi^{\mu}$ is a spacetime-dependent infinitesimal gauge transformation. Varying the action \eqref{eq. 0.1} w.r.t the metric perturbation $h_{\mu\nu}$, we obtain the following equation of motion:
\begin{equation}\label{eq. 0.3}
\begin{split}
\Box h_{\mu\nu}  - \partial_{\lambda}\partial_{\mu}h_{ \ \nu}^{\lambda} - \partial_{\lambda}
\partial_{\nu}h_{ \ \mu}^{\lambda} + \eta_{\mu\nu}\partial_{\lambda}\partial_{\sigma}h^{\lambda
\sigma} + \partial_{\mu}\partial_{\nu}h - \eta_{\mu\nu}\Box h = m^{2}(h_{\mu\nu} - \eta_{\mu\nu}h) \, .
\end{split}
\end{equation} 
These are coupled partial differential equations, and the underlying physics is not transparent.  To extract the physics, we need to do a series of transformations: First, using the fact that the left-hand side of the above equation consists of the linearised form of the Einstein tensor whose divergence vanishes, we obtain the following relation:
\begin{equation}\label{eq. 0.4}
m^{2}(\partial^{\mu}h_{\mu\nu} - \partial_{\nu}h) = 0.
\end{equation}
Since $m \neq 0$, substituting the above result back into the equations of motion gives
the following equation
\begin{equation}\label{eq. 0.5}
\Box h_{\mu\nu} - \partial_{\mu}\partial_{\nu}h = m^{2}(h_{\mu\nu} - \eta_{\mu\nu}h).
\end{equation}
Second, taking the trace of the above expression \textit{w.r.t} the Minkowski metric we obtain $h = 0$ 
which means that the field $h_{\mu\nu}$ is traceless. Using the traceless condition and Eq.~(\ref{eq. 0.4}) leads to 
$\partial^{\mu}h_{\mu\nu} = 0$. This implies that 
$h_{\mu\nu}$ is transverse. Lastly, using the transverse and traceless relations of $h_{\mu\nu}$ in Eq.~(\ref{eq. 0.5}), we get the following wave equation:
\begin{equation}\label{eq. 0.6}
(\Box - m^{2})h_{\mu\nu} = 0.
\end{equation} 
Thus, the equations of motion (\ref{eq. 0.3}) leads to a 
massive wave equation, transverse condition, and traceless constraint:
\begin{equation}\label{eq. 0.7}
(\Box - m^{2})h_{\mu\nu} = 0, \ \partial^{\mu}h_{\mu\nu} = 0, \ h = 0.
\end{equation}
Although the above three equations are exactly equivalent to (\ref{eq. 0.3}), it is 
easier to count the number of degrees of freedom (DoF) by using the above set of equations. 
The first equation is the evolution of a symmetric tensor field $h_{\mu\nu}$ and implies 
$10$ DoF in $D = 4$. The traceless condition reduces one degree of freedom by forming a single constraint on the system. The transverse condition leads to $4$ more constraints. Hence, the Fierz-Pauli action consists of $5$ DoF in 4-D spacetime. Coupling the Fierz-Pauli action with matter leads to singular behaviour in the limit $m \to 0$. However, using the Stuckelberg mechanism, it is possible to show that theory is well defined and does not lead to any ghost DOF. Detailed calculation can be seen in Appendix \ref{App:MassiveGrav}.

From Eq.~\eqref{S.29b} we infer that a massive graviton  is equivalent to a filtered graviton coupled to the energy-momentum tensor $T_{\mu\nu}$ and a scalar with mass $m$ coupled with gravitational strength to the trace of the energy-momentum tensor $T$. The scalar is the longitudinal mode responsible for the vDVZ discontinuity~\citep{Hinterbichler:2011tt}. Due to this feature, linearlized massive gravity has gained attention as a potential modification of GR, particularly in the context of cosmology, especially, as it can lead to alternative explanations for observed phenomena such as cosmic acceleration, structure formation, and the behavior of gravitational waves \citep{dubovsky2004phases, dubovsky2005cosmological,
de2011cosmology, Hinterbichler:2011tt, koyama2011self, gumrukccuouglu2012cosmological2, kobayashi2012new, 
gumrukccuouglu2012gravitational1, fasiello2012cosmological,
de2013nonlinear, 
fasiello2013cosmological,  gumrukccuouglu2013cosmological, 
comelli2014cosmology, kenna2020stable}.

From the binary Neutron Star merger event (GW170817) and its electromagnetic counterpart (GRB 170817A), stringent bounds have been placed on the graviton mass. The speed difference between GWs and light is constrained to \( |\Delta v| < 10^{-15}c \), implying the graviton mass \( m_g < 1.2 \times 10^{-22} \, \mathrm{eV/c^2} \). Current constraints from solar system, cosmological, and GW observations place an upper limit to be \( m_g < 1.2 \times 10^{-22} \, \mathrm{eV/c^2} \)~\citep{
Goldhaber:1974wg, Goldhaber:2008xy,
Talmadge:1988qz, will1998bounding, will2018solar,
 visser1998mass,  
Finn:2001qi,  
 gruzinov2005graviton, dvali2007degravitation, 
 dubovsky2010signatures, 
 zakharov2016constraining,
de2017graviton, desai2018limit,  brax2018signatures, rana2018bounds,
gupta2018limit, miao2019bounding, bernus2019constraining, shao2020new, bernus2020constraint, de2021minimal,  LIGOScientific:2021sio}. 
 
However, challenges arise from the fact that massive gravity theories do not 
intrinsically elucidate the origin of the graviton mass, leading to considerable skepticism.  Recently, this discrepancy has been addressed by \cite{Mandal:2024nhv}  by generalizing to an arbitrary spacetime. Specifically, they showed that spontaneous symmetry breaking of matter
coupled to background geometry leads to the non-zero mass of the spin-2 modes even though
we start with the linearised Einstein-Hilbert action. Moreover, the effective action in a spin-2
field turns out to be the extended Fierz-Pauli action with mass deformation parameter being 
$\alpha = {1}/{2}$. Further, assuming $U(1)$ spontaneous 
symmetry breaking at the dark sector with massive dark photons, they provided a upper bound on 
the mass of spin-2 modes in  which is consistent with the LIGO and other
observations~\citep{Mandal:2024nhv}.

\subsection{Parity violating theories of gravity}
GR preserves parity symmetry. In other words, under reflection of spatial coordinates, the laws of physics remain unchanged. The metric tensor and Einstein's field equations inherently respect parity, as these equations are derived from parity-preserving principles. Parity violation is well-known in the weak interaction of the Standard Model of particle physics (e.g., in neutrino physics)~\citep{Kobayashi:1973fv,BaBar:2001pki}. 

Recent large scale structure data indicate parity violation on cosmological scales. Analyses of galaxy spins have hinted at asymmetries potentially linked to parity-violating physics~\citep{Motloch:2021mfz}. Similarly, investigations into the four-point correlation function (4PCF) of galaxies have highlighted the sensitivity of this statistic to parity-violating signals in large-scale structure data~\citep{Cahn:2021ltp,Philcox:2022hkh,Hou:2022wfj}. This raises the possibility that gravity, as a fundamental interaction, might also exhibit parity-violating effects under certain conditions. Future surveys, including DESI and Euclid, promise improved constraints and may shed light on the nature and origin of these intriguing signals~~\citep{2018-Amendola.others-LRR,Scolnic:2019apa,Fan:2020muj}.

Parity-violating gravity theories have been extensively developed and studied across multiple frameworks. Broadly, these theories are classified into two categories. The first involves modifications to the Einstein-Hilbert action within the standard Riemannian geometry, introducing parity violation by incorporating additional terms, such as the Chern-Simons scalar coupling. The second category arises, where deviations from GR are achieved by altering the fundamental geometric structure, such as using torsion or non-metricity, to generate parity-violating effects~\citep{RevModPhys.48.393, Ni:2009fg}. Experiments traditionally designed to test Lorentz symmetry for ordinary matter have shown sensitivity to effects from torsion and non-metricity, offering an indirect laboratory-based approach to probe these alternative gravitational effects~\citep{Will:2018bme}.

By using an action formalism with the vierbein to incorporate spinors, it is possible to investigate the influence of torsion and non-metricity on the behavior of laboratory test particles. This approach allows for the calculation of trajectories and Hamiltonians that can be compared against experimental results. Notably, current experimental data have established constraints on the components of both the torsion and non-metricity tensors~\citep{PhysRevLett.100.111102, Lehnert:2013jsa, PhysRevD.95.084033}. These findings draw on analogies between certain effects of torsion on matter and the terms in the action that break CPT and Lorentz symmetries, utilizing data from previous experiments conducted within the EFT framework~\citep{RevModPhys.83.11}. 

In the rest of this subsection, we will focus on the first category. We will discuss two models --- dynamical Chern-Simons gravity and non-minimally coupled parity violating models\footnote{Some extensions of HL gravity also break parity~\citep{Herrero-Valea:2023zex}.}. 

\subsubsection{Dynamical Chern-Simons gravity}

The most extensively studied parity-violating modified gravity theory is \emph{Chern-Simons gravity}. This theory introduces a gravitational Chern-Simons term coupled to a pseudo-scalar field, modifying the standard framework of General Relativity. 
Unlike scalar fields, pseudo-scalars change sign under parity transformations (reflections), i. e.,
$\vartheta(r, \pi - \theta, \phi + \pi) = -\vartheta(r, \theta, \phi)$.
Chern-Simons (CS) gravity is a notable example of this approach, inspired by similar modifications in electrodynamics. It introduces a parity-violating pseudo-scalar field that couples with the contraction of the Riemann tensor and its dual:
\[
{}^*R^\tau_{\sigma\mu\nu} = \frac{1}{2} \epsilon_{\mu\nu}^{\quad\alpha\beta} R^\tau_{\sigma\alpha\beta}.
\]
CS gravity has gained attention as it emerges as a low-energy limit of certain string theories and loop quantum gravity frameworks~\citep{Alexander:2009tp}. The action is of the form
\begin{eqnarray}
S &=& \int d^4x \, \sqrt{-g} \left[\frac{R}{2\kappa^2} + \frac{\alpha}{4}\vartheta{}^*RR - \frac{\beta}{2}\left(\nabla\vartheta\right)^2 - \frac{\beta}{2}V\left(\vartheta\right)\right] \, .
\label{csaction}
\end{eqnarray}
In geometric units ($G = c = 1$), $ \vartheta $ is chosen to be dimensionless, which leads to $ \left[\alpha\right] = [L]^2 $ and $ \left[\beta\right] $ is dimensionless and $ {}^*RR $ is referred to as Pontryagin density quantifying the extent to which local Lorentz invariance is violated and is given by:
\begin{eqnarray}\label{pontryagin}
{}^*RR &=& \frac{1}{2} R_{\mu\nu\rho\sigma}\epsilon^{\mu\nu\alpha\beta}R^{\rho\sigma}_{\quad\alpha\beta}
\end{eqnarray}
CS theories are broadly of two types: The first type is $ \vartheta = {\rm constant}$, with no kinetic and potential term~\cite{Jackiw2003}. The second type is $ \vartheta $ is a dynamical field - dynamical CS (dCS)~\cite{Smith:2007jm}.

In spherically symmetric spacetimes, the Pontryagin density vanishes, reducing the theory effectively to GR with a minimally coupled scalar field and its potential. Due to the Birkhoff theorem, the Schwarzschild solution is the only stable configuration in such scenarios~\citep{Bekenstein:1998aw}. Consequently, Schwarzschild spacetime remains a solution for both  versions of CS gravity~\citep{Jackiw2003}. In contrast, constructing axisymmetric solutions in CS gravity is more challenging because the Pontryagin density does not vanish for such spacetimes. Nevertheless, perturbative methods have been employed to derive axisymmetric solutions from spherically symmetric ones by expanding in the spin parameter~\citep{Yagi_Spin2_PhysRevD.86.044037,Konno10.1143/PTP.122.561}. To date, no fully consistent solution representing a fast-spinning Kerr-like black hole exists in either dynamical or non-dynamical CS gravity theories. This limitation underscores the complexity of extending CS gravity to highly spinning configurations.

There are no extra intrinsic degrees of freedom of the gravitational field in this modification, except the two usual massless spin-2 degrees of freedom. In the literature, one usually assumes $ V\left(\vartheta\right) = 0 $, then, the field equations of (\ref{csaction}) lead to
\begin{eqnarray}
R_{\mu\nu} &=& - 2\kappa^2\alpha C_{\mu\nu} + \kappa^2\beta\vartheta_{;\mu}\vartheta_{;\nu} \label{beom1} \\
\Box\vartheta &=& -\frac{\alpha}{4\beta} {}^*RR \label{beom2} \\
C_{\mu\nu} &=& \frac{1}{2} \left[\vartheta_{;\sigma} \left(\epsilon^\sigma_{\,\,\,\mu}{}^{\alpha\beta}R_{\nu\beta;\alpha} + \epsilon^\sigma_{\,\,\,\nu}{}^{\alpha\beta}R_{\mu\beta;\alpha}\right) + \vartheta_{;\tau\sigma}\left({}^*R^\tau_{\,\,\,\mu}{}^\sigma_{\,\,\,\nu} + {}^*R^\tau_{\,\,\,\nu}{}^\sigma_{\,\,\,\mu}\right)\right] \label{cotton0}
\end{eqnarray}
where $ C_{\mu\nu} $ is the Cotton tensor~\citep{Jackiw2003}. Linearizing (\ref{beom1}) about a Minkowski background, and choosing the transverse-traceless gauge, one obtains the radiative part of the perturbed metric as
\begin{eqnarray}
h_{ab} \left(t, r\right) &=&\frac{1}{2\pi}\int_{p}\left(\begin{array}{cc}
h_+(t) - ip\dot{\vartheta}h_\times(t) & h_\times(t) + ip\dot{\vartheta}h_+(t) \\ 
h_\times(t) + ip\dot{\vartheta}h_+(t) & -h_+(t) + ip\dot{\vartheta}h_\times(t)
\end{array} \right) e^{i\textbf{p}\cdot \textbf{r}} dp\label{pc}
\end{eqnarray}
Hence, CS modifications lead to the plus and cross carrying different intensities, as well as imparting a circular polarization to the linearly polarized modes of GR. This potentially can lead to leptogenesis in the early universe~\citep{Alexander:2004us}. 

An interesting consequence of dCS theory is the phenomenon of \emph{vacuum birefringence}. While circularly polarized GWs still propagate at the speed of light in CS theories, parity violation introduces frequency-dependent velocity variations. This means that different frequency components of the waves travel at slightly different speeds in vacuum, resulting in a dispersive effect. Such birefringence, arising from the parity-violating nature of CS gravity, could be detectable by gravitational wave observatories, offering a potential observational signature of these modifications~\citep{Callister:2023tws}.

Recently, \cite{Srivastava:2021imr} analytically computed the fundamental mode ($n=0$) QNM frequencies for a slowly rotating black-hole solution in dCS gravity accurate to linear order in spin ($\chi)$ and quadratic order in the CS coupling parameter ($\alpha$). The authors showed that dCS corrections are potentially observable when the final black-hole mass is less than $15 M_{\odot}$. Hence, the future BNS events can potentially distinguish dCS and GR~\cite{LIGOScientific:2017vwq}.
They showed that for $\tilde{\alpha}=0.1$ the ratio of the imaginary parts of the dCS correction to the purely GR correction in the first QNM frequency (for the polar sector) is $0.263$. Also for a certain range of parameters, the dCS corrections make the magnitude of the imaginary part of the first QNM of the fundamental mode smaller, thereby decreasing the decay rate.

\subsubsection{Non-minimally coupled parity violating models}

As has been highlighted in this review, many contemporary challenges in high-energy physics and cosmology are tied to the nature of gravitational interactions. For instance, DM and DE remains undetected except through its gravitational effects, underscoring the importance of understanding and probing gravitational phenomena. One of the key questions is to  systematically investigate the effects of non-minimally coupling the matter fields with gravity, especially the parity violating terms. 

Recent advancements have introduced systematic frameworks, such as the effective field theory of gravity, to identify these non-minimal coupling terms~\citep{Falkowski:2023hsg}. Specifically, a general EFT framework for gravity coupled to the Standard Model of particle physics was systematically developed~\citep{2019-Ruhdorfer.etal-JHEP}. They showed that the first non-trivial gravity operators appear at mass-dimension 6 and are shown to couple exclusively to the Bosonic sector of the Standard Model. Further, they showed that no new gravity-related operators emerge at mass-dimension 7 
and operators at mass dimensions 8 include SM fermions. Additionally, couplings between the scalar Higgs field and SM gauge bosons emerge only at this dimension, highlighting a higher-order interaction structure.

\cite{Kushwaha:2020nfa} showed that \emph{magnetogenesis and baryogenesis are two sides of the same coin} by limiting to mass dimension $6$ operators coupling to the EM field. They considered the following action:
\begin{eqnarray}\label{eq:Model_action}
S &=  -\frac{1}{2 \kappa^2}\int d^4x \sqrt{-g} \, R \, +  \int d^4x \sqrt{-g} \left[  \frac{1}{2} \partial_{\mu}\phi \partial^{\mu}\phi -  V(\phi) \right] 
-\frac{1}{4} \int d^4x \, \sqrt{-g} \, F_{\mu\nu} F^{\mu\nu} - \frac{\sigma}{M^2} \,\int d^4x \, \sqrt{-g} \, R_{\rho\sigma}\,^{\alpha\beta} F_{\alpha\beta} \, \tilde{F}^{\rho\sigma}
\end{eqnarray}
where $R_{\rho\sigma}\,^{\alpha\beta}$ is the Riemann tensor, $A_{\mu}$ is the four-vector potential of the EM field, $F_{\mu\nu} = \nabla_{\mu}A_{\nu} - \nabla_{\nu}A_{\mu} $ and $\tilde{F}^{\rho\sigma} = \frac{1}{2} \epsilon^{\mu\nu\rho\sigma}F_{\mu\nu} $ is the dual of $F_{\mu\nu}$. $\epsilon^{\mu\nu\rho\sigma} = \frac{1}{\sqrt{-g}}\, \eta^{\mu\nu\rho\sigma}$ is a fully antisymmetric tensor, $\eta^{\mu\nu\rho\sigma}$ is Levi-Civita symbol whose values are $\pm1$ and we set $\eta^{0123} = 1 = - \eta_{0123}$.  $M$ is the energy scale, which sets the scale for the breaking of conformal invariance and parity invariance of the EM field. They showed that the model can generate sufficient primordial helical magnetic fields at all observable scales.

Refining \cite{1996-Davidson-PLB} criteria they showed that the helical field generated during the end stages of inflation can explain the baryogenesis~\citep{Kushwaha:2022twx}. Specifically, they identified a key missing ingredient in Davidson's condition to generate baryon asymmetry due to the primordial magnetic field.  \cite{Kushwaha:2022twx} explicitly showed that the presence of primordial helical fields leads to the non-zero Chern-Simons number and, eventually, the change in the Fermion number. 

\section{Beyond metric theories of gravity}
\label{sec:MGModels-ClassC}

The motivation for exploring beyond metric theories of gravity stems from both theoretical and observational challenges to GR. These theories aim to address unresolved issues, such as reconciling gravity with quantum mechanics and explaining phenomena like cosmic acceleration and galactic rotation curves. By extending or modifying the fundamental framework of GR, these alternate theories create opportunities for deeper insights into the nature of gravity and the universe. Furthermore, these theories offer novel predictions that can guide future experimental tests and refine our understanding of gravitational interactions.

In this section, we consider two classes of such theories --- Scalar-tensor, and Tensor-Vector-Scalar (TeVeS) theories --- that have been useful to explain astrophysical and cosmological observations. 

\subsection{Scalar-Tensor theory of gravity}

Prior to Einstein, Nordstr$\ddot{o}$m attempted a \emph{scalar theory} by promoting the Newtonian potential to a Lorentz scalar, but its non-geometric nature failed to uphold the EP, a key aspect of GR. Dissatisfied, Einstein developed a dynamic spacetime geometry, which later gained empirical support from various observations, including those in cosmology.
At first glance, scalar-tensor theories might appear to revisit the outdated idea of scalar gravity, but they represent a sophisticated extension of GR. Rather than simply combining scalar and tensor fields, these theories utilize a \emph{nonminimal coupling term}, allowing the scalar field to intricately influence the spacetime dynamics.

Scalar-tensor theories trace their origins to Jordan, who envisioned embedding a four-dimensional curved spacetime within a five-dimensional flat manifold. He demonstrated that a scalar field could serve as a constraint for projective geometry, enabling the formulation of a space and time-dependent Newton's constant~\citep{1937-Dirac-nat}. This idea extended Dirac's argument for a variable gravitational constant, moving beyond GR's fixed constant framework~\citep{1938-Dirac-PRSA}. Jordan also explored connections to the five-dimensional theories proposed by Kaluza and Klein, which unify gravity and electromagnetism.

Building on these foundations, Jordan formulated a general Lagrangian describing a scalar field in a four-dimensional curved spacetime~\citep{2005-Brans-Talk}:
\begin{equation}\label{Jordan-Lagrangian}
\mathcal{L}_{J} = \sqrt{- g}\Bigg[\varphi_{J}^{\gamma}\left(R - \omega_{J}\frac{1}{\varphi_{J}^{2}}g^{\mu\nu}\partial_{\mu}\varphi_{J}\partial_{\nu}\varphi_{J}\right) + L_{(M)}(\varphi_{J}, \Psi)\Bigg],    
\end{equation}
where $\varphi_{J}(x)$ is Jordan’s scalar field, while $\gamma$ and $\omega_{J}$ are the two constants in this theory, also $\Psi$ represents the collective matter fields. The introduction of the non-minimal coupling term, $\varphi_{J}^{\gamma}R$, the first term on the right-hand side, marked the birth of the scalar-tensor theory. The term $L_{(M)}(\varphi_{J}, \Psi)$ is the matter Lagrangian, which in principle can depend on the scalar field, as well.

The expression in the second term on the right-hand side of the equation (\ref{Jordan-Lagrangian}) bears resemblance to a kinetic term of $\varphi_{J}$. Requiring this term to possess the correct mass dimension of 4 leads to the conclusion that $\varphi_{J}$ has a mass dimension of $2/\gamma$. Consequently, $\varphi_{J}^{\gamma}$, which multiplies $R$ in the first term on the right-hand side of (\ref{Jordan-Lagrangian}), holds a mass dimension of 2, akin to $G^{-1}$. This confirms that the first two terms of (\ref{Jordan-Lagrangian}) do not incorporate any dimensional constants. This assertion stands for all values of $\gamma$, although this ``invariance" under a change of $\gamma$ may not apply if $\varphi_{J}$ is part of the matter Lagrangian in general. Brans and Dicke introduced a scalar field $\varphi$ defined as~\citep{2005-Brans-Talk},
\begin{equation}
\varphi = \varphi_{J}^{\gamma}.
\end{equation}
This simplification of (\ref{Jordan-Lagrangian}) is based on the above observation that the specific choice of $\gamma$ is inconsequential. This approach was deemed justified as they insisted on decoupling the matter component of the Lagrangian, $\sqrt{- g}L_{\text{matter}}$, from $\varphi(x)$ as a means of adhering to their requirement for the WEP to be upheld, in contrast to Jordan's model.
In this manner, they proposed the Jordan-Brans-Dicke Lagrangian~\citep{1961-Brans.Dicke-Phys.Rev.}
\begin{equation}
\mathcal{L}_{BD} = \sqrt{- g}\left(\varphi R - \omega\frac{1}{\varphi}g^{\mu\nu}\partial_{\mu}\varphi\partial_{\nu}\varphi + 16\pi L_{(M)}(\Psi)\right).    
\label{eq:LJBD}
\end{equation}
Variation of the above action w.r.t the metric tensor leads to the following 
modified Einstein's equation:
\begin{equation}
\label{eq:EOM-JBD}
R_{\mu\nu} - \frac{1}{2}g_{\mu\nu}R - \frac{\omega}{\varphi^{2}}\left(\partial_{\mu}\varphi\partial_{\nu}\varphi - \frac{1}{2}g_{\mu\nu}g^{\alpha\beta}\partial_{\alpha}\varphi\partial_{\beta}\varphi\right) - \frac{1}{\varphi}(\nabla_{\mu}\nabla_{\nu}\varphi - g_{\mu\nu}\Box\varphi) = \frac{8\pi}{\varphi}T_{\mu\nu}^{(M)}.  
\end{equation}
Varying the action \eqref{eq:LJBD} w.r.t $\varphi$ leads to the following scalar field equation:
\begin{equation}
\Box\varphi - \frac{g^{\mu\nu}\partial_{\mu}\varphi\partial_{\nu}\varphi}{2\varphi} + \frac{\varphi}{2\omega}R = 0.    
\end{equation}
The above equation relates $\varphi$ to the geometry via the Ricci scalar. 
This implies that besides the metric, the scalar field determines the gravity
Therefore, this theory should be interpreted as adding a scalar field to the metric to produce the total gravitational field. 
On the other hand, combining the previous two equations, we obtain the following resultant equation
\begin{equation}
\Box\varphi = \frac{8\pi}{2\omega + 3}T^{(M)} \, .    
\end{equation}
The strongest constraint on $\omega$ comes from the Cassini mission measurement of the Shapiro time delay and it has put a bound on $\omega > 40,000$~\citep{2003Bertotti,Mariani:2023rca}.

Numerous scalar-tensor theories have been developed over time creating the need for a unified framework to systematically address them~\citep{Clifton:2011jh}. Such a framework must avoid pathological behaviors, such as ghost instabilities, while encompassing a broad range of scalar-tensor theories as special cases. In this context, the Galileon theory~\citep{Nicolis:2008in},was introduced in flat spacetime as a scalar field theory free from the Ostrogradsky ghost instability \eqref{App:HDTheories}, despite involving higher-derivative terms in its Lagrangian~\citep{Woodard:2006nt} . This theory was later extended to include dynamical gravity and was generalized to accommodate additional parameters~\citep{Deffayet:2009wt,Deffayet:2011gz}. The resulting framework, known as the generalized Galileon theory, was shown by 
\cite{Kobayashi:2011nu} to be equivalent to the Horndeski theory~\citep{Horndeski:1974wa}, originally constructed in the 1970s.

\begin{table}[!htb]
\resizebox{\textwidth}{!}{
\begin{tabular}{||l|l||} 
\hline \hline  Special Cases  & Description of the theory \\
\hline \hline $\mathcal{L} = G_{4}(\varphi)R + G_{2}(\varphi, X)$ & K-essence coupled non-minimally to gravitation
\\
& (includes $f(R, \varphi)$, DBI, BD, Chameleon, Dilaton)
\\
\hline $\mathcal{L} = G_{3}(\varphi, X)\Box\varphi$ & Kinetic-gravity braiding/G-inflation
\\
\hline $G_{5} \neq 0$ & Non-minimal coupling to Einstein tensor
\\
\hline $G_{2} = 8\xi^{(4)}X^2(3 - \mathrm{ln}X), G_{3} = 4\xi^{(3)}X(7 - 3 \mathrm{ln}X),$ & Non-minimal field coupling to Gauss-Bonnet term
\\
$G_{4} = 4\xi^{(2)}X(2 - \mathrm{ln}X), G_{5} = -4\xi^{(1)}\mathrm{ln}X.$ & $(\xi ^{(n)} := \partial^n\xi/\partial\varphi^n)$
\\
\hline $\mathcal{L} = \sum_{i = 1}^{5} c_{i}\mathcal{L}_{i},$ & Covariant Galileon (covariant scalar field
\\
$\mathcal{L}_{1} = M^3\varphi, \mathcal{L}_{2} = (\nabla\varphi)^2, \mathcal{L}_{3} = (\Box\varphi)(\nabla\varphi)^2/M^3,$ & respecting Galilean symmetry in flat spacetime)
\\
$\mathcal{L}_{4} = (\nabla\varphi)^2[2(\Box\varphi)^2 - 2\varphi_{\mu\nu}\varphi^{\mu\nu} - R(\nabla\varphi)^2/2]/M^6,$ & 
\\
$\mathcal{L}_{5} = (\nabla\varphi)^2 [(\Box\varphi)^3 - 3(\Box\varphi)\varphi_{\mu\nu}\varphi^{\mu\nu} + 2\tensor{\varphi}{_\mu^\nu}\tensor{\varphi}{_\nu^\rho}\tensor{\varphi}{_\rho^\mu}$ & 
\\
\ \ \ \ \ \ \ $ -6\varphi_{\mu}\varphi^{\mu\nu}\varphi^{\rho}G_{\nu\rho}]/M^9.$ &
\\
\hline \hline
\end{tabular}
}
    \caption{Scalar-tensor theories contained in the Horndeski action \eqref{Eq:Horndeski action} \cite{Deffayet:2009wt, 2010-Kobayashi-Phys.Rev.Lett., Kobayashi:2019hrl, Unnikrishnan:2013rka,2010-DeFelice-Phys.Rev.Lett.}. Source: \cite{Bansal:2024bbb}.}
    \label{Table: Horndeski special cases}
\end{table}

The Horndeski theory represents the most general covariant scalar-tensor framework involving a single scalar field, where the Euler-Lagrange equations remain second-order in the derivatives of both the metric and the scalar field. %
\begin{eqnarray}
\label{Eq:Horndeski action}
S &=& \int \mathrm{d}^{4}x\sqrt{- {g}}
\left[\frac{1}{2 \kappa^2} {G}_{4}(\varphi,  {X}) {R} +  {G}_{2}(\varphi,  {X}) -  {G}_{3}(\varphi,  {X}) {\Box}\varphi
 + \frac{1}{2 \kappa^2} {G}_{4 {X}}[( {\Box}\varphi)^{2} -  {\nabla}_{\mu} {\nabla}_{\nu}\varphi {\nabla}^{\mu} {\nabla}^{\nu}\varphi] +  {G}_{5}(\varphi,  {X}) {G}_{\mu\nu} {\nabla}^{\mu} {\nabla}^{\nu}\varphi 
 \right. \\
& & \left.
- \frac{1}{6} {G}_{5 {X}}(\varphi,  {X})\left\{ ( {\Box}\varphi)^{3} + 2 {\nabla}_{\mu} {\nabla}^{\nu}\varphi {\nabla}_{\nu} {\nabla}^{\alpha}\varphi {\nabla}_{\alpha} {\nabla}^{\mu}\varphi - 3 {\nabla}_{\mu} {\nabla}_{\nu}\varphi {\nabla}^{\mu} {\nabla}^{\nu}\varphi {\Box}\varphi\right\}
\right] \, .
\nonumber
\end{eqnarray}
where $ {X} = - {g}^{\mu\nu} {\nabla}_{\mu}\varphi {\nabla}_{\nu}\varphi\,/2$, $ {\Box}\varphi =  {g}^{\mu\nu} {\nabla}_{\mu} {\nabla}_{\nu}\varphi$. 
$ {G}_{2},  {G}_{3},  {G}_{4}$ and $ {G}_{5}$ are arbitrary functions of $\varphi$ and $ {X}$. Different subclasses of Horndeski theories exist, depending on the form of these functions. Some of these cases are outlined in Table~\eqref{Table: Horndeski special cases}. In the above action, the terms containing $ {G}_{4}$ and $ {G}_{5}$ have quadratic and cubic order second derivative terms, respectively.

While Horndeski and Lovelock theories (discussed in Sec. \eqref{sec:lovelock}) extend GR and share similarities in maintaining second-order field equations, Lovelock theory generalizes gravity through higher-dimensional curvature invariants, whereas Horndeski theory focuses on scalar-tensor interactions in four-dimensional spacetimes. There are some crucial differences between the two theories in-terms of motivation, applicability and goals: 
First, Lovelock gravity is motivated by a geometric approach to generalizing Einstein's field equations in higher dimensions. However, Horndeski theory is developed to explore the dynamics of scalar-tensor theories while maintaining second-order equations. 
Second, Lovelock gravity is more focused on higher-dimensional theories of gravity and their geometric consequences. However, Horndeski theory is primarily used in the context of 4-D space-time, especially for exploring cosmological phenomena such as inflation and dark energy.

In the rest of this subsection, we discuss applications of Horndeski theory in  BHs and late-time cosmology.

\subsubsection{Implications for primordial black holes}

\cite{Hawking:1975vcx} showed that quantum effects in the background of a body collapsing to a Schwarzschild black hole leads to the emission of a thermal radiation at a characteristic temperature:
\begin{equation}
\label{eq:H-temp}
T_{\rm H} = \left(\frac{\hbar c}{k_{_{B}}} \right) \frac{\kappa}{2 \pi} = 
\left(\frac{\hbar c^3}{G k_{_{B}}}\right) \frac{1}{8 \pi M}\, ,
\end{equation}
where $G$ is the Newton's constant in four dimensions, $k_{_{B}}$ is
the Boltzmann constant, and $M$ is the mass of the black hole.  The factor of proportionality between temperature and surface gravity (and as such
between entropy and area) gets fixed in Hawking's derivation
thus leading to the Bekenstein-Hawking area law~\citep{Bekenstein:1973ur}:
\begin{equation} 
\label{eq:BH-law}
S_{_{\rm BH}} = \left(\frac{k_{_{B}}}{4}\right) 
\frac{A_{\rm H}}{L_P^2} \, .
\end{equation}
This raises several pertinent questions: How does a pure state evolve into a mixed (thermal) state? Is there a information loss due to the formation of black-hole and Hawking process? Does the usual quantum mechanics need to be modified in the context of black-holes? For a detailed review, see \citep{Mathur:2008wi}. Understanding the final stages of BH evaporation is crucial for addressing the above questions as the temperature increases rapidly as the black hole loses mass during its evaporation. This rapid rise necessitates the inclusion of high-energy corrections in the theoretical framework to accurately describe the late stages of BH evaporation.
This is particularly relevant for primordial black holes (PBHs) in the mass range \( 10^{17} - 10^{23} \, \text{g} \), where the associated Hawking temperature \eqref{eq:H-temp} becomes significantly high. PBHs have garnered renewed attention as potential candidates for dark matter (DM)~\cite{Carr:2020xqk}.  

Numerical studies by Page have demonstrated that $90\%$ of Hawking radiation originates from \( s \)-waves \cite{Page:1976df}. For a spherically symmetric spacetime in 4-D, focusing on \( s \)-waves effectively reduces the problem to a 2-D gravity model, which is particularly relevant for studying Hawking radiation dynamics. Hence, 2-D dilaton models are used as toy models that can replicate 4-D BHs~\citep{Jackiw:1984je, Grumiller:2002nm}. For instance,  CGHS (Callan-Giddings-Harvey-Strominger) model \cite{Callan:1992rs} is an exactly solvable classical 2-D dilaton gravity model.  It is described by the following action:
\[
S_{\text{CGHS}} = \frac{1}{4\pi} \int d^2x \sqrt{-g} e^{-2\varphi} \left[ \mathcal{R} + 4 \nabla_\mu \varphi \nabla^\mu \varphi + 4\lambda^2 \right],
\]
where \( \varphi \) is the dilaton field, \( \mathcal{R} \) is the 2-D Ricci scalar, and \( \lambda^2 \) represents a cosmological constant. However, it is not possible to study the end-stages of this 2-D graviton model as it has never been obtained from a 4-D gravity action. 

Recently, \cite{Mandal:2023kpu} derived the above 2-D dilaton model from the 4-D Horndeski action \eqref{Eq:Horndeski action}. The authors developed a systematic approach to incorporate higher-derivative corrections into the CGHS model to examine the late stages of black hole evaporation. A key advantage of this framework is its avoidance of Ostrogradsky instability \eqref{App:HDTheories}. Consequently, the Horndeski action \eqref{Eq:Horndeski action} serves as a robust tool to analyze higher-derivative effects on the \( s \)-wave contribution of 4-D Hawking radiation. The work lays the groundwork for exploring Hawking radiation effects in PBHs within the mass range \( 10^{16} - 10^{17} \, \text{g} \). Observational constraints, such as the positron annihilation rate inferred from INTEGRAL's Galactic 511 keV line measurements, already impose significant restrictions on PBHs in this mass range \cite{DeRocco:2019fjq, Keith:2021guq}. However, these constraints typically assume that the Hawking flux remains consistent with that of larger black holes. However, higher-derivative corrections could introduce substantial deviations for PBHs in this regime, potentially refining constraints on their role as dark matter candidates.

\subsubsection{Implications for resolving the current tensions in cosmology}

As discussed in the Introduction, with increasing observational precision and tighter constraints on cosmological and model parameters, several significant cosmological tensions have surfaced, notably the \( H_0 \) and \( \sigma_8 \) discrepancies between early and late Universe observations~\cite{2022-Brout.others-Astrophys.J.,2022-Perivolaropoulos.Skara-NewAstron.Rev.,2016-Riess.others-Astrophys.J.,2019-Riess.etal-Astrophys.J.,2023-Gupta-Mon.Not.Roy.Astron.Soc.}. Although the \(\Lambda\)CDM model fits the data well overall, estimated parameter values show tensions of about \(3\sigma\). This has prompted the development and extensive study of alternative models~\cite{
2021-DiValentino.etal-Class.Quant.Grav.,Kamionkowski:2022pkx}.

These tensions provide an opportunity to explore non-gravitational interactions between DE and DM~\citep{Wang:2016lxa}. Proposed dark sector interaction models suggest that energy transfer between DE and DM can mitigate the \( H_0 \) tension, though the \(\sigma_8\) tension typically remains unresolved~\citep{2021-DiValentino.etal-Class.Quant.Grav.,Schoneberg:2021qvd,2024-Wang-Rept.Prog.Phys.,2024-Giare-}. In most cases, these interactions are introduced phenomenologically, with interaction strengths (\(Q_\nu\)) specified arbitrarily due to limited knowledge of the dark sector. 

Recently, \cite{2021-Johnson.Shankaranarayanan-Phys.Rev.D} have explored these interactions using a classical field-theoretic framework, involving a coupled canonical scalar field model for DE and DM with an arbitrary coupling function. 
Under the conformal transformation (see details in Appendix \ref{App:ConformalTransf} and ~\cite{2022-Johnson.etal-JCAP}): 
\begin{equation}
\label{eq:conftrans}
{g}_{\mu \nu}=\Omega^{2} \tilde{g}_{\mu \nu},
\quad \mbox{where} \quad
\Omega^{2}= F(\tilde{R},\tilde{\chi}) \equiv \frac{\partial f(\tilde{R}, \tilde{\chi})}{\partial \tilde{R}} 
\end{equation}
and a field redefinition, the following action in the Jordan frame 

\begin{equation}
\label{eq:fRaction}
S_{J}=\int d^{4}x\sqrt{-\tilde{g}}\left[\frac{1}{2 \kappa^2} f(\tilde{R},\tilde{\chi})- \frac{1}{2}   \tilde{g}^{\mu \nu} \tilde{\nabla}_{\mu} \tilde{\chi} \tilde{\nabla}_{\nu} \tilde{\chi} -V(\tilde{\chi})\right] 
\end{equation}
(where $f(\tilde{R},\tilde{\chi})$ is an arbitrary, smooth function of Ricci scalar, and scalar field $\tilde{\chi}$, and $V(\chi)$ is the self-interaction potential of the scalar field $\tilde{\chi}$) takes the following form in the Einstein frame: 
\begin{equation}
 \label{eq:Scde}
S = \int d^4x \sqrt{-g}\left(\frac{1}{2 \kappa^2}R-\frac{1}{2}g^{\mu \nu}\nabla_{\mu}\phi \nabla_{\nu}\phi - U(\phi)-\frac{1}{2} e^{2\alpha(\phi)}g^{\mu \nu}\nabla_{\mu}\chi \nabla_{\nu}\chi -e^{4 \alpha(\phi)} V(\chi) \right).
\end{equation}
where, 
\[
 U = \frac{F\tilde{R}-f}{2\kappa^2 F^{2}} \, .
\]
and $\alpha(\phi)$ denotes the interaction between dark energy and dark matter. Defining the dark matter fluid by specifying the four velocity energy density and pressure
\begin{equation}
\label{eq:dm4v}
u_{\mu} = -\left[-g^{\alpha \beta} \nabla_{\alpha}\chi \nabla_{\beta}\chi \right]^{-\frac{1}{2}} \nabla_{\mu} \chi
\end{equation}
\begin{equation}
\label{eq:dmrhop}
p_{m}  = -\frac{1}{2}e^{2 \alpha}\left[g^{\mu \nu} \nabla_{\mu} \chi \nabla_{\nu} \chi + e^{2\alpha}V(\chi) \right], \quad
\rho_{m}  = -\frac{1}{2}e^{2 \alpha}\left[g^{\mu \nu} \nabla_{\mu} \chi \nabla_{\nu} \chi - e^{2\alpha}V(\chi) \right].
\end{equation} 
Then the interaction function in the field theory and fluid descriptions are given by
\begin{equation}
\label{eq:interaction02}
Q_{\nu}^{\rm (F)} = -e^{2\alpha(\phi)} \alpha_{,\phi}(\phi) \nabla_{\nu} \phi \left[ \nabla^{\sigma} \chi \nabla_{\sigma} \chi + 4 e^{2\alpha(\phi)} V(\chi) \right] =  -\alpha_{,\phi}(\phi) \nabla_{\nu}\phi (\rho_m  - 3 p_m) 
\end{equation}
They showed that a one-to-one mapping between the field theory description and fluid description of the interacting dark sector described above exist \emph{only} for this form of interaction function.

Recently, \cite{Bansal:2024bbb} extended the analysis for Horndeski action~\eqref{Eq:Horndeski action}~\citep{Horndeski:1974wa,Kobayashi:2019hrl}. Using standard conformal and extended conformal transformations~\citep{1993-Bekenstein-Phys.Rev.D,2013-Zumalacarregui.etal-PRD,2015-vandeBruck.etal-JCAP} 
on the Horndeski action  ~\eqref{Eq:Horndeski action}, the authors derived interaction terms for a broader class of models. They showed that the above one-to-one mapping between the field theory description and fluid description of the interacting dark sector described is valid for a large scale of models. For the extended conformal coupling they obtained  generalized interaction terms, revealing significant physical implications for energy and momentum transfer within the dark sector. Recent studies suggest that momentum transfer in the dark sector could simultaneously address the $H_0$ and $\sigma_8$ tensions~\citep{Chamings:2019kcl,2020-Amendola.Tsujikawa-JCAP}. While these previous approaches are largely phenomenological, the Horndeski gravity can provide a single framework to 
potentially solve $H_0$ and $\sigma_8$ tensions. 
 
\subsection{TeVeS --- Modified Newtonian dynamics (MOND)}

These models are proposed as an alternative to dark matter, combining scalar and vector fields to modify gravity on galactic scales. MOND which is a specific case of TeVeS is an alternative paradigm to Newtonian dynamics and GR, aiming to explain the observed mass discrepancies in the Universe without invoking the dark matter required by standard gravitational theories.

MOND proposes deviations from standard dynamics at accelerations below a characteristic threshold, \( a_0 \), which introduces a new scale to the study of galactic and cosmic systems. This acceleration threshold fundamentally alters the way gravitational interactions behave at low accelerations~\citep{Milgrom1983}. A key aspect of MOND is its invariance under space-time scaling transformations in the low-acceleration regime, a symmetry not present in standard Newtonian or relativistic frameworks. The theory has successfully predicted several empirical laws governing galactic dynamics, extending the framework of classical Keplerian laws. These predictions, many involving \( a_0 \), revealed previously unrecognized correlations in galactic behavior and have been consistently validated by increasingly precise observational data~\citep{milgrom2002mond,Milgrom:2008rv,milgrom2013testing}.

MOND suggests that Newtonian gravity breaks down for accelerations below \( a_0 \), where the gravitational force transitions from the familiar inverse-square law to a non-linear regime. This modification addresses the observed anomalies in galaxy rotation curves—discrepancies that standard dynamics attribute to the presence of dark matter. The MOND framework offers a direct explanation for these anomalies without requiring unseen mass components. Formally, MOND introduces a relationship between the gravitational acceleration \( a \) and the Newtonian acceleration \( a_N \) such that~\citep{Milgrom1983,  milgrom1998modified,Milgrom:2008rv}:
\begin{equation}
\mu\left(\frac{a}{a_0}\right) a = \frac{GM}{r^2},
\end{equation}
where, \( \mu(x) \) is an interpolation function that transitions between Newtonian behavior (\( \mu(x) \to 1 \)) at high accelerations and MONDian behavior (\( \mu(x) \sim x \)) at low accelerations,  $a_0$ is the threshold acceleration, $G$ is the gravitational constant, $M$ is the mass of the central object, and $r$ is the radial distance from the object. In this framework, the acceleration $a_0$ introduces a fundamental scale that governs the dynamics of galaxies and large-scale structures in the Universe. The key idea is that at low accelerations, typical of the outskirts of galaxies, the gravitational dynamics deviate from the Newtonian expectations, leading to observed phenomena such as flat rotation curves without requiring the presence of dark matter.

A key phenomenological prediction of MOND  is the baryonic Tully-Fisher relation, which emerges from the MOND mass-asymptotic-speed relation (MASR). This relation connects the asymptotic rotational velocity of a galaxy, \( V_{\infty} \) (which MOND predicts to be constant), to the galaxy’s total baryonic mass, \( M \), through the equation~\citep{Famaey2012, bugg2015mond}:
\begin{equation}
V_{\infty}^{4} = M G a_0 \, .
\end{equation}
Another significant feature of MOND is the mass-discrepancy-acceleration relation (MDAR), also known as the radial acceleration relation (RAR). This relation describes a precise correlation between the observed mass discrepancies and the internal accelerations of galactic systems. It forms the foundation for MOND’s predictions of galaxy rotation curves and the dynamics of other galactic systems, a correlation that has been confirmed by numerous studies.

In general, MOND provides an accurate description of the observed dynamics of individual galaxies, spanning from dwarf galaxies to massive spirals, ellipticals, and dwarf spheroidals, as well as galaxy groups. This is achieved by relying solely on the distribution of visible matter, without the need to invoke dark matter. The fundamental laws of galactic dynamics predicted by MOND, supplemented by a few general assumptions unrelated to MOND itself, align well with observational data. The constant \( a_0 \) plays multiple, independent roles in these laws, further underscoring the theory’s internal consistency and reliability. However, it faces challenges when applied to galaxy clusters. In the dark matter paradigm, dark matter contributes significantly to the mass in clusters, influencing the dynamics. In contrast, MOND struggles to explain the observed mass distribution of galaxy clusters without additional effects, such as the external field effect. While MOND can explain individual galaxies, its application to large-scale structure is less successful compared to the CDM paradigm~\citep{mortlock2001gravitational}. Recently, \cite{Mandal:2025hbe} have shown that MOND-like theories emerge from Q-Balls (a non-topological solitonic objects in quantum field theoruies).

\section{Conclusions}
\label{sec:Conclusions}

General Relativity (GR), while remarkably successful, faces challenges in reconciling with quantum mechanics, explaining dark matter and dark energy, and describing extreme environments. These limitations, coupled with some observational hints, motivate the exploration of modified gravity theories.

This review classified modified gravity theories based on the GR principles they preserve or violate: (1) metric theories preserving local Lorentz invariance (LLI) and gauge invariance (GI); (2) metric theories breaking gauge invariance, LLI, or parity; and (3) beyond-metric theories violating the Einstein Equivalence Principle (EEP).  This classification highlights the core assumptions of GR challenged by these alternatives.  For each class, we discussed representative theories, their current status, limitations, and implications for cosmology and black hole physics. While these modified theories offer potential solutions to some GR challenges, many open questions remain.  Nevertheless, they introduce new features that future observations can exploit to probe gravity more effectively.

Upcoming astronomical, cosmological, and gravitational wave missions are set to rigorously test gravity across a range of curvature scales, from weak-field cosmological environments to extreme strong-field regimes near black holes. The \emph{Square Kilometre Array}  will test gravity on both cosmological and astrophysical scales by studying pulsar timing arrays, gravitational waves, and the large-scale distribution of matter, offering insights into phenomena like dark energy and potential deviations from GR~\citep{Weltman:2018zrl, Katz:2024ayw}. The \emph{Euclid Space Telescope} and \emph{Nancy Grace Roman Telescope} will probe gravity at cosmological scales by mapping dark matter and dark energy through gravitational lensing and galaxy clustering, exploring the universe's expansion history and structure formation~\citep{2018-Amendola.others-LRR, Scolnic:2019apa,Fan:2020muj}. 

Gravitational wave observatories, including the LISA~\citep{LISA:2022kgy} and next-generation ground-based detectors like the Einstein Telescope and Cosmic Explorer, will focus on strong-field tests of GR~\citep{Branchesi:2023, Evans:2023-CE}. These missions will measure low-frequency and high-frequency gravitational waves emitted by BH mergers and other cosmic phenomena, providing unprecedented precision to evaluate GR's validity in extreme spacetime curvatures. The following two decades promise to be an exciting era for gravitational research, as these missions will rigorously test gravity across a wide range of regimes, from the vast cosmological structures of the universe to the most compact and extreme astrophysical objects. The \emph{multimessenger signatures} of these exotic astrophysical events hold the potential to uncover groundbreaking new physics and transform our understanding of the fundamental forces that shape the cosmos.

Testing gravity in the multimessenger era transcends the verification of established theories --- it is a pursuit of the unknown. Each observed gravitational wave, photon, and cosmic particle opens a new window into the universe, offering insights into phenomena like the mysterious nature of black holes and the elusive properties of the dark universe. As we extend the horizons of both observational capabilities and theoretical frameworks, we are not merely honoring Einstein's remarkable legacy. Instead, we are driving the next great revolution in physics, advancing our understanding of the cosmos and the fundamental principles that govern it

\begin{ack}[Acknowledgements]

The authors thank P. Bansal, I. Chakraborty, A. Chowdhury, P. G. Christopher, J. P. Johnson, A. Kushwaha, T. Parvez and S. Xavier for comments on the earlier version of the draft. This work is supported by SERB-CRG grant.
\end{ack}

\seealso{article title}

\begin{appendices}
\renewcommand{\thechapter}{A} 
\renewcommand{\thesection}{A\arabic{section}}
\counterwithin*{equation}{section} 
\renewcommand\theequation{\thesection\arabic{equation}} 

\section{Ostrogradskian instability of higher derivative theories}
\label{App:HDTheories}
Since Newton's time, all fundamental theories in physics have been established on the basis of equations of motion that do not encompass terms with more than second-time derivatives of the dynamical variables. In the framework of Lagrangian formalism, this restriction entails that the Lagrangian can exclusively function as a representation of the dynamical variable and its first-time derivative, thereby yielding a phase space of 2N dimensions per canonical coordinate for N particles. Nevertheless, the Lagrangian is capable of accommodating higher derivative terms if they can be formulated as a complete time derivative of a certain function, thereby leaving the equations of motion unaltered. 

Conversely, the incorporation of higher derivative terms has historically been pursued in an endeavor to amend fundamental theories, with the aim of eliminating theoretical complexities or aligning them with observational data."

Even in gravity, higher curvature theories were proposed immediately after Einstein’s field equations by Weyl and Eddington. One of the motivations for generalized field equations was the regularization in
order to tame divergences arising from the assumption of taking the electron as a point-like particle and this motivation endured also in the pre-renormalization era of the quantum field theory. Effective quantum field theories endowed with higher derivative order to tame divergences arising from the assumption of taking the electron as a point-like particle and this motivation endured also in the pre-renormalization era of the quantum field theory. Effective quantum field theories endowed with higher derivatives. Even in gravity, higher curvature theories were proposed immediately after Einstein’s field equations by Weyl and Eddington. One of the motivations for generalized field equations was the regularization in
order to tame divergences arising from the assumption of taking the electron as a point-like particle and this motivation endured also in the pre-renormalization era of the quantum field theory. Effective quantum field theories endowed with higher derivative order to tame divergences arising from the assumption of taking the electron as a point-like particle and this motivation endured also in the pre-renormalization era of the quantum field theory. Effective quantum field theories endowed with higher derivatives.

The pioneering formalism for Lagrangians involving more than one time derivative was originally devised by Ostrogradski. We will commence by revisiting the development of the Hamiltonian for a conventional Lagrangian, before delving into Ostrogradski's approach to formulating the Hamiltonian for a Lagrangian with higher derivatives, along with its field theoretical extension.

\subsection{Canonical formulation of lower derivative theories}

For the sake of mathematical simplicity, we start by considering a one dimensional time-independent system whose action is given by
\begin{equation}
S[q] = \int dt \ L(q, \dot{q}),    
\end{equation}
whose variation leads to the Euler-Lagrange equation
\begin{equation}
\frac{\partial L}{\partial q} - \frac{d}{dt}\frac{\partial L}{\partial \dot{q}} = 0.    
\end{equation}
Unless otherwise stated, we assume that the above-mentioned Lagrangian is non-degenerate which means that the expression $p = \frac{\partial L}{\partial \dot{q}}$ depends on $\dot{q}$ so that we can invert it to solve $\dot{q}$ in terms of $q$ and the conjugate momentum $p$. For multiple generalized coordinates $\{q^{a}\}$, the non-degenerate (or non-singular) condition is equivalent to saying that the determinant of the Hessian matrix $M_{ab} = \frac{\partial^{2}L}{\partial\dot{q}^{a}\partial\dot{q}^{b}}$ is non-zero. It is also important to note that constructing the Hamiltonian for the singular and non-singular theories is different. 

Since two pieces of initial value data are required, the phase space is 2-dimensional and consequently, there should be two canonical coordinates, denoted by $Q$ and $P$. These can simply be defined as
\begin{equation}
Q \equiv q, \ P \equiv \frac{\partial L}{\partial \dot{q}}.    
\end{equation}
On the other hand, due to the non-degeneracy condition of Lagrangian, the above phase transformation is invertible and as a result, $\dot{q}$ can be expressed in terms of $Q$ and $P$. This means that there exists a function $v(Q, P)$ such that the following condition is satisfied
\begin{equation}
\frac{\partial L}{\partial\dot{q}}\Big|_{q = Q, \ \dot{q} = v} = P.   
\end{equation}
The canonical Hamiltonian is obtained by doing a Legendre transformation in the following manner
\begin{equation}
H \equiv P\dot{q} - L = Pv(Q, P) - L(Q, v(Q, P)).    
\end{equation}
The canonical evolution equations are given by
\begin{equation}
\begin{split}
\dot{Q} & \equiv \frac{\partial H}{\partial P} = v + P\frac{\partial v}{\partial P} - \frac{\partial L}{\partial\dot{q}}\frac{\partial v}{\partial P} = v\\
\dot{P} & \equiv - \frac{\partial H}{\partial Q} = - P\frac{\partial v}{\partial Q} + \frac{\partial L}{\partial Q} + \frac{\partial L}{\partial\dot{q}}\frac{\partial v}{\partial Q} = \frac{\partial L}{\partial Q}.
\end{split}    
\end{equation}
This clearly shows that the Hamiltonian generates time evolution. Moreover, if the Lagrangian is independent of time explicitly, then $H$ is a conserved quantity, i.e., the energy of the system.

\subsection{Canonical formulation of higher derivative theories}

In this section, we consider a system whose action is given by the following
\begin{equation}
S[q] = \int dt \ L(q, \dot{q},\ldots, q^{(N)}),    
\end{equation}
where $q^{(n)} = \frac{d^{n}q}{dt^{n}}$. The variation of the above action leads to the following relation
\begin{equation}
\delta S[q] = \int dt \Big[\frac{\partial L}{\partial q}\delta q + \frac{\partial L}{\partial\dot{q}}\delta\dot{q} + \frac{\partial L}{\partial\ddot{q}}\delta\ddot{q} + \ldots + \frac{\partial L}{\partial q^{(N)}}\delta q^{(N)}\Big].    
\end{equation}
After doing integration by parts, we may convert the terms of the form
\begin{equation}
\frac{\partial L}{\partial q^{(n)}}\delta q^{(n)},    
\end{equation}
into terms proportional to $\delta q$ and eliminating the surface terms (using the assumption that $\delta q$ vanishes there), we find the generalized Euler-Lagrange equation
\begin{equation}
\sum_{j = 0}^{N}\left(- \frac{d}{dt}\right)^{j}\frac{\partial L}{\partial q^{(j)}} = 0,    
\end{equation}
and this contains the term $q^{(2N)}$. Hence the canonical phase space must contain $N$ coordinates and $N$ conjugate momenta. These are given by Ostrogradski as
\begin{equation}
Q_{i} \equiv \left(\frac{d}{dt}\right)^{i - 1}q, \ P_{i} \equiv \sum_{j = i}^{N}\left( - \frac{d}{dt}\right)^{j - i}\frac{\partial L}{\partial q^{(j)}}, \ i = 1, 2, \ldots, N    
\end{equation}
If the non-degeneracy condition holds, the action’s dependence on $q^{(N)}$ cannot be eliminated by partial integration. Due to non-degeneracy, we can solve $q^{(N)}$ in terms of $P_{N}, q$ and the first $N - 1$ derivatives of $q$. That is, there exists a function
$\mathcal{A}(Q_{1}, Q_{2}, \ldots, Q_{N}, P_{N})$ such that the following condition holds
\begin{equation}
\frac{\partial L}{\partial q^{(N)}}\Big|_{q^{(i - 1)} = Q_{i}, \ q^{(N)} = \mathcal{A}} = P_{N}.    
\end{equation}
Therefore, Ostrogradski’s Hamiltonian takes the form
\begin{equation}\label{Hamiltonian in higher derivative}
\begin{split}
H & = \sum_{i = 1}^{N}P_{i}q^{(i)} - L\\
 & = P_{1}Q_{2} + P_{2}Q_{3} + \ldots + P_{N - 1}Q_{N} + P_{N}\mathcal{A} - L(Q_{1}, \ldots, Q_{N}, \mathcal{A}).
\end{split}
\end{equation}
Like earlier, it can be checked explicitly
\begin{equation}
\dot{Q}_{i} \equiv \frac{\partial H}{\partial P_{i}}, \ \dot{P}_{i} \equiv - \frac{\partial H}{\partial Q_{i}}, \ i = 1,2, \ldots, N .   
\end{equation}
Hence, there is no doubt that Ostrogradski’s Hamiltonian generates time evolution.

The implication of the Hamiltonian (\ref{Hamiltonian in higher derivative}) is that it may not consistently exhibit positive values. This follows from the fact that $H$ is linear in $P_{1}, P_{2}, \ldots, P_{N-1}$ and it might be bounded only for $P_{N}$. Consequently, Ostrogradski's findings suggest that higher derivative theories, wherein higher derivative terms (exceeding first-time derivative in the action) cannot be eliminated via partial integration, are intrinsically unstable due to the Hamiltonian's linear reliance on the conjugate momenta. In practical terms, this signifies that energy can be altered without limitation by traversing varying directions within the $2N$-dimensional phase space. In the literature, this result is widely stated as ``the Hamiltonian is not bounded from below". The fact that energy can take both positive and negative values without any bounds is the main source of difficulty in higher derivative theories. Since no special form for the Lagrangian is assumed in getting (\ref{Hamiltonian in higher derivative}), the generality of this result should be emphasized: the energy is not positive valued for non-degenerate higher derivative theories. Moreover, the above result cannot be changed by any kind of interaction terms or by adjusting the parameters.

Generalization of the above formalism to higher spatial dimensions is trivial. In this case, we have a copy of the above-mentioned formulas for each spatial dimension. The spatial higher dimensional action is given by
\begin{equation}
S = \int dt \ L(x^{\alpha}, \dot{x}^{\alpha}, \ldots, x^{\alpha (N)}), 
\end{equation}
where $x^{\alpha (n)} = \frac{d^{n}x^{\alpha}}{dt^{n}}$ and the index $\alpha$ runs over $1, 2, \ldots, m$ for $m$ spatial dimensions. As a result, the Euler-Lagrange equations in $m$ dimensions is
\begin{equation}
\sum_{j = 0}^{N}\left(- \frac{d}{dt}\right)^{j}\frac{\partial L}{\partial x^{\alpha (j)}} = 0, \ \alpha = 1, 2, \ldots, m.    
\end{equation}
The Ostrogradskian canonical coordinates and conjugate momenta are expressed as
\begin{equation}
\begin{split}
Q_{i}^{\alpha} & \equiv \left(\frac{d}{dt}\right)^{i - 1}x^{\alpha}\\
P_{i}^{\alpha} & \equiv \sum_{j = i}^{N}\left(- \frac{d}{dt}\right)^{j - i}\frac{\partial L}{\partial x^{\alpha (j)}}, \ i = 1, 2, \ldots, N, \ \alpha = 1, 2, \ldots, m.
\end{split}    
\end{equation}
Like earlier, if the non-degeneracy condition is satisfied, there should be a function $\mathcal{A}^{\alpha}(Q_{1}^{\alpha}, Q_{2}^{\alpha}, \ldots, Q_{N}^{\alpha}, P_{N}^{\alpha})$ such that
\begin{equation}
\frac{\partial L}{\partial x^{\alpha (N)}}\Big|_{x^{\alpha (i - 1)} = Q_{i}^{\alpha}, \ x^{\alpha (N)} = \mathcal{A}^{\alpha}} = P_{N}^{\alpha}.    
\end{equation}
Thus, the higher spatial dimensional Ostrogradskian Hamiltonian can be expressed as
\begin{equation}
\begin{split}
H & = \sum_{\alpha = 1}^{m}\sum_{i = 1}^{N}P_{i}^{\alpha}x^{\alpha (i)} - L\\
 & = \sum_{\alpha = 1}^{m}(P_{1}^{\alpha}Q_{2}^{\alpha} + \ldots + P_{N - 1}^{\alpha}Q_{N}^{\alpha} + P_{N}^{\alpha}\mathcal{A}^{\alpha}) - L(Q_{1}^{\alpha}, \ldots, Q_{N}^{\alpha}, \mathcal{A}^{\alpha}),
\end{split}
\end{equation}
where the evolution equations are
\begin{equation}
\dot{Q}_{i}^{\alpha} \equiv \frac{\partial H}{\partial P_{i}^{\alpha}}, \ \dot{P}_{i}^{\alpha} = - \frac{\partial H}{\partial Q_{i}^{\alpha}}, \ i = 1, 2, \ldots, N, \ \alpha = 1, 2, \ldots, m.    
\end{equation}

\subsection{Higher derivative harmonic oscillator}

In this section, we consider an example, namely a higher derivative harmonic oscillator, which was examined by Pais and Uhlenbeck in detail \citep{PhysRev.79.145}. This model is described by the following Lagrangian 
\begin{equation}
L = - \frac{gm}{2\omega^{2}}\ddot{q}^{2} + \frac{m}{2}\dot{q}^{2} - \frac{m\omega^{2}}{2}q^{2},    
\end{equation}
where $m$ is the particle mass, $\omega$ is the frequency and $g$ is a small positive real number that can be considered as a coupling constant. The Euler-Lagrange equation for this second-order Lagrangian is given by
\begin{equation}
\frac{\partial L}{\partial q} - \frac{d}{dt}\frac{\partial L}{\partial\dot{q}} + \frac{d^{2}}{dt^{2}}\frac{\partial L}{\partial\ddot{q}} = 0,    
\end{equation}
which leads to the following equation of motion
\begin{equation}\label{EOM of higher derivative HO}
m\left(\frac{g}{\omega^{2}}q^{(4)} + \ddot{q} + \omega^{2}q\right) = 0.    
\end{equation}
The general solution of the above equation is given by
\begin{equation}\label{general solution}
q(t) = A_{+}\cos(k_{+}t) + B_{+}\sin(k_{+}t) + A_{-}\cos(k_{-}t) + B_{-}\sin(k_{-}t),
\end{equation}
where the two frequencies $k_{\pm}$ are given by
\begin{equation}
k_{\pm} \equiv \omega\sqrt{\frac{1 \mp \sqrt{1 - 4g}}{2g}},    
\end{equation}
where $0 < g < 1/4$. Note that in the limit $g \rightarrow 0$, we obtain $k_{+} = \omega$ (i.e. usual harmonic oscillator) while $k_{-}$ diverges. Also, note that the solution of (\ref{EOM of higher derivative HO}) is no more pure oscillations if $g$ is equal or greater than $1/4$. For example, the case $g = 1/4$ corresponds to the equal frequency case $(\omega_{1} = \omega_{2} = \omega)$ of the Pais-Uhlenbeck oscillator for which the solution includes terms like $\sin(\omega t), \cos(\omega t), t\sin(\omega t)$ and $t\cos(\omega t)$. From the expressions, it is obvious that this kind of solution is not stable since the last two terms have a runaway character.

In the above analysis, $k_{+}$ and $k_{-}$ modes denote positive and negative energy excitations accordingly to which we come later. In terms of the quantities $q_{0}^{(n)} = \frac{d^{n}}{dt^{n}}q(t)\Big|_{t = 0}$, the constants in (\ref{general solution}) can be expressed as
\begin{equation}
\begin{split}
A_{+} & = \frac{k_{-}^{2}q_{0} + \ddot{q}_{0}}{k_{-}^{2} - k_{+}^{2}}, \ B_{+} = \frac{k_{-}^{2}\dot{q}_{0} + q_{0}^{(3)}}{k_{+}(k_{-}^{2} - k_{+}^{2})}\\
A_{-} & = \frac{k_{+}^{2}q_{0} + \ddot{q}_{0}}{k_{+}^{2} - k_{-}^{2}}, \ B = \frac{k_{+}^{2}\dot{q}_{0} + q_{0}^{(3)}}{k_{-}(k_{+}^{2} - k_{-}^{2})}.
\end{split}    
\end{equation}
On the other hand, the conjugate momenta are given by
\begin{equation}
\begin{split}
P_{1} & = m\dot{q} + \frac{gm}{\omega^{2}}q^{(3)} \implies q^{(3)} = \frac{\omega^{2}P_{1} - m\omega^{2}Q_{2}}{gm}\\
P_{2} & = - \frac{gm}{\omega^{2}}\ddot{q} \implies \ddot{q} = - \frac{\omega^{2}P_{2}}{gm},
\end{split}
\end{equation}
where $Q_{2} = \dot{q}$. The Hamiltonian can be written as
\begin{equation}\label{Hamiltonian}
\begin{split}
H & = P_{1}Q_{2} - \frac{\omega^{2}}{2gm}P_{2}^{2} - \frac{m}{2}Q_{2}^{2} + \frac{m\omega^{2}}{2}Q_{1}^{2}\\
 & = \frac{gm}{\omega^{2}}\dot{q}q^{(3)} - \frac{gm}{2\omega^{2}}\ddot{q}^{2} + \frac{m}{2}\dot{q}^{2} + \frac{m\omega^{2}}{2}q^{2}\\
 & = \frac{m}{2}\sqrt{1 - 4g}k_{+}^{2}(A_{+}^{2} + B_{+}^{2}) - \frac{m}{2}\sqrt{1 - 4g}k_{-}^{2}(A_{-}^{2} + B_{-}^{2}).
\end{split}    
\end{equation}
Using Noether’s theorem, one may also check that $H$ is really the conserved quantity corresponding to the energy. Since there is a time translation symmetry in this theory, i.e., the action is invariant under the transformation $t \rightarrow t' = t + \delta t$, there should be an associated conserved quantity which is nothing but the energy of the system. From the last expression in (\ref{Hamiltonian}), it is seen that the ``+" modes carry positive energy whereas the ``-" modes carry negative energy. 

\subsection{Quantization of higher derivative harmonic oscillator}

Let us start by denoting the ``empty" state wavefunction (``vacuum'' state in quantum field theory) by $\Omega(Q_{1}, Q_{2})$, which is the minimum excitation for both negative and positive energy states. We define the positive energy lowering operator as $a$, the positive energy raising operator as $a^{\dagger}$, the negative energy lowering operator as $b$, and the negative energy raising operator as $b^{\dagger}$. Therefore, in order to find the vacuum/empty state wavefunction, one should solve the following equations
\begin{equation}\label{def vacuum}
a\ket{\Omega} = 0, \ b\ket{\Omega} = 0.    
\end{equation}
Since the solution (\ref{general solution}) can be expressed in terms of complex exponentials, the raising and lowering operators can easily be extracted for quantization in the following manner
\begin{equation}\label{re-expressed solution}
q(t) = \frac{1}{2}(A_{+} + iB_{+})e^{- ik_{+}t} + \frac{1}{2}(A_{+} - iB_{+})e^{ik_{+}t} + \frac{1}{2}(A_{-} + iB_{-})e^{- ik_{-}t} + \frac{1}{2}(A_{-} - iB_{-})e^{ik_{-}t}.   
\end{equation}
Now the ladder operators can be constructed explicitly in terms of $A_{+}, A_{-}, B_{+}, B_{-}$, which were just constants in the classical analysis. Since the $k_{+}$ mode carries positive energy, the lowering operator for positive energy excitations must be proportional to the $e^{-i k_{+}t}$ term and one can easily conclude from (\ref{re-expressed solution}) that
\begin{equation}
\begin{split}
a & \propto A_{+} + iB_{+}\\
 & \propto \frac{mk_{+}}{2}(1 + \sqrt{1 - 4g})Q_{1} + iP_{1} - k_{+}P_{2} - \frac{im}{2}(1 - \sqrt{1 - 4g})Q_{2}.
\end{split}    
\end{equation}
Similarly, since the $k_{-}$ mode carries negative energy, its lowering operator must be proportional to the $e^{ik_{-}t}$ term
\begin{equation}
\begin{split}
b & \propto A_{-} - iB_{-}\\
 & \frac{mk_{-}}{2}(1 - \sqrt{1 - 4g})Q_{1} - iP_{1} - k_{-}P_{2} + \frac{im}{2}(1 + \sqrt{1 - 4g})Q_{2}.
\end{split}    
\end{equation}
Now substituting $P_{i} = - i\frac{\partial}{\partial Q_{i}}$, the two coupled equations follow from (\ref{def vacuum}) can be expressed as
\begin{equation}
\begin{split}
\Big[\frac{mk_{+}}{2}(1 + \sqrt{1 - 4g})Q_{1} + \frac{\partial}{\partial Q_{1}} + ik_{+}\frac{\partial}{\partial Q_{2}} - \frac{im}{2}(1 - \sqrt{1 - 4g})Q_{2}\Big]\Omega(Q_{1}, Q_{2}) & = 0\\
\Big[\frac{mk_{-}}{2}(1 - \sqrt{1 - 4g})Q_{1} - \frac{\partial}{\partial Q_{1}} + ik_{-}\frac{\partial}{\partial Q_{2}} + \frac{im}{2}(1 + \sqrt{1 - 4g})Q_{2}\Big]\Omega(Q_{1}, Q_{2}) & = 0.
\end{split}
\end{equation}
and the unique solution of the above partial differential equations is given by
\begin{equation}
\Omega(Q_{1}, Q_{2}) = N\exp\left(- \frac{m\sqrt{1 - 4g}}{2(k_{+} + k_{-})}(k_{+}k_{-}Q_{1}^{2} + Q_{2}^{2}) - i\sqrt{g}mQ_{1}Q_{2}\right). 
\end{equation}

In order to have a sensible quantum theory, we must have a normalizable wavefunction which demands
\begin{equation}
\braket{\Omega|\Omega} < \infty.    
\end{equation}
The above condition holds in this case since the non-oscillating part of the wavefunction $\Omega(Q_{1}, Q_{2})$ is decaying. As a result, any normalized state can be built from the vacuum state $\ket{\Omega}$ by acting with the desired number of $a^{\dagger}$ and $b^{\dagger}$ operators
\begin{equation}
\ket{N_{+}, N_{-}} \equiv \frac{(a^{\dagger})^{N_{+}}}{\sqrt{N_{+}!}}\frac{(b^{\dagger})^{N_{-}}}{\sqrt{N_{-}!}}\ket{\Omega}, 
\end{equation}
where $N_{+}$ and $N_{-}$ label the positive and negative energy states, respectively. The commutation relation between ladder operators is given by
\begin{equation}
[a, a^{\dagger}] = [b, b^{\dagger}] = 1,   
\end{equation}
and the Hamiltonian is expressed as
\begin{equation}
H\ket{N_{+}, N_{-}} = (N_{+}k_{+} - N_{-}k_{-})\ket{N_{+}, N_{-}}.    
\end{equation}
The unboundedness of the Hamiltonian in the quantized non-degenerate higher derivative model stems from the ability of both positive $N_{+}$ and negative $N_{-}$ energy excitations to assume arbitrary values. This observation aligns with classical analysis, as the canonical structure remains consistent across classical and quantum theory. Thus, we have established that the Hamiltonian of a higher derivative theory does not exhibit positive values through a straightforward example. This indicates that interacting higher derivative theories are inherently unstable. The preceding analysis pertains to the classical behavior of the higher derivative harmonic oscillator. Although some individuals might speculate that quantization could rectify the instability stemming from an unbounded Hamiltonian, it is noteworthy that, unlike the case of the Hydrogen atom, instability persists even after quantization.

\section{Derivation of modified Einstein's equations for $f(R)$}
\label{App:f(R)}

The field equations can be obtained by the variation of the above action \textit{w.r.t} the metric $g_{\mu\nu}$. It is more convenient to vary the action \textit{w.r.t} the inverse metric $g^{\mu\nu}$. Taking the variation of the action, we obtain
\begin{equation}
\delta S = (\delta S)_{1} + (\delta S)_{2} + (\delta S)_{3},    
\end{equation}
where
\begin{equation}
\begin{split}
(\delta S)_{1} & = \frac{1}{2\kappa^{2}}\int d^{4}x \delta\sqrt{-g} f(R)\\
(\delta S)_{2} & = \frac{1}{2\kappa^{2}}\int d^{4}x \sqrt{-g} \delta f(R)\\
(\delta S)_{3} & = \int d^{4}x \delta\mathcal{L}_{M}(g_{\mu\nu}, \psi_{M}).
\end{split}    
\end{equation}
Using the following identity 
\begin{equation}
\delta\sqrt{-g} = - \frac{1}{2\sqrt{-g}}\delta g = - \frac{1}{2}\sqrt{-g} g_{\mu\nu}\delta g^{\mu\nu},   
\end{equation}
we may now write the first term $(\delta S)_{1}$ as
\begin{equation}
(\delta S)_{1} = \frac{1}{2\kappa^{2}}\int d^{4}x \sqrt{-g} \left(- \frac{1}{2}g_{\mu\nu}f(R)\right)\delta g^{\mu\nu}.   
\end{equation}
By applying the chain rule in the second term $(\delta S)_{2}$, we obtain
\begin{equation}
\begin{split}
(\delta S)_{2} & = \frac{1}{2\kappa^{2}}\int d^{4}x \sqrt{-g} f_{,R}(R)\delta(g^{\mu\nu}R_{\mu\nu})\\
 & = \frac{1}{2\kappa^{2}}\int d^{4}x \sqrt{-g} f_{,R}(R)(\delta g^{\mu\nu} R_{\mu\nu} + g^{\mu\nu}\delta R_{\mu\nu}),
\end{split}    
\end{equation}
where $f_{,R} \equiv \frac{df}{dR}$ is the first derivative of $f$ \textit{w.r.t} the Ricci scalar $R$. The first term is already expressed in terms of the variation of the inverse metric. For the second term, however, we need to find the variation of the Ricci tensor $\delta R_{\mu\nu}$ in terms of the inverse metric. Recall that the Ricci tensor is obtained by the contraction of the Riemann tensor, where the latter is defined by
\begin{equation}
R_{ \ \mu\lambda\nu}^{\rho} = \partial_{\lambda}\Gamma_{ \ \nu\mu}^{\rho} + \Gamma_{ \ \lambda\sigma}^{\rho}\Gamma_{ \ \nu\mu}^{\sigma} - (\lambda\leftrightarrow \nu).    
\end{equation}
Therefore, we first need to find the variation of the Riemann tensor. This can be computed in a convenient way \textit{w.r.t} the variation of the connection. The variation $\delta\Gamma_{ \ \mu\nu}^{\rho}$ is a difference of two connections, thus is a tensor, and we can compute its covariant derivative as
\begin{equation}
\nabla_{\lambda}(\delta\Gamma_{ \ \mu\nu}^{\rho}) = \partial_{\lambda}\delta\Gamma_{ \ \mu\nu}^{\rho} + \Gamma_{ \ \lambda\sigma}^{\lambda}\delta\Gamma_{ \ \mu\nu}^{\sigma} - \Gamma_{ \ \lambda\nu}^{\sigma}\delta\Gamma_{ \ \mu\sigma}^{\rho} - \Gamma_{ \ \lambda\mu}^{\sigma}\delta\Gamma_{ \ \sigma\nu}^{\rho}.    
\end{equation}
Therefore, we can now write the variation of the Riemann tensor as
\begin{equation}
\delta R_{ \ \mu\lambda\nu}^{\rho} = \nabla_{\lambda}(\delta\Gamma_{ \ \mu\nu}^{\rho}) - \nabla_{\nu}(\delta\Gamma_{ \ \lambda\mu}^{\rho}).   
\end{equation}
By contracting the first and third indices in the above equation, we obtain the variation of the Ricci tensor in terms of the variation of the connection
\begin{equation}
\delta R_{\mu\nu} = \delta R_{ \ \mu\lambda\nu}^{\lambda} = \nabla_{\lambda}(\delta\Gamma_{ \ \mu\nu}^{\lambda}) - \nabla_{\nu}(\delta\Gamma_{ \ \lambda\mu}^{\lambda}).    
\end{equation}
Therefore, the second term $(\delta S)_{2}$ can be expressed in terms of the covariant derivative of the connection as
\begin{equation}
g^{\mu\nu}\delta R_{\mu\nu} = \nabla_{\sigma}(g^{\mu\nu}\delta\Gamma_{ \ \mu\nu}^{\sigma} - g^{\mu\sigma}\delta\Gamma_{ \ \lambda\mu}^{\lambda}).    
\end{equation}
Since we aimed to express all of the variations in terms of variation of the inverse metric, we now take the covariant derivative expressed in the following
\begin{equation}
\Gamma_{ \ \beta\gamma}^{\alpha} = \frac{1}{2}g^{\alpha\lambda}(\partial_{\beta}g_{\gamma\lambda} + \partial_{\gamma}g_{\lambda\beta} - \partial_{\lambda}g_{\beta\gamma}).    
\end{equation}
As a result, we have
\begin{equation}
\delta\Gamma_{ \ \mu\nu}^{\sigma} = - \frac{1}{2}(g_{\lambda\mu}\nabla_{\nu}\delta g^{\lambda\sigma} + g_{\lambda\nu}\nabla_{\mu}\delta g^{\lambda\sigma} - g_{\mu\alpha}g_{\nu\beta}\nabla^{\sigma}\delta g^{\alpha\beta}).    
\end{equation}
Therefore, finally, we obtain the following relation
\begin{equation}
g^{\mu\nu}\delta R_{\mu\nu} = \nabla_{\sigma}(g_{\mu\nu}\nabla^{\sigma}\delta g^{\mu\nu} - \nabla_{\lambda}\delta g^{\sigma\lambda}).   
\end{equation}
By implementing the above, we obtain the following relation
\begin{equation}
(\delta S)_{2} = \frac{1}{2\kappa^{2}}\int d^{4}x \sqrt{-g} f_{,R}(R)[\delta g^{\mu\nu} R_{\mu\nu} + \nabla_{\sigma}(g_{\mu\nu}\nabla^{\sigma}\delta g^{\mu\nu} - \nabla_{\lambda}\delta g^{\sigma\lambda})].    
\end{equation}
By performing two integrations by parts on the last two terms and ignoring the surface terms, we finally obtain the $(\delta S)_{2}$ in terms of the variation of the inverse metric perturbation $g^{\mu\nu}$ as
\begin{equation}
(\delta S)_{2} = \frac{1}{2\kappa^{2}}\int d^{4}x \sqrt{-g} (f_{,R}(R)R_{\mu\nu} + g_{\mu\nu}\Box f_{,R}(R) - \nabla_{\mu}\nabla_{\nu}f_{,R}(R))\delta g^{\mu\nu},    
\end{equation}
where the d'Alembertian is defined as $\Box = g^{\mu\nu}\nabla_{\mu}\nabla_{\nu}$. Putting the computed $(\delta S)_{1}, \ (\delta S)_{2},$ and $(\delta S)_{M}$ together, we can express the total variation of the action as,
\begin{equation}
\delta S = \frac{1}{2\kappa^{2}}\int d^{4}x \sqrt{-g} \left(f_{,R}(R)R_{\mu\nu} - \frac{1}{2}g_{\mu\nu}f(R) + g_{\mu\nu}\Box f_{,R}(R) - \nabla_{\mu}\nabla_{\nu}f_{,R}(R)\right)\delta g^{\mu\nu} + \int d^{4}x \delta\mathcal{L}_{M}(g_{\mu\nu}, \psi_{M})    
\end{equation}
We now take the variation \textit{w.r.t} the inverse metric 
\begin{equation}
- \frac{2}{\sqrt{-g}}\frac{\delta S}{\delta g^{\mu\nu}} = - \frac{1}{\kappa^{2}}\left(f_{,R}R_{\mu\nu} - \frac{1}{2}g_{\mu\nu}f(R) + g_{\mu\nu}\Box f_{,R}(R) - \nabla_{\mu}\nabla_{\nu}f_{,R}(R)\right) + \frac{-2}{\sqrt{-g}}\frac{\delta\mathcal{L}_{M}}{\delta g^{\mu\nu}} = 0.    
\end{equation}
Rearranging the terms above leads to Eq.~\eqref{f(R) field equations}.

\section{Conformal transformation: Einstein and Jordan frames} 
\label{App:ConformalTransf}
%

In this appendix, we discuss the mathematical framework to show that it is possible to write the non-linear action (\ref{f(R) action with matter}) in Ricci scalar $R$ to a more conventional form similar to GR by performing a Weyl transformation which is of the following form
\begin{equation}
\tilde{g}_{\mu\nu} = \Omega^{2}g_{\mu\nu},
\end{equation}
where $\Omega^{2}$ is known as the conformal factor, and the tilde represents quantities in the Weyl-transformed frame.

In order to rewrite the $f(R)$ action in terms of Weyl frame quantities, we express the Ricci scalar of the original frame $R$ in terms of the Ricci scalar of the Weyl frame $\tilde{R}$. Since $\tilde{g}^{\mu\lambda}\tilde{g}_{\lambda\nu} = g^{\mu\lambda}g_{\lambda\nu} = \delta_{\nu}^{\mu}$, we obtain the relation $\tilde{g}^{\mu\nu} = \Omega^{-2}g^{\mu\nu}$. Moreover, given that the determinant of the metric is a multilinear function of its column, we also have the following relation $\tilde{g} = \Omega^{8}g$. Let us now consider the transformation of covariant derivatives. First, we consider that $\nabla_{\mu}$ is the covariant derivative associated with $g_{\mu\nu}$, and $\tilde{\nabla}_{\mu}$ is the covariant derivative associated with $\tilde{g}_{\mu\nu}$. Therefore, it can be shown that $\nabla_{\mu} - \tilde{\nabla}_{\mu}$ defines a tensor of type $(1, 2)$ which we denote by $\mathcal{C}_{ \ \mu\nu}^{\lambda}$. As a consequence, we have the following relation
\begin{equation}
\nabla_{\mu}\omega_{\nu} = \tilde{\nabla}_{\mu}\omega_{\nu} - \mathcal{C}_{ \ \mu\nu}^{\lambda}\omega_{\lambda}.
\end{equation}
Using the metric compatibility condition $\nabla_{\lambda}g_{\mu\nu} = 0$, we get a unique expression for $\mathcal{C}_{ \ \mu\nu}^{\lambda}$
\begin{equation}\label{def C-components}
\mathcal{C}_{ \ \mu\nu}^{\lambda} = \frac{1}{2}g^{\lambda\sigma}(\tilde{\nabla}_{\mu}g_{\nu\sigma} + \tilde{\nabla}_{\nu}g_{\mu\sigma} - \tilde{\nabla}_{\sigma}g_{\mu\nu}).
\end{equation}
On the other hand, since $\tilde{\nabla}_{\lambda}\tilde{g}_{\mu\nu} = 0$, we obtain the following relation
\begin{equation}
\tilde{\nabla}_{\lambda}g_{\mu\nu} = \tilde{\nabla}_{\lambda}(\Omega^{-2}\tilde{g}_{\mu\nu}) = -2\Omega^{-3}\tilde{g}_{\mu\nu}\tilde{\nabla}_{\lambda}\Omega.
\end{equation}
Implementing the above relation into (\ref{def C-components}), we have 
\begin{equation}
\mathcal{C}_{ \ \mu\nu}^{\lambda} = \tilde{g}_{\mu\nu}\tilde{\nabla}^{\lambda}\log\Omega - 2\delta_{(\mu}^{\lambda}\tilde{\nabla}_{\nu)}\log\Omega.
\end{equation}
Now we try to express the Ricci scalar $R$, associated with $\nabla_{\mu}$ in terms of Ricci scalar $\tilde{R}$, associated with $\tilde{\nabla}_{\mu}$. In order to do that first we start by relating their Riemann tensors based on the definition
\begin{equation}
R_{ \ \nu\sigma\mu}^{\rho} = \tilde{R}_{ \ \nu\sigma\mu}^{\rho} - 2\tilde{\nabla}_{[\mu}\mathcal{C}_{ \ \sigma]\nu}^{\rho} + 2\mathcal{C}_{ \ \nu[\mu}^{\lambda}\mathcal{C}_{ \ \sigma]\lambda}^{\rho}.
\end{equation}  
Writing the above relation explicitly, we obtain the following relation
\begin{equation}
\begin{split}
R_{ \ \nu\sigma\mu}^{\rho} & = \tilde{R}_{ \ \nu\sigma\mu}^{\rho} - \Big[2\tilde{g}_{\nu[\sigma}\tilde{\nabla}_{\mu]}\tilde{\nabla}^{\rho}\log\Omega - 2\delta_{[\sigma}^{\rho}\tilde{\nabla}_{\mu]}\tilde{\nabla}_{\nu}\log\Omega + 2\delta_{[\mu}^{\rho}\tilde{\nabla}_{\sigma]}\log\Omega\tilde{\nabla}_{\nu}\log\Omega\\
 & + 2\tilde{g}_{\nu[\sigma}\tilde{\nabla}_{\mu]}\log\Omega\tilde{\nabla}^{\rho}\log\Omega + 2\delta_{[\sigma}^{\rho}\tilde{g}_{\mu]\nu}\tilde{\nabla}^{\lambda}\log\Omega\tilde{\nabla}_{\lambda}\log\Omega\Big].
\end{split}
\end{equation}
Contracting over $\rho$ and $\sigma$ indices, we obtain the following relation between Ricci scalars in two different frames
\begin{equation}
R_{\mu\nu} = \tilde{R}_{\mu\nu} + \tilde{g}_{\mu\nu}\tilde{\Box}\log\Omega + 2\tilde{\nabla}_{\mu}\tilde{\nabla}_{\nu}\log\Omega - 2\tilde{g}_{\mu\nu}\tilde{\nabla}_{\lambda}\log\Omega\tilde{\nabla}^{\lambda}\log\Omega + 2\tilde{\nabla}_{\mu}\log\Omega\tilde{\nabla}_{\nu}\log\Omega.
\end{equation}
Now contracting the above relation with $g^{\mu\nu}$, we have the following relation
\begin{equation}\label{Ricci scalar relation}
R = \Omega^{2}(\tilde{R} + 6\tilde{\Box}\log\Omega - 6\tilde{g}^{\mu\nu}\tilde{\nabla}_{\mu}\log\Omega\tilde{\nabla}_{\nu}\log\Omega).
\end{equation}
Further, we can also compare the geodesics in the original and the Weyl frame. Assuming that the tangent vector $v^{\mu}$ is parallel-transported by an affinely parametrized geodesic \textit{w.r.t} $\nabla_{\mu}$,
\begin{equation}
v^{\mu}\nabla_{\mu}v^{\nu} = 0.
\end{equation}
Using the previous results in the above, we may write
\begin{equation}
\begin{split}
v^{\mu}\tilde{\nabla}_{\mu}v^{\nu} & = v^{\mu}\nabla_{\mu}v^{\nu} + v^{\mu}\mathcal{C}_{ \ \mu\lambda}^{\nu}v^{\lambda}\\
 & = v^{\mu}v^{\lambda}[\tilde{g}_{\mu\lambda}\tilde{\nabla}^{\nu}\log\Omega - \delta_{\mu}^{\nu}\tilde{\nabla}_{\lambda}\log\Omega - \delta_{\lambda}^{\nu}\tilde{\nabla}_{\mu}\log\Omega]\\
 & = (v^{\mu}v^{\lambda}g_{\mu\lambda})g^{\nu\sigma}\nabla_{\sigma}\log\Omega - 2v^{\nu}v^{\lambda}\nabla_{\lambda}\log\Omega.
\end{split}
\end{equation}
From the above relation, we can see that only in the case of null geodesics, the first of the last line in the above expression vanishes and therefore, it shows the geodesics coincide in both frame. However, the last term clearly shows that $\tilde{\nabla}_{\mu}$ geodesics are not affinely parametrized.

It is convenient to rewrite the action (\ref{f(R) action with matter}) in the following form
\begin{equation}
S = \int d^{4}x\sqrt{-g}\left(\frac{1}{2\kappa^{2}}f_{,R}(R)R - \mathcal{U}\right) + \int d^{4}x \ \mathcal{L}_{M}(g_{\mu\nu}, \psi_{M}),
\end{equation}
where
\begin{equation}
\mathcal{U} = \frac{f_{,R}R - f}{2\kappa^{2}}.
\end{equation}
We may now implement the Ricci scalar associated with $g_{\mu\nu}$, from (\ref{Ricci scalar relation}), to express it in terms of the Ricci scalar $\tilde{R}$ associated with $\tilde{g}_{\mu\nu}$ and the Weyl factor $\Omega$
\begin{equation}\label{action after weyl transf}
\begin{split}
S & = \int d^{4}x \sqrt{-\tilde{g}}\Bigg[\frac{1}{2\kappa^{2}}f_{,R}\Omega^{-2}(\tilde{R} + 6\tilde{g}^{\mu\nu}\tilde{\nabla}_{\mu}\tilde{\nabla}_{\nu}\log\Omega - 6\tilde{g}^{\mu\nu}\tilde{\nabla}_{\mu}\log\Omega\tilde{\nabla}_{\nu}\log\Omega)\\
 & - \Omega^{-4}\mathcal{U}\Big] + \int d^{4}x \ \mathcal{L}_{M}(\Omega^{-2}\tilde{g}_{\mu\nu}, \psi_{M}).
\end{split}
\end{equation}
Now by setting the conformal factor $\Omega^{2} = f_{,R}$ and introducing a new scalar field $\kappa\phi = \sqrt{3/2}\log f_{,R}$, we finally obtain an action that is linear in $\tilde{R}$
\begin{equation}\label{Einstein action}
\begin{split}
S & = \int d^{4}x \sqrt{-\tilde{g}}\left(\frac{1}{2\kappa^{2}}\tilde{R} - \frac{1}{2}\tilde{g}^{\mu\nu}\tilde{\nabla}_{\mu}\phi\tilde{\nabla}_{\nu}\phi - V(\phi)\right)\\
 & + \int d^{4}x \mathcal{L}_{M}(f_{R}^{-1}(\phi)\tilde{g}_{\mu\nu}, \psi_{M}),
\end{split}
\end{equation} 
where
\begin{equation}
V(\phi) = \frac{f_{,R}(\phi)R - f}{2\kappa^{2}f_{,R}(\phi)}.
\end{equation}
Here we want to point out the fact that the first integral of equation (\ref{action after weyl transf}) is a total derivative $\tilde{\nabla}_{\mu}\tilde{\nabla}_{\nu}\log\Omega$, and can easily be converted to a surface term through integration by and therefore ignored. From the above action, we note that the non-minimal coupling to the metric is removed. As a result, the Einstein's equations for the conformal metric $\tilde{g}_{\mu\nu}$ takes the conventional form. Therefore, the conformal frame is known as the Einstein frame whereas the original frame with metric $g_{\mu\nu}$ is called the Jordan frame.

Taking the variation of the action (\ref{Einstein action}) \textit{w.r.t} $\phi$, we obtain
\begin{equation}
- \partial_{\mu}\left(\frac{\partial(\sqrt{-\tilde{g}}\mathcal{L}_{\phi})}{\partial(\partial_{\mu}\phi)}\right) + \frac{\partial(\sqrt{-\tilde{g}}\mathcal{L}_{\phi})}{\partial\phi} + \frac{\partial\mathcal{L}_{M}}{\partial\phi} = 0
\end{equation}
which in our case reduces to
\begin{equation}
\tilde{\Box}\phi - V_{,\phi} + \frac{1}{\sqrt{-\tilde{g}}}\frac{\partial\mathcal{L}_{M}}{\partial\phi} = 0,
\end{equation}
where the d'Alembertian operator in the Einstein frame is defined by $\tilde{\Box} = \tilde{g}^{\mu\nu}\tilde{\nabla}_{\mu}\tilde{\nabla}_{\nu}$ and $V_{,\phi}$ is the derivative of the field potential, defined in the action, \textit{w.r.t} the scalar field $\phi$. We can also express the energy-momentum tensor of the matter in the Einstein frame
\begin{equation}
\tilde{T}_{\mu\nu}^{(M)} = - \frac{2}{\sqrt{-\tilde{g}}}\frac{\delta\mathcal{L}_{M}}{\delta\tilde{g}^{\mu\nu}} = \frac{1}{f_{,R}}\left(- \frac{2}{\sqrt{-g}}\frac{\delta\mathcal{L}_{M}}{\delta g^{\mu\nu}}\right) = \frac{T_{\mu\nu}^{(M)}}{f_{,R}}.
\end{equation}
On the other hand, we also obtain the following relation
\begin{equation}
\begin{split}
\frac{\delta\mathcal{L}_{M}}{\delta\phi} & = \frac{\delta\mathcal{L}_{M}}{\delta g^{\mu\nu}}\frac{\delta g^{\mu\nu}}{\delta\phi} = \frac{1}{f_{,R}}\frac{\delta\mathcal{L}_{M}}{\delta g^{\mu\nu}}\frac{\partial(f_{,R})\tilde{g}^{\mu\nu}}{\partial\phi}\\
 & = - \sqrt{-\tilde{g}}\frac{f_{,R\phi}}{2f_{,R}}\tilde{T}_{\mu\nu}^{(M)}\tilde{g}^{\mu\nu} = - \sqrt{-\tilde{g}}\frac{f_{,R\phi}}{2f_{,R}}\tilde{T}.
\end{split}
\end{equation}
In order to quantify the coupling between the field $\phi$ and matter, we define the following quantity
\begin{equation}
Q \equiv - \frac{f_{,R\phi}}{2\kappa f_{,R}}.
\end{equation}
Considering the definition of the field $\phi$ in the $f(R)$ theory, we obtain $Q = - \frac{1}{\sqrt{6}}$. Using this definition for the coupling, the scalar field equation can be re-expressed as
\begin{equation}
\tilde{\Box}\phi - V_{,\phi} + \kappa Q\tilde{T} = 0.
\end{equation} 
From the above equation, it is evident that the field $\phi$ is directly coupled to matter.

Like earlier, the energy-momentum tensor of the scalar field in the Einstein-frame can also be obtained by considering the variation of the scalar field action \textit{w.r.t} the inverse metric $\tilde{g}^{\mu\nu}$
\begin{equation}
\tilde{T}_{\mu\nu}^{(\Phi)} = - \frac{2}{\sqrt{-\tilde{g}}}\frac{\delta(\sqrt{-\tilde{g}}\mathcal{L}_{\phi})}{\delta\tilde{g}^{\mu\nu}} = \partial_{\mu}\phi\partial_{\nu}\phi - \tilde{g}_{\mu\nu}\left(\frac{1}{2}\tilde{g}^{\alpha\beta}\partial_{\alpha}\phi\partial_{\beta}\phi + V(\phi)\right).
\end{equation}
Considering all the previous result, we can now write the metric field equations in Einstein frame which looks the following
\begin{equation}
\tilde{G}_{\mu\nu} = \kappa^{2}(\tilde{T}_{\mu\nu}^{(M)} + \tilde{T}_{\mu\nu}^{(\phi)}),
\end{equation}
where $\tilde{G}_{\mu\nu}$ is the Einstein tensor associated with the conformal metric and is expressed as
\begin{equation}
\tilde{G}_{\mu\nu} = \tilde{R}_{\mu\nu} - \frac{1}{2}\tilde{g}_{\mu\nu}\tilde{R}.
\end{equation}
\section{Gupta-Feynman formalism}
\label{App:GuptaFeynman}

The Gupta-Feynman formalism treats the metric as a regular tensor field, not a geometric construct. By defining $h_{\mu \nu}$ to be the deviation of the metric from the Minkowski metric and treating $h_{\mu \nu}$ as the dynamical variable. One can analyze the classical and quantum aspects of the field $h_{\mu \nu}$. The general form for the Lagrangian for gravity is constructed by summing over all the possible products of derivatives of field tensor~ $h_{\mu \nu}$, by putting arbitrary coefficients in front of term. Since gravitation is a highly nonlinear theory, one expects an action filled with self-interaction in terms of ever-increasing order. However, considering the analogy of Maxwells electromagnetism, which contains the field in second order in Lagrangian, we can write the action up to the second order in $h_{\mu \nu}$, excluding the bounbary terms as:
\begin{equation}\label{FeyGupActon}
S=\int d\tau \left[  a~\partial^{\sigma}h^{\mu \nu}\partial_{\sigma} h_{\mu \nu}+ b~\partial_{\mu}h^{\mu\nu}\partial_{\sigma}h^{\sigma}_{\nu}+
c~\partial_{\nu}h^{\mu \nu}\partial_{\mu}h^{\sigma}_{\sigma}+
d~\partial_{\mu}h^{\nu}_{\nu}\partial^{\mu}~h^{\sigma}_{\sigma}-
\lambda~T^{\mu\nu}h_{\mu \nu} \right]
\end{equation}
where $a$, $b$, $c$, $d$, $\lambda$ are constants and the interaction term is assumed as $-\lambda h_{\mu \nu}T^{\mu \nu}$. To fix the undetermined coefficients; vary the action~\eqref{FeyGupActon} w.r.t $h_{\mu \nu}$. It will give differential equation relating the field derivatives and the source tensor $T_{\mu \nu}$. Demanding that the divergence of the tensor $T_{\mu \nu}$ vanishes, will give the coefficients as;~$a=-d=1/2,b=-c=-1$. Thus, the Lagrangian density for the action~\eqref{FeyGupActon} becomes:
\begin{equation}\label{FeyGupLgrn}
 \mathcal{L} =\frac{1}{2}~\partial^{\sigma}h^{\mu \nu}~\partial_{\sigma} h_{\mu \nu}-\partial_{\mu}h^{\mu\nu}~\partial_{\sigma}h^{\sigma}_{\nu} +
 \partial_{\nu}h^{\mu \nu}~\partial_{\mu}h^{\sigma}_{\sigma}-
 \frac{1}{2}~\partial_{\mu}h^{\nu}_{\nu}~\partial^{\mu}~h^{\sigma}_{\sigma} 
  -\lambda~~T^{\mu\nu}~h_{\mu \nu}.
\end{equation}
The EOM for matter free regions of spacetime is,
\begin{equation}\label{FeyGupLEOM}
G^{(L)}_{\mu\nu}\equiv
\partial^{\sigma}\partial_{\sigma}h_{\mu \nu}-\left(
\partial^{\sigma}\partial_{\nu}h_{\mu \sigma}+
\partial^{\sigma} \partial_{\mu} h_{\nu \sigma}\right)+
\partial_{\mu}\partial_{\nu}h^{\sigma}_{~~\sigma}+
\eta_{\mu\nu} \partial^{\sigma} \partial^{\rho} h_{\sigma\rho}-
\eta_{\mu\nu} \partial^{\rho} \partial_{\rho} h^{\sigma}_{~~\sigma}=0.
\end{equation}
 The sole reason for retaining the stress-energy tensor term is for appealing conservation energy momentum  to fix the undetermined coefficients. Once it is done, we no longer require this term. Implimenting the transformation \eqref{GUPtrans} on \eqref{FeyGupLEOM} we will get,
\begin{equation}\label{FGEOMGUP}
G^{(L)}_{\mu\nu}+\gamma~\mathcal{C}_{\mu \nu}=0
\end{equation}
where $\mathcal{C}_{\mu \nu}$ defined as:
\begin{equation}
\mathcal{C}_{\mu \nu}\equiv
\Box^2 h_{\mu \nu}
- \Box \left( \partial^{\sigma}\partial_{\nu} \right) h_{\mu \sigma} 
- \Box \left( \partial^{\sigma} \partial_{\mu} \right) h_{\nu\rho}
+ \Box \left( \partial_{\mu}\partial_{\nu} \right) h^{\sigma}_{\sigma}
+ \eta_{\mu\nu}\Box \left(  \partial_{\sigma} \partial_{\rho}\right) h^{\sigma\rho}
-\eta_{\mu\nu}\Box^2  h^{\sigma}_{\sigma}.
\end{equation}
\citep{Nenmeli:2021orl} succesfully mapped the \eqref{FGEOMGUP} with the EOM of stelle gravity with suitable choice on parameters $\alpha$ and $\beta$ in \eqref{ch2eqnQGMatr}. It is accumblish by Linearizing the Stelle gravity EOM in \eqref{app:StelleEOM01} about the Minkowski background leads to:
\begin{align}\label{steleMikLinr}
&\frac{G^{(L)}_{\mu\nu}}{\kappa^2}-2\left(\alpha- 2\beta \right)
\left[
	\Box (\partial_{\mu}\partial_{\nu} ) h^{\sigma}_{\sigma}
	-(\partial_{\mu}\partial_{\nu}\partial^{\rho}\partial_{\lambda} )h^{\lambda}_{\rho}
	 \right]+
\left(\alpha-4\beta\right)\eta_{\mu \nu}~\left[ 
								\Box^2 h^{\sigma}_{~\sigma}-
								\Box( \partial_{\lambda}\partial_{\rho})h^{\lambda\rho}  							\right]
	  \nonumber\\&
~~~~~~~+\alpha~\left[  
			\Box^2 h_{\mu \nu}
			+\Box(\partial_{\mu}\partial_{\nu})h^{\sigma}_{\sigma}
 	        -\Box(\partial_{\sigma}\partial_{\nu})h^{\sigma}_{\mu} 
			-\Box (\partial_{\sigma}\partial_{\mu}) h^{\sigma}_{\nu}
\right]=0
\end{align}
In particular, by setting $\alpha=2\beta$ in \eqref{steleMikLinr} will lead:
\begin{align}\label{steleMikLinrSpcase}
&\frac{G^{(L)}_{\mu\nu}}{\kappa^2}+\alpha\left[\Box^2 h_{\mu \nu}+\Box(\partial_{\mu}\partial_{\nu})h^{\sigma}_{\sigma} -\Box(\partial_{\sigma}\partial_{\nu})h^{\sigma}_{\mu} -\Box (\partial_{\sigma}\partial_{\mu}) h^{\sigma}_{\nu}-\eta_{\mu \nu}\left(  \Box^2 h^{\sigma}_{\sigma}-\Box( \partial_{\lambda}\partial_{\rho})h^{\lambda\rho}   \right)  \right]=0
\end{align}
Thus, comparing the Eq.~\eqref{FGEOMGUP} and Eq.~\eqref{steleMikLinrSpcase}, we case see that $\alpha =\gamma/\kappa^2$ case, both equations become similar. Thereby establishes a relationship between the coefficients of stelle gravity coupling constants as $\alpha=\gamma/\kappa^2$, and $\beta=2\gamma/\kappa^2$, As a result, the masses of Yukawa bosons coincide equal to $(2\gamma)^{-1/2}$.

\section{Massive gravity coupled to matter field}
\label{App:MassiveGrav}

The action of linearised massive gravity theory coupled to the source term in the Minkowski background is given by:
\begin{equation}\label{S.1}
S = S_{FP} + \int d^{D}x  \kappa h_{\mu\nu}T^{\mu\nu} \, ,
\end{equation}
where  $S_{FP}$ is the Fierz-Pauli action \eqref{eq. 0.1} and $T^{\mu\nu}$ is the energy-momentum tensor of the source. In order to isolate the DoF, it is helpful to introduce the Stueckelberg fields in a series of steps \citep{Ruegg:2003ps}.  The first step is to demand that the above action preserves the gauge symmetry --- $\delta h_{\mu\nu} = \partial_{\mu}\xi_{\nu} + \partial_{\nu}\xi_{\mu}$ --- introduces the first Stueckelberg field ($V_{\mu}$). Under the following transformation
\begin{equation}\label{S.2}
h_{\mu\nu} \rightarrow h_{\mu\nu} + \partial_{\mu}V_{\nu} + \partial_{\nu}V_{\mu},
\end{equation} 
the action (\ref{S.1}) takes the following form:
\begin{equation}\label{S.3}
\begin{split}
S & = \int d^{D}x \Big[\mathcal{L}_{m = 0} - \frac{1}{2}m^{2}(h_{\mu\nu}h^{\mu\nu} - h^{2})
- \frac{1}{2}m^{2}F_{\mu\nu}F^{\mu\nu}\\
 & - 2m^{2}(h_{\mu\nu}\partial^{\mu}V^{\nu} - h\partial_{\mu}V^{\mu}) + \kappa h_{\mu\nu}
 T^{\mu\nu} - 2\kappa V_{\mu}\partial_{\nu}T^{\mu\nu}\Big],
\end{split}
\end{equation}
where $F_{\mu\nu} = \partial_{\mu}V_{\nu} - \partial_{\nu}V_{\mu}$. 
[Note that under the transformation 
$\mathcal{L}_{m = 0}$ remains invariant; however, other terms in action (\ref{S.1}) are not invariant.] The above action now has a gauge symmetry, which is defined
by
\begin{equation}\label{S.4}
\delta h_{\mu\nu} = \partial_{\mu}\xi_{\nu} + \partial_{\nu}\xi_{\mu}, \ \delta V_{\mu} = - 
\xi_{\mu}.
\end{equation}
At this point, one might think to consider scaling $V_{\mu} \rightarrow V_{\mu}/m$ 
to normalize the vector kinetic term, then take the $m \rightarrow 0$ limit. However, in this 
scenario, the resulting massless graviton and massless photon would yield 4 DoF (in 4 dimensions). Hence, $m \rightarrow 0$ is singular, leading to the loss of one of the original 5 DoF. 

The second step is to introduce another Stueckelberg field $\phi$:
\begin{equation}\label{S.5}
V_{\mu} \rightarrow V_{\mu} + \partial_{\mu}\phi.
\end{equation}
Under this transformation, the action (\ref{S.3}) becomes:
\begin{equation}\label{S.6}
\begin{split}
S & = \int d^{D}x \Big[\mathcal{L}_{m = 0} - \frac{1}{2}m^{2}(h_{\mu\nu}h^{\mu\nu} - h^{2})
- \frac{1}{2}m^{2}F_{\mu\nu}F^{\mu\nu}\\
 & - 2m^{2}(h_{\mu\nu}\partial^{\mu}V^{\nu} - h\partial_{\mu}V^{\mu}) - 2m^{2}(h_{\mu\nu}
 \partial^{\mu}\partial^{\nu}\phi - h\Box\phi)\\
 & + \kappa h_{\mu\nu}T^{\mu\nu} - 2\kappa V_{\mu}\partial_{\nu}T^{\mu\nu} + 2\kappa\phi
 \partial_{\mu}\partial_{\nu}T^{\mu\nu}\Big].
\end{split}
\end{equation}
The above action has two independent gauge symmetries:
\begin{equation}\label{S.7}
\begin{split}
\delta h_{\mu\nu} & = \partial_{\mu}\xi_{\nu} + \partial_{\nu}\xi_{\mu}, \ \delta V_{\mu} = 
- \xi_{\mu}\\
\delta V_{\mu} & = \partial_{\mu}\Lambda, \ \delta\phi = - \Lambda.
\end{split}
\end{equation}
Rescaling $V_{\mu} \rightarrow \frac{1}{m}V_{\mu}, \ \phi\rightarrow\frac{1}{m^{2}}\phi$ in the above action (\ref{S.6}) leads to:
\begin{equation}\label{S.8}
\begin{split}
S & = \int d^{D}x \Big[\mathcal{L}_{m = 0} - \frac{1}{2}m^{2}(h_{\mu\nu}h^{\mu\nu} - h^{2})
- \frac{1}{2}F_{\mu\nu}F^{\mu\nu}\\
 & - 2m(h_{\mu\nu}\partial^{\mu}V^{\nu} - h\partial_{\mu}V^{\mu}) - 2(h_{\mu\nu}
 \partial^{\mu}\partial^{\nu}\phi - h\Box\phi)\\
 & + \kappa h_{\mu\nu}T^{\mu\nu} - \frac{2}{m}\kappa V_{\mu}\partial_{\nu}T^{\mu\nu} + 
 \frac{2}{m^{2}}\kappa\phi\partial_{\mu}\partial_{\nu}T^{\mu\nu}\Big] \, .
\end{split}
\end{equation}
This action is invariant under the gauge transformations:
\begin{equation}\label{S.9}
\begin{split}
\delta h_{\mu\nu} & = \partial_{\mu}\xi_{\nu} + \partial_{\nu}\xi_{\mu}, \ \delta V_{\mu} = 
- m\xi_{\mu}\\
\delta V_{\mu} & = \partial_{\mu}\Lambda, \ \delta\phi = - m\Lambda.
\end{split}
\end{equation}

The third step is to impose conservation of energy-momentum tensor and take the limit $m \rightarrow 0$ in action (\ref{S.8}). This leads to:
\begin{equation}\label{S.10}
\begin{split}
S & = \int d^{D}x \Big[\mathcal{L}_{m = 0} 
- \frac{1}{2}F_{\mu\nu}F^{\mu\nu} - 2(h_{\mu\nu}\partial^{\mu}\partial^{\nu}\phi - h\Box\phi) + \kappa h_{\mu\nu}T^{\mu\nu}
 \Big] \, .
\end{split}
\end{equation}
The above action represents a scalar-tensor-vector theory with massless vector and tensor fields. 
Interestingly, the vector is decoupled, but the scalar is kinetically mixed with the tensor.

The fourth step is to un-mix the scalar and tensor DoF at the expense of the minimal coupling to $T_{\mu\nu}$. The following field redefinition achieves this:
\begin{equation}\label{S.11}
h_{\mu\nu} = h_{\mu\nu}' + \pi\eta_{\mu\nu},
\end{equation}
where $\pi$ is an arbitrary scalar field. The above transformation can be interpreted as a linear version of the conformal transformation. The change in the massless spin-2 part is given by
\begin{equation}\label{S.12}
\mathcal{L}_{m = 0}[h] = \mathcal{L}_{m = 0}[h'] + (D - 2)\Big[\partial_{\mu}\pi\partial^{\mu}h'
- \partial_{\mu}\pi\partial_{\nu}h^{'\mu\nu} + \frac{1}{2}(D - 1)\partial_{\mu}\pi\partial^{\mu}
\pi\Big].
\end{equation} 
This is nothing but the linearisation of the effect of a conformal transformation on the 
Einstein-Hilbert action. Setting $\pi = 
\frac{2}{D - 2}\phi$, the action (\ref{S.10}) reduces to:
\begin{equation}\label{S.13}
S = \int d^{D}x \Big[\mathcal{L}_{m = 0}[h'] - \frac{1}{2}F_{\mu\nu}F^{\mu\nu} - \frac{D - 1}
{D - 2}\partial_{\mu}\phi\partial^{\mu}\phi + \kappa h_{\mu\nu}'T^{\mu\nu} + \frac{2}{D - 2}
\kappa\phi T\Big],
\end{equation}
where $T$ is the trace of the energy-momentum tensor. As a result, it is now easy to see that the 
gauge transformations now look like
\begin{equation}\label{S.14}
\begin{split}
\delta h_{\mu\nu} & = \partial_{\mu}\xi_{\nu} + \partial_{\nu}\xi_{\mu}, \ \delta V_{\mu} = 
0\\
\delta V_{\mu} & = \partial_{\mu}\Lambda, \ \delta\phi = 0.
\end{split}
\end{equation}
It is easy to see that in $D = 4$, the above action has  five DoF ---- two DoF carried by the canonical massless graviton, two DoF carried by the canonical massless vector, and one DoF carried by the canonical massless scalar.
To confirm or infirm that these extra DoF do not introduce ghost degrees of freedom causing instabilities, we return to the action \eqref{S.6}.  

Fifth step is to apply the transformation $h_{\mu\nu} = 
h_{\mu\nu}' + \frac{2}{D - 2}\phi\eta_{\mu\nu}$ in the action \eqref{S.6}. Using the conservation of energy-momentum ($\partial_{\mu} T^{\mu\nu} = 0$), we get:
\begin{equation}\label{S.15}
\begin{split}
S & = \int d^{D}x \Big[\mathcal{L}_{m = 0}[h'] - \frac{1}{2}m^{2}(h_{\mu\nu}'h^{'\mu\nu} - 
h^{'2}) - \frac{1}{2}F_{\mu\nu}F^{\mu\nu} + 2\frac{D - 1}{D - 2}\phi\left(\Box + \frac{D}
{D - 2}m^{2}\right)\phi\\
 & - 2m \left( h_{\mu\nu}'\partial^{\mu}V^{\nu} - h'\partial_{\mu}V^{\mu}
- 2  \frac{D - 1}{D - 2} \phi\partial_{\mu}V^{\mu} \right)  + 
2\frac{D - 1}{D - 2} m^{2}h'\phi + 
\kappa h_{\mu\nu}'T^{\mu\nu} + \frac{2}{D - 2}
 \phi T\Big].
\end{split}
\end{equation}
The above action shows that the scalar, vector, and tensor are all coupled. This translates into the following gauge symmetries of the action: 
%
\begin{equation}\label{S.16}
\begin{split}
\delta h_{\mu\nu}' & = \partial_{\mu}\xi_{\nu} + \partial_{\nu}\xi_{\mu} + \frac{2}{D - 2}m
\Lambda\eta_{\mu\nu}, \ \delta V_{\mu} = - m\xi_{\mu}\\
\delta V_{\mu} & = \partial_{\mu}\Lambda, \ \delta\phi = - m\Lambda.
\end{split}
\end{equation}
The last step is to decouple the scalar, vector, and tensor fields in the action \eqref{S.15} by removing the gauge redundancies on the vector and tensor fields. Imposing the following Lorentz-like gauge conditions
\begin{equation}\label{S.17}
\begin{split}
S_{GF1} & = - \int d^{D}x \left(\partial^{\nu}h_{\mu\nu}' - \frac{1}{2}\partial_{\mu}h'
 + mV_{\mu}\right)^{2}\\
S_{GF2} & = - \int d^{D}x \left(\partial_{\mu}V^{\mu} + m\left(\frac{1}{2}h' + 2\frac{D - 1}
{D - 2}\phi\right)\right)^{2},  
\end{split}
\end{equation}
in the above action, removes the first term in the second line of  Eq.~\eqref{S.15}: 
%
\begin{equation}\label{S.18}
\begin{split}
S + S_{GF1} + S_{GF2} & = \int d^{D}x \Big[\frac{1}{2}h_{\mu\nu}'(\Box - m^{2})h^{'\mu\nu}
- \frac{1}{4}h'(\Box - m^{2})h' + V_{\mu}(\Box - m^{2})V^{\mu}\\
 & + 2\frac{D - 1}{D - 2}\phi(\Box - m^{2})\phi + \kappa h_{\mu\nu}'T^{\mu\nu} + 
 \frac{2}{D - 2}\kappa\phi T\Big].
\end{split}
\end{equation}
From the above action, we can read the propagators of the tensor, vector, and scalar field in momentum space is
\begin{equation}\label{S.19}
\begin{split}
\frac{- i}{p^{2} + m^{2}}\Big[\frac{1}{2}(\eta_{\alpha\sigma}\eta_{\beta\lambda} + 
\eta_{\alpha\lambda}\eta_{\beta\sigma}) - \frac{1}{D - 2}\eta_{\alpha\beta}\eta_{\sigma\lambda}
\Big], \ \frac{1}{2}\frac{-i\eta_{\mu\nu}}{p^{2} + m^{2}}, \ \frac{(D - 2)}{4(D - 1)}
\frac{-i}{p^{2} + m^{2}}.
\end{split}
\end{equation}
This confirms that the $\alpha = 1$ theory has no ghost DoF.

As discussed in the previous two sections, the Fierz-Pauli action ($\alpha = 1$) contains 5 DoF --- two helicity-2 modes, two helicity-1 modes, and one scalar DoF. To explicitly see the degravitation mechanism, we first need to isolate the observable DoF (helicity-2 modes) from the other degrees of freedom in the action \eqref{S.1}. After this, we integrate the extra DoF from the action. 

To do this, we start with the action (\ref{S.3}) that includes one Stuekelberg vector field $A_{\mu}$:
\begin{equation}\label{S.20}
S = \int d^{D}x \Big[\mathcal{L}_{m = 0} - \frac{1}{2}m^{2}(h_{\mu\nu}h^{\mu\nu}
- h^{2}) - \frac{1}{2}m^{2}F_{\mu\nu}F^{\mu\nu} - 2m^{2}(h_{\mu\nu}\partial^{\mu}
V^{\nu} - h\partial_{\mu}V^{\mu}) + \kappa h_{\mu\nu}T^{\mu\nu}\Big].
\end{equation}
In the above action, $h_{\mu\nu}$ is a helicity-2 field containing the two observable DoF, while $V_{\mu}$ contains the other three.  

The second step is to substitute the generalized Lorenz-gauge condition: 
\begin{equation}\label{S.22}
\partial_{\mu}V^{\mu} = \frac{h}{2} - \mathcal{N} \, ,
\end{equation}
in the above action, leading to:
\begin{equation}\label{S.21}
\begin{split}
S & = \int d^{D}x \Bigg[\mathcal{L}_{m = 0} + m^{2}\Big[ - \frac{1}{2}h_{\mu\nu}h^{\mu\nu}
 + \frac{1}{4}h^{2} + V_{\mu}\Box V^{\mu} + \mathcal{N}(h - \mathcal{N})\\
 & - V^{\mu}\left(\partial_{\mu}h - 2\partial^{\nu}h_{\mu\nu} + 2\partial_{\mu}\mathcal{N}
 \right)\Big] + \kappa h_{\mu\nu}T^{\mu\nu}\Bigg].
\end{split}
\end{equation} 
where $\mathcal{N}$ is the auxiliary scalar field. It is easy to check that substituting \eqref{S.22} in the above action leads to the action (\ref{S.20}). This shows that both the actions (\ref{S.20}) and (\ref{S.21}) are equivalent. Varying the above action w.r.t $V_{\mu}$, we get: 
\begin{equation}\label{S.23}
\Box V_{\mu} = \left(\frac{1}{2}\partial_{\mu}h - \partial^{\nu}h_{\mu\nu} + 
\partial_{\mu}\mathcal{N}\right).
\end{equation}
The third step is to rewrite the above expression using the inverse  operator ($\Box^{-1}$):
\begin{equation}\label{S.23a}
V_{\mu} = \frac{1}{\Box}\left(\frac{1}{2}\partial_{\mu}h - \partial^{\nu}h_{\mu\nu} + 
\partial_{\mu}\mathcal{N}\right).
\end{equation}
and substituting in the action (\ref{S.21}). This results in the following on-shell action
\begin{equation}\label{S.24}
S = \int d^{D}x \Big[\frac{1}{2}h_{\mu\nu}\left(1 - \frac{m^{2}}{\Box}\right)\mathcal{E}
^{\mu\nu,\alpha\beta}h_{\alpha\beta} - 2\mathcal{N}\frac{1}{\Box}(\partial_{\mu}\partial
_{\nu}h^{\mu\nu} - \Box h) + \kappa h_{\mu\nu}T^{\mu\nu}\Big],
\end{equation}
where $\mathcal{E}^{\mu\nu,\alpha\beta}$ is the second order differential operator
\begin{equation}\label{S.25}
\mathcal{E}^{\mu\nu,\alpha\beta} = \left(\frac{1}{2}(\eta^{\mu\alpha}\eta^{\nu\beta} + 
\eta^{\mu\beta}\eta^{\nu\alpha}) - \eta^{\mu\nu}\eta^{\alpha\beta}\right)\Box - 2\partial
^{(\mu}\partial^{(\alpha}\eta^{ \ \beta)\nu)} + \eta^{\alpha\beta}\partial^{\mu}
\partial^{\nu} + \eta^{\mu\nu}\partial^{\alpha}\partial^{\beta}.
\end{equation}
The fourth step is to use the following linearized conformal transformation
\begin{equation}\label{S.26}
h_{\mu\nu} = h_{\mu\nu}' + \frac{2}{D - 2}\frac{1}{\Box - m^{2}}\mathcal{N}\eta_{\mu\nu},
\end{equation}
the action (\ref{S.24}) reduces to:
\begin{equation}\label{S.27}
S = \int d^{D}x \Big[\frac{1}{2}h_{\mu\nu}'\left(1 - \frac{m^{2}}{\Box}\right)\mathcal{E}
^{\mu\nu,\alpha\beta}h_{\alpha\beta}' + 2\frac{D - 1}{D - 2}\mathcal{N}\frac{1}{\Box - m^{2}}
\mathcal{N} + \kappa h_{\mu\nu}'T^{\mu\nu} + \frac{2}{D - 2}\frac{1}{\Box - m^{2}}\mathcal{N}
T\Big].
\end{equation}
The last step is to substitute the following redefined auxiliary field variable
\begin{equation}\label{S.28}
\mathcal{N}' = \frac{1}{\Box - m^{2}}\mathcal{N},
\end{equation}
in the action (\ref{S.27}). This leads to:
\begin{equation}\label{S.29}
S = \int d^{D}x \Big[\frac{1}{2}h_{\mu\nu}'\left(1 - \frac{m^{2}}{\Box}\right)\mathcal{E}
^{\mu\nu,\alpha\beta}h_{\alpha\beta}' + 2\frac{D - 1}{D - 2}\mathcal{N}'(\Box - m^{2})
\mathcal{N}' + \kappa h_{\mu\nu}'T^{\mu\nu} + \frac{2}{D - 2}\kappa\mathcal{N}'T\Big].
\end{equation}

It is crucial here to note that the equation of motion of metric perturbations 
$h_{\mu\nu}'$ in Fourier space can be expressed in the following form
\begin{equation}\label{S.29b}
\mathcal{E}^{\mu\nu,\alpha\beta}(k)h_{\alpha\beta}'(k) = \kappa \frac{k^{2}}{k^{2} + m^{2}} 
T^{\mu\nu}(k),
\end{equation}
which shows that the presence of a mass in the theory effectively reduces the strength of the metric perturbations at low energies. The left-hand side of the above equation is the same as GR.
\end{appendices}

\bibliographystyle{Harvard}
\input{References.bbl}
\end{document}